\documentclass[12pt,a4paper,oneside]{article}
\usepackage{graphicx,amssymb,amsmath,color}
\usepackage[linkcolor={blue},citecolor={red},colorlinks=true]{hyperref}
\usepackage{cite}
\usepackage{relsize}
\usepackage{enumerate}
\usepackage[margin=3 cm]{geometry}
\usepackage{slashed}
\usepackage{multirow}
\usepackage[normalem]{ulem}
\usepackage[dvipsnames, x11names]{xcolor}

\def\beq{\begin{equation}}
\def\eeq{\end{equation}}
\def\bea{\begin{eqnarray}}
\def\eea{\end{eqnarray}}

\def\bit{\begin{itemize}}
\def\eit{\end{itemize}}

\def\l{\left}
\def\r{\right}

\def\baa{\begin{array}}
\def\eaa{\end{array}}

\def\sl#1{\mathord{\not\mathrel{{\mathrel{#1}}}}}
\def\d{\partial}

\def\simgt{\mathrel{\lower2.5pt\vbox{\lineskip=0pt\baselineskip=0pt
           \hbox{$>$}\hbox{$\sim$}}}}
\def\simlt{\mathrel{\lower2.5pt\vbox{\lineskip=0pt\baselineskip=0pt
           \hbox{$<$}\hbox{$\sim$}}}}
\newcommand{\vev}[1]{ \langle {#1} \rangle }

\def\bfc{\begin{figure}\begin{center}}
\def\efc{\end{center}\end{figure}}
\def\nn{\nonumber\\}

\definecolor{chromeyellow}{rgb}{1.0, 0.65, 0.0}
\definecolor{darkcoral}{rgb}{0.8, 0.36, 0.27}
\definecolor{cadmiumgreen}{rgb}{0.0, 0.42, 0.24}

\begin{document}

\begin{flushright}
\hspace{3cm} 
SISSA  12/2022/FISI\\
TU-1157
\end{flushright}
\vspace{.6cm}
\begin{center}

\hspace{-0.4cm}{\Large \bf 
Ultra$-$relativistic bubbles from the  simplest Higgs portal and their cosmological consequences\\}

\vspace{1cm}{Aleksandr Azatov$^{a,b,c,1}$, Giulio Barni$^{a,b}$, Sabyasachi Chakraborty$^{a,b,c}$, Miguel Vanvlasselaer$^{a,b,c,2}$, Wen Yin$^{d,3}$,  }
\\[7mm]
 {\it \small

$^a$ SISSA International School for Advanced Studies, Via Bonomea 265, 34136, Trieste, Italy\\[0.15cm]
$^b$ INFN - Sezione di Trieste, Via Bonomea 265, 34136, Trieste, Italy\\[0.1cm]
$^c$ IFPU, Institute for Fundamental Physics of the Universe, Via Beirut 2, 34014 Trieste, Italy\\[0.1cm]
$^d$ Department of Physics, Tohoku University,  
Sendai, Miyagi 980-8578, Japan \\[0.1cm] }
\end{center}

\bigskip \bigskip \bigskip

\centerline{\bf Abstract} 
\begin{quote}

We analyze the phase transitions in the minimal extension of the SM with a real singlet scalar field. The novelty of our study is that we identify and analyze in details the region of parameter space where the first order phase transition can occur and in particular when the bubbles with true vacuum can reach relativistic velocities. This region is interesting since it can lead to the new recently discussed baryogenesis and Dark Matter production mechanisms. We fully analyze different models for the production of Dark Matter and baryogenesis as well as the possibilities of  discovery at the current and future experiments.

\end{quote}

\vfill
\noindent\line(1,0){188}
{\scriptsize{ \\ E-mail:
\texttt{$^1$\href{mailto:aleksandr.azatov@NOSPAMsissa.it}{aleksandr.azatov@sissa.it}},
\texttt{$^2$\href{giulio.barni@NOSPAMsissa.it}{giulio.barni@sissa.it}},
\texttt{$^1$\href{mailto:sachakr@NOSPAMsissa.it}{sachakr@sissa.it}},
\texttt{$^2$\href{miguel.vanvlasselaer@NOSPAMsissa.it}{miguel.vanvlasselaer@sissa.it}},
\texttt{$^3$\href{yin.wen.b3@tohoku.ac.jp}{yin.wen.b3@tohoku.ac.jp}}
}}

\newpage

\newpage

\section{Introduction}

The origin of Dark Matter (DM) and matter antimatter asymmetry are one of the most important unresolved puzzles of the early Universe cosmology. In this paper we will investigate in detail the novel mechanisms proposed in  \cite{Azatov:2021irb,Azatov:2021ifm, Baldes:2021vyz} for DM  production  and baryon asymmetry generation during the first order phase transitions (FOPT).

The FOPT are known to be very useful for constructing  baryogenesis models since the process is out-of-equilibrium as required by one of the Sakharov's conditions~\cite{Sakharov:1967dj}. Applications of this mechanism to the electroweak phase transition lead to the seminal scenarios of electroweak baryogenesis \cite{Kuzmin:1985mm,Shaposhnikov:1986jp}. In the Standard Model (SM), electroweak phase transition is expected to be a smooth crossover  \cite{Kajantie:1996qd,DOnofrio:2015gop}. However, this is not the case for various Beyond Standard Model (BSM) frameworks where successful scenarios of electroweak baryogenesis can be built (see for a recent review \cite{Morrissey:2012db}). The connection between DM and phase transitions is less direct. However, there are class of models where these two phenomenon are strongly related (see for example Ref\cite{Schramm:1984bt,Baker:2016xzo,Baker:2018vos,Heurtier:2019beu,Cohen:2008nb,Bian:2018mkl,Hui:1998dc,Chung:2011hv,Chung:2011it, Hambye:2018qjv,Baldes:2020kam,Shelton:2010ta,Petraki:2011mv,Baldes:2017rcu,Hall:2019rld,Falkowski:2012fb,PhysRevLett.125.151102,Chway:2019kft,Marfatia:2020bcs}).

The focus of this paper will be electroweak FOPT with very relativistic bubbles. In this context, collision between bubbles and particles in the plasma, as was shown recently \cite{Vanvlasselaer:2020niz}, can lead to the production of new states with mass scale significantly larger than the scale of the phase transition\footnote{Note that this effect is different from the mechanism presented in\cite{Falkowski:2012fb,Katz:2016adq} where the heavy fields are produced by the collisions between different bubble walls.}. These heavy states can serve as DM candidates \cite{Azatov:2021ifm}  or their production and subsequent decay \cite{Azatov:2021irb,Baldes:2021vyz}, if accompanied with $C$, $CP$  and baryon number violating interactions can lead to baryon asymmetry generation. The successful realization of this scenario require bubble expansion with ultra-relativistic velocities. Such bubble wall motion is known to be possible and is expected within certain classes of the potentials (\cite{Baratella:2018pxi,DelleRose:2019pgi,Craig:2020jfv,Ellis:2019oqb,Azatov:2020nbe}). However in the 
case of electroweak phase transition, the requirement of ultra relativistic velocities leads to 
very non-trivial constraints on the effective potential. In this  paper we analyze in detail the simplest extension of the SM that can lead 
to a FOPT: real singlet scalar field. There have been numerous studies of phase transition in 
this type of scenario \cite{Espinosa:2007qk,Noble:2007kk,Ashoorioon:2009nf,Damgaard:2013kva,Alanne:2014bra, Huang:2016cjm, Beniwal:2017eik,Barger:2007im,Espinosa:2011ax,Kozaczuk:2019pet,  Wang:2022dkz}. The Lagrangian can be restricted further by considering  $Z_2$ 
\cite{Kurup:2017dzf} or $Z_3$ symmetries\cite{Kang:2017mkl,Niemi:2021qvp}. Interestingly such a real singlet scalar field can appear in 
composite Higgs models \cite{Gripaios:2009pe} (see also \cite{Espinosa:2011eu,DeCurtis:2019rxl} for studies pertaining to phase transition). However, most of the above mentioned works did not analyze in detail the region of parameter space with relativistic bubbles since slow bubble wall velocities are required \cite{Kuzmin:1985mm,Shaposhnikov:1986jp} for the usual electroweak baryogenesis. We fill this gap by analyzing in detail phase transitions in the real singlet extension of SM with an approximate $Z_2$ discrete symmetry, focusing on the parameter space which results in very fast bubble expansions. 
Anticipating the results of this paper we found  that the most promising scenario for relativistic bubbles is the case where the phase transition occurs in two steps: the first  transition is related to the $Z_2$ symmetry breaking while keeping EW symmetry unbroken and in the second one the Higgs field acquires its  vacuum expectation value (vev) which comes together with $Z_2$ symmetry restoration.

This will permit us to evaluate the region of parameter space where Baryogenesis via relativistic bubble walls is viable and to put constraints on the mass scale of the dark sector responsible for $CP$ and $B$ violation. Implications for non-thermal  DM production from bubble wall plasma particle collisions will also be considered.

The  article is organised as follows: in section \ref{sec:potential}, we review the singlet extension of the SM and write the two field potential with thermal corrections. In section \ref{sec:fopt}, we remind the reader of the basics of FOPTs in early Universe and present the computation of the terminal velocity of the bubble wall in the ultra-relativistic regime, for the generic and for SM. In section \ref{sec:PPT}, we qualitatively discuss the different  patterns of phase transitions and argue that only a two-steps PT can yield ultra relativistic bubble wall motion. In section \ref{sec:num_results} we present a numerical study of the two-step PT analyzing in detail the region with fast bubble motion. In section \ref{sec:prod}, we draw the consequences of our previous results for the production of heavy DM and Baryogenesis and in section \ref{sec:GW}, we comment on the GW signal induced by such strong transitions. Finally, in section \ref{sec:conclusion}, we summarize and conclude. \\

\section{Review of the singlet extension of the SM } 
\label{sec:potential}
Let us start by reviewing the effective potential of the  SM with a real scalar field ($s$). The scalar potential will be given by
\bea
V(\mathcal{H}, s) = -\frac{m_h^2}{2}(\mathcal{H}^\dagger \mathcal{H}) +\lambda (\mathcal{H}^\dagger \mathcal{H})^2 - \frac{m_s^2}{4} s^2 +\frac{\lambda_{hs}}{2} s^2 (\mathcal{H}^\dagger \mathcal{H}) +\frac{\lambda_s}{4}s^4\;, 
\label{eq:scalar_pot}
\eea 
where $\mathcal{H}$ is the SM Higgs doublet and $m_h \approx 125$ GeV is the physical mass of the Higgs. For simplicity, we will impose $Z_2$ symmetry on the potential to avoid the terms with odd  powers of ${s}$ field. As a result, when $\vev{s}=0$, we avoid any scalar mixing terms  which are constrained by the recent Higgs signal strength measurements~\cite{CMS:2020xrn}. The Higgs doublet can be decomposed as usual 
\bea 
\mathcal{H}^T = \bigg(G^+, \frac{h+iG^0}{\sqrt{2}}\bigg)\;, 
\eea
where $h$ is the usual Higgs boson getting a vev which is given by $v_{EW}=\sqrt{m_h^2/2\lambda} \approx 246$ GeV if $\vev{s} \equiv v_s=0$. 

\subsection{Coleman-Weinberg potential}
Next, we take into account 1-loop corrections which are encapsulated by the Coleman-Weinberg (CW) potential 
\cite{Weinberg:1973am} 
\bea
V_{CW} = \sum_{i=Z,h,W,t}\frac{n_i(-1)^{F}}{64\pi^2}\left[M_i^4 \bigg(\log \frac{M_i^2}{M^2_{i0}}-\frac{3}{2}\bigg)+ 2 M^2_{i}M^2_{i0}\right]\;.
\label{eq:CW_cur_off_reg}
\eea 
In this expression, $M_i$ stand for the masses depending on the Higgs and singlet fields values, $M_i\equiv M_i(h,s)$, and $M_{i0}$ are the field values from the tree-level vev, $M_{i0}\equiv M_i(v_{EW},0)$. Then one can easily check that this potential corresponds to the following renormalization conditions  
\begin{align}
&V_{\rm eff}=V_0+V_{CW},\nn
&   \frac{dV_{ \rm eff}}{dh}\bigg|_{(h,s)=(v_{EW}, 0)} = 0, \qquad
   \frac{d^2V_{\rm eff}}{dh^2}\bigg|_{(h,s)=(v_{EW}, 0)} = m^2_{h}\;.
   \label{eq:condition}
\end{align}
Eq.(\ref{eq:CW_cur_off_reg}) cannot be used directly for contributions of the Goldstone bosons,  since their masses vanish in the true vacuum and the CW potential is IR divergent. However the solution to this issue was  suggested in Ref.~\cite{Delaunay:2007wb}  which emphasized that the physical Higgs mass is defined at $p^2=m^2_h$ and the effective potential at $p^2=0$ (see for details Ref.~\cite{Delaunay:2007wb}). Thus it would be better to use modified renormalization condition
\bea
\frac{d^2V_{\rm eff}}{dh^2}\bigg|_{(h,s)=(v_{EW}, 0)} = m_{h}^2 -\Sigma\left(p^2=m_{h}^2\right)+\Sigma\left(0\right)\;,
\eea
 where the differences of self-energies are taking into account the running of self-energy from $p^2=0$ to $p^2=m_h^2$. In this case the IR divergences in $\Sigma(0)$ due to the virtual Goldstone bosons are cancelled against the IR divergences of $V_{\rm eff}$. In practice, at the end of this renormalization procedure, the contribution of the Goldstone bosons is given by :
\bea 
V^{GB}_{CW}(h) = \sum_{i=G}\frac{n_G}{64\pi^2}M_{G^{\pm,0}}^4(h) \bigg[\log \frac{M_G^2}{m_h^2}-\frac{3}{2}\bigg]\;. 
\eea 
The number of d.o.f, masses of various particles and the scalar mass matrix as a functions of $h,s$ are given by:
\begin{align}
& \qquad \qquad \qquad \qquad n_{W^{\pm}} = 6, \qquad  n_Z = 3, \qquad  n_{G^{\pm,0}}=3, \qquad n_t = -12\;, \nn 
&   M^2_{W^{\pm}}(h) = \frac{g^2h^2}{4}, \quad M^2_Z(h) = \frac{g^2+g'^2}{4} h^2, \quad M^2_{G^{\pm,0}}(h) = -\frac{m_h^2}{2} + \lambda h^2,\quad M^2_t(h) = \frac{y^2_t h^2}{2} 
\nonumber\\
&M^2(h,s) = \begin{pmatrix}
3\lambda h^2 + \frac{\lambda_{hs}}{2}s^2-\frac{m_h^2}{2} & \lambda_{hs}sh
\\
\lambda_{hs}sh & 3\lambda_s s^2+\frac{\lambda_{hs}}{2} h^2 -\frac{ m_s^2}{2}\;
\end{pmatrix}\;.
\label{eq:mass_matrix}
\end{align}
 
 As a side remark, in the  region where the Higgs $h\to 0$ and $s\sim O(v_{EW})$, there will be two scales involved in the problem: the value of the singlet field $s$ and the masses of the SM particles $M_{W,Z,t}(h\to 0,s) \to 0 \ll s$. This type of two scale potential has been studied in the past \cite{Einhorn:1983fc}, by using two different renormalisation scales. It was concluded that resummation is needed when the $\log\left(M_i(0,s)/s\right)$ is large enough to cancel the loop suppression. Although we have two largely separated scales, we have checked that for our region of the parameter space, such a resummation is not necessary.

\subsection{Finite temperature potential}
The temperature and the density effects  can be taken into account by complementing the zero temperature potential with thermal corrections~(see for example \cite{Curtin:2016urg, Quiros:1999jp}),
 	\bea
 	V(T,M_i) = V^{T=0}_{\rm eff}(M_i)+ V_T(M_i)\;.
 	\label{eq:thermal_pot1}
 	\eea
 In Eq.\eqref{eq:thermal_pot1}, $V^{T=0}_{\rm eff}(M_i)$ is the potential we computed in the previous subsection and the thermal potential $V_T(M_i)$ is given by
 	\begin{align}
 	V_T(M_i(h,s)) &= \sum_{i \in B}\frac{n_i}{2\pi^2}T^4 J_B\bigg(\frac{M_i^2(h,s)}{T^2}\bigg)- \sum_{i \in F}\frac{n_i}{2\pi^2}T^4 J_F\bigg(\frac{M_i^2(h,s)}{T^2}\bigg)\;, 
 	\nn 
 	J_{B/F}(y^2) &= \int \limits_0^{\infty} dx\ x^2\log \Big[1\mp\exp{(-\sqrt{x^2+y^2})}\Big]\;.
 	\label{eq:thermal_pot}
 	\end{align}
 	However, to save computation time, we use the approximate expansion of the function $J_{B/F}(y^2)$ as given in  Ref.~\cite{Curtin:2016urg}:
 	\begin{align}
 	    J_B(y^2)&=\begin{cases}
 	    -\dfrac{\pi^4}{45}+\dfrac{\pi^2}{12}y^2-\dfrac{\pi}{6}y^3-\dfrac{y^4}{32}\log\bigg[\dfrac{y^2}{16\pi^2\exp[3/2-2\gamma]}\bigg]\;, & y^2\ll 1\;, \\
 	    - \mathlarger{\sum}_{n = 1}^{m >3}\dfrac{1}{n^2}y^2K_2(y n)\;, & y^2 \gg 1\;,
 	    \end{cases}\nn
 	    J_F(y^2)&=\begin{cases}
 	    \dfrac{7\pi^4}{360}-\dfrac{\pi^2}{24}y^2-\dfrac{y^4}{32}\log\bigg[\dfrac{y^2}{\pi^2\exp[3/2-2\gamma]}\bigg]\;, & y^2\ll 1\;, \\
 	- \mathlarger{\sum}_{n = 1}^{m >3}\dfrac{(-1)^n}{n^2}y^2K_2(y n)\;, & y^2 \gg 1\;.
 	    \end{cases}
 	    	\label{expansion_thermal}
 	\end{align}
 where $\gamma \approx  0.577$ is the \emph{Euler constant},	and $K_2(z)$ are the second-kind Bessel function. To account for dangerously divergent higher loops due to the Daisy diagrams at finite temperature, we follow the so-called ``Truncated-Full-Dressing" procedure \cite{Curtin:2016urg}. Doing so, the full one-loop potential becomes 
 	\bea
&& 	V(h,s, T)= V_{\rm tree}(h,s)+ 
 	\nn
 	&&\sum_i\bigg[ V_{CW}\bigg(M_i^2(h,s)+\Pi_i(h,T)\bigg) + V_T\bigg(M_i^2(h,s)+\Pi_i(h,T)\bigg)\bigg],
 	\eea
 	where $\Pi_i(T)$ are the thermal masses of various degrees of freedom. In the real singlet extension of the SM\cite{Curtin:2016urg}, the expressions of the thermal masses read 
 
\begin{align} 
&\text{Scalar:} \qquad \Pi_h(T)= T^2\bigg(\frac{3g^2}{16}+\frac{g'^2}{16}+\frac{\lambda}{2}+\frac{y_t^2}{4}+\frac{\lambda_{hs}}{24}\bigg)\;, \qquad
\Pi_s(T) = T^2\bigg(\frac{\lambda_{hs}}{6}+\frac{\lambda_{s}}{4}\bigg)\;,
\\
&\text{Gauge:}  \qquad \Pi^L_{g}(T) = T^2 \text{diag}\bigg(\frac{11}{6}g^2, \frac{11}{6}(g^2+g'^2)\bigg)\;, \qquad
\Pi^T_{g}(T) =0\;, 
\label{eq:thermal_masses}
\end{align}
where $\Pi_g^L(T)$ denote the thermal mass of the \emph{longitudinal} mode of the gauge bosons, while transverse modes $\Pi^T_{g}(T)$ are protected by gauge invariance and thus do not receive a mass at leading order in perturbation theory.

\section{First order phase transitions with relativistic bubbles}
\label{sec:fopt}
Starting with the effective potential derived in the previous section, we can proceed to the analysis of the phase transition. FOPT happens when the minima of the potential corresponding to  different phases are separated by a potential barrier and the transition happens via bubble nucleation. The probability of nucleation of a bubble is given by \cite{Coleman:1977py,Linde:1980tt,Linde:1981zj}
\bea 
\Gamma(T) \simeq \Gamma_3+\Gamma_4= T^4\bigg(\frac{S_3}{2\pi T}\bigg)^{3/2} e^{-S_3(T)/T}+\frac{1}{R_0^4}\l(\frac{S_4}{2\pi}\r)e^{-S_4}\;,
\label{eq:nucleation_rate}
\eea 
where $S_3,S_4$ are $O(3,4)$ bounce actions and $R_0$ is the initial bubble radius. Bubble nucleation is characterised by a critical temperature $T_{\rm crit}$, defined as the point when the two phases of the system have vacua with the same energy. Below $T_{\rm crit}$ phase transition becomes energetically possible. The probability to find a specific point of the Universe to be in the false vacuum is given by~\cite{Ellis:2019oqb,PhysRevLett.44.963.2, osti_6848527}:
\bea 
P_f[T] = \exp [-I(T)]\;, \qquad I(T) \equiv \frac{4\pi}{3} \int^{T_c}_T \frac{dT_1\Gamma(T_1) v_w^3}{T_1^4 H(T_1)} \bigg[\int_T^{T_1} \frac{dT_2}{H(T_2)}\bigg]^3.
\label{eq:nucleation}
\eea 
In Eq.(\ref{eq:nucleation_rate}) the strongest dependence on the temperature comes from the $\Gamma(T)\propto \exp{(-S_3/T)}$, so that the  quantity $I(T)$ is mostly controlled by the ratio $\Gamma\left(T\right)/H^4\left(T\right)$ and an order one fraction of the volume of the Universe will be in the true vacuum when $\Gamma[T]\sim H^4[T]$. The temperature that satisfies this condition is
coined as the nucleation temperature $T_{\rm nuc}$ or, to phrase differently, the nucleation temperature $T_{\rm nuc}$ corresponds to the  appearance of roughly one bubble of true vacuum per Hubble volume. Assuming the scale of the phase transition to be $O(100)$ GeV we get
\bea 
\Gamma\sim H^4,~~~
H(T)^2 = \frac{1}{3M^2_{\rm pl}}\big( \rho_{\rm rad}+ \rho_{\rm vac}+ \rho_{\rm wall}\big), 
 \quad \rho_w \approx 0,\quad  \rho_{\rm rad} = \frac{30 g_\star }{\pi^2}T^4,\\
 \quad \rho_{\rm vac} = \Delta V,\quad \Rightarrow \quad
\frac{S_3}{T_{\rm nuc}} = {3 \over 2} \log \left(\frac{S_3}{2 \pi T_{\rm nuc}}\right) + 4 \log\left(\frac{T_{\rm nuc}}{ H}\right)\;.
\eea 
In case we want to be more precise about the temperature when the phase transition completes, we can define the so-called \emph{percolation temperature} $T_{\rm per}$ as the temperature when around $\sim$30 \% of the space has been converted to the true phase
\bea 
I(T \equiv T_{\rm per}) = 0.34\;, \qquad \text{(percolation temperature)}.
\label{eq:condi_per}
\eea 
If the condition in Eq.\eqref{eq:condi_per} is not fulfilled, $I(T)<0.34$, then the bubbles do not percolate. 

At last the energy released during the PT is traditionally  quantified by the ratio between the energy stored in the vacuum and in the plasma at the moment of the transition 
\bea 
\alpha \equiv \frac{\Delta V}{\rho_{\rm rad}} \qquad \text{(strength parameter)}.
\label{eq:alpha}
\eea 

\subsection{Velocity of the EW bubble} 
\label{app:dynamics}
Let us proceed to the computation of the bubble wall velocity $v_w$. The dynamics of the bubble wall motion is controlled by the driving force due to the potential differences between the false and true vacuum
\bea
\Delta V\equiv V_{\rm false}-V_{\rm true}\;,
\eea
and the friction due to finite temperature effects. The calculation of the friction is generically a very complicated problem, however for the ultra-relativistic bubble motion at leading order (LO - tree level) very simple expressions have been obtained\cite{Bodeker:2009qy, Dine:1992wr} for the pressure force from friction,
\bea
\Delta \mathcal{P}_{\text{LO}} \simeq \sum_i g_i c_i \frac{\Delta M_i^2}{24}T_{\text{nuc}}^2\;,
\label{eq:LOpresLO}
\eea
where $\Delta M_i^2$ is the change in the mass of the particle $i$ during the PT, $c_i=1\left(1/2\right)$ for bosons (fermions) and $g_i$ is the number of d.o.f of the incoming particle. Eq.\eqref{eq:LOpresLO} assumes that the masses of the particles outside of the bubble are less than the temperature, otherwise the friction will have an additional Boltzmann suppression $\propto \exp[-M_{\rm false}/T]$.
Interestingly the production of the heavy particles can also contribute to the friction at the same order \cite{Vanvlasselaer:2020niz}
\bea
\Delta \mathcal{P}^{\rm mixing}_{\text{LO}} \propto 
v^2 T^2 \Theta(\gamma_w T_{\rm nuc}- M^2_{ \rm heavy}/v)\;,
\label{eq:pressure-mix}
\eea
where $v$ is the vev of the Higgs field and we review the heavy particle production later in the section \ref{sec:heavypart-prod}.
One can see that the friction (pressure from plasma on the bubble wall) becomes velocity independent so that permanent accelerating (runaway) behavior of the bubble expansion becomes possible. However for theories where the gauge bosons receive a mass during the phase transition, this is not the case and the effect with multiple gauge boson emissions,  leads to the additional contribution (NLO- Next to Leading order) to the pressure which scales as \cite{Bodeker:2017cim,Vanvlasselaer:2020niz,Gouttenoire:2021kjv,Baldes:2020kam}\footnote{There is a claim that this friction scales as $\gamma^2 T^4$\cite{Hoeche:2020rsg} see \cite{Vanvlasselaer:2020niz,Gouttenoire:2021kjv} for criticism.}
\bea
\Delta\mathcal{P}_{\text{NLO}} &\propto & \sum_{a} g_a g_{gauge}^3\gamma_w T_{\text{nuc}}^3 v \;.
\label{eq:LOpresNLO}
\eea
At this point we can see that bubbles will keep accelerating till the moment when NLO friction becomes large enough to balance the driving force
\bea
\Delta V=\Delta {\cal P}_{\rm LO}+\Delta {\cal P}^{\rm mixing}_{\rm LO}
+\Delta {\cal P}_{\rm NLO},
\eea
which will set the $\gamma_w^{\rm terminal}$ the bubble can reach,
\bea
 \gamma_{w}^{\text{terminal}} \sim \l(\frac{\Delta V- \Delta{\cal P}_{\rm LO}-\Delta {\cal P}^{\rm mixing}_{\rm LO}}{T_{\rm nuc}^3  v}\r).
\eea
Note that we can be as well in the situation where the percolation (bubble collision) starts before the terminal $\gamma_w^{\rm terminal}$ is reached.

\subsection{Friction forces during the electroweak phase transitions}
\label{sec:friction}
Let  us apply the discussion of the previous section to the EW phase transition (EWPT). Our main interest will be  the possibility of heavy particle production which can later source  baryogenesis and DM production  following the ideas in \cite{Vanvlasselaer:2020niz,Azatov:2021ifm,Azatov:2021irb,Baldes:2021vyz}. The heavy particle production (see also the discussion in section~\ref{sec:heavypart-prod}) happens due to the collision between the plasma particles and the bubble wall. The typical center of mass energy for such process will be roughly $\sim\sqrt{\gamma_w T v_{EW}}$, thus $\gamma_w$  will be controlling the maximal mass of the new fields that can be produced. To calculate $\gamma_w$ in the context of the EWPT, we need to know the forces acting on the bubble wall.

The LO friction from the top, $Z$, and $W$  takes the form: 
\bea 
\Delta \mathcal{P}^{SM}_{\rm LO} \approx T_{\text{nuc}}^2v_{EW}^2 \bigg(\frac{y_t^2 }{8}+ \frac{g^2+g'^2}{32}+ \frac{g^2}{16}\bigg) \approx 0.17\; T_{\text{nuc}}^2v_{EW}^2.
\label{eq:LOpresSM}
\eea 
There is also a contribution to the LO friction from the singlet and Higgs scalars, however it depends on the masses of these fields in the false vacuum and we find it to be numerically subleading compared to the estimate  in Eq.\eqref{eq:LOpresSM}\footnote{These contributions are smaller due to the number of d.o.f. and possible Boltzmann suppression factors $\propto \exp[-M_{\rm false}/T].$}. In our numerical calculation we took this additional contribution into account, however, for the current discussion it is sufficient  to use Eq.\eqref{eq:LOpresSM}. This gives a rough condition on the nucleation temperature for the transition to become ultra-relativistic
\bea 
\Delta V > 0.17\; T_{\text{nuc}}^2v_{EW}^2 \qquad \text{(relativistic wall condition)}.
\label{eq:relat_cond}
\eea 
The computation of the NLO friction in the SM has been carried out in \cite{Gouttenoire:2021kjv} and the following approximate expression has been derived: 
\begin{align}
\Delta \mathcal{P}^{SM}_{\rm NLO}  
\approx  \bigg[\sum_{abc}\nu_a g_a \beta_c C_{abc}\bigg]\frac{\kappa \zeta(3)}{\pi^3}\alpha  M_Z(v_{EW}) \gamma_w \log \frac{M_Z(v_{EW})}{\mu}  T_{\text{nuc}}^3\;,
\label{eq:LLpressure}
\end{align}
where $\nu_a= 1 (3/4)$ for $a$ a boson (fermion), $g_a$ is the number of degrees of freedom of $a$ and $M_{g,i}(v_{EW})$ is the mass of the gauge boson inside the bubble. $C_{abc}$ represents the coupling which appears in the vertex (see appendix  \ref{sec:sum}), $\beta_{c=Z^0} = 1$, and $\beta_{c=W^{\pm}}=\cos \theta_W= M_W/M_Z$,  $\alpha = e^2(v_{EW})/4\pi \sim 1/128$ is the electromagnetic fine structure constant.  The renormalization scale $\mu$ has to be understood as the lower cut-off for the integration over soft momentum, typically the thermal mass $\mu \propto \alpha_i^{1/2}T_{\text{nuc}}$\cite{Vanvlasselaer:2020niz}. The $\kappa$  factor  is introduced  \cite{Gouttenoire:2021kjv} to account for the contributions of the both reflected and transmitted bosons and is approximately equal to $\kappa \sim 4$. The sum in the square brackets is approximately equal to  
\bea 
\bigg[\sum_{abc}\nu_a g_a \beta_c C_{abc}\bigg] \approx 157\;,
\label{eq:sum_final_result}
\eea 
see appendix \ref{sec:sum} for details. At this point we can compute the terminal wall velocity by balancing the pressure against the driving force
\bea
&&\Delta V - \Delta \mathcal{P}^{SM}_{\rm LO} = \Delta \mathcal{P}^{SM}_{\rm NLO} (\gamma_w=\gamma^{\rm terminal}_w)\quad \text{(Terminal velocity criterion)}\\
\nn
&&\Rightarrow
\gamma^{\rm terminal }\sim 6 \times \l(\frac{\Delta V-\Delta\mathcal{P}_{\rm LO}^{ SM}}{(100 {\rm~ GeV})^4}\r)\l(\frac{100 {\rm GeV}}{T_{\rm nuc}}\r)^3\frac{1}{\log \frac{M_z}{g T}}.
\label{eq:terminal_velo}
\eea
Taking into account that $\left(\Delta V-\Delta{\cal P}_{\rm LO}\right)/(100 {\rm~ GeV})^4\lesssim \mathcal{O}(1)$, we can see that the bubbles will become ultra-relativistic, \textit{i.e.} $\gamma^{\rm terminal }\gg 1$ only if $T_{\rm nuc}$ is significantly lower than the scale of the phase transition $\sim 100$ GeV.

\section{Phase transition in the singlet extension}
\label{sec:PPT}
After the preparatory discussion in the previous sections \ref{sec:potential} and \ref{sec:fopt}, we can proceed to the analysis of phase transition in the model with the singlet field, Eq.(\ref{eq:scalar_pot}). Our analysis will be focused on the region of parameter space with relativistic bubble expansion, for other studies of phase transition in SM plus $Z_2$ real singlet scalar, see Refs.~\cite{Cline:2012hg,Kurup:2017dzf,Curtin:2014jma,Vaskonen:2016yiu}.

In the previous section \ref{sec:fopt}, we have seen that the velocity of the bubble expansion  is fixed by the balance between the friction from the plasma and the driving force. At low temperatures the friction is suppressed  (see Eq.(\ref{eq:LOpresSM})-(\ref{eq:LLpressure})) so that we expect the bubbles to become relativistic (large Lorentz $\gamma^{\rm terminal}_w$ factor). 

Let us  check whether low nucleation temperatures are feasible in the singlet extension. We will assume that $Z_2$ remains unbroken in the true vacuum in order to avoid to constraints from the Higgs-scalar mixing (see for example \cite{Buttazzo:2015bka}). Then in the model with $Z_2$ odd singlet, the phase transition can occur in two ways:  one-step $(\vev{h}=0, \vev{s}=0)\to (v_{EW},0)$ and two-steps\footnote{We will see later that at temperatures $T\ll T_c$ Coleman-Weinberg potential  can shift a little bit the false vacuum position to  $(\delta v_h,v_s)$ where $\delta v_h \ll v_s, v_h$.} $(\vev{h}=0, \vev{s}=0)\to (0,\vev{s}|_{\neq 0}) \to(v_{EW},0)$ \cite{Espinosa:2011ax,Curtin:2014jma,Kurup:2017dzf}, and each of these phase transitions can be first or the second order. We review both of these scenarios of phase transitions in order to understand in what case it is possible to obtain relativistic bubbles.

\subsection{ One-step phase transition:}
This case has been largely studied in the literature \cite{Curtin:2014jma,Kurup:2017dzf} and we will not provide a complete description of it. In this scenario the singlet never gets a vev, and all of its effect reduce to the additional contributions to the Higgs potential from Coleman-Weinberg terms  and thermal corrections. However it turns out that relativistic bubbles are very unlikely for such phase transitions (see also results in \cite{Kurup:2017dzf}). In the limit when $|m_s|\ll T_c $ we can show analytically that this is indeed the case. Near the origin $h\to 0$,  the potential in $h$ direction is  dominated by the $\propto h^2$ terms
\bea 
V_{\text{dominant}}(h\to 0, s=0, T) &=& \frac{m_{\rm eff}^2(T)}{2}h^2+\dots
\eea 
The effective mass $m_{\rm eff}^2$ include the tree-level terms and the leading thermal contributions
and is approximately equal to 
\bea 
m_{\rm eff}^2(T) \simeq -m_h^2+ T^2\bigg(\frac{m_h^2}{4 v_{EW}^2}+\frac{ y_t^2}{4}+ \frac{g^2+g'^2}{16}+ \frac{g^2}{8}+\frac{\lambda_{hs}}{24}\bigg)>0\;.
\label{Eq:mass}
\eea 
The FOPT can happen only if $m_{\rm eff}^2(T)>0$.
Then assuming perturbative values of the coupling $\lambda_{hs}$  we can estimate the lowest  temperature where the FOPT can occur  to be $T^{nuc}_{\text{min}}\gtrsim 100$ GeV.  Comparing this value with the discussion in the section \ref{sec:friction} we can see that the bubble  wall velocities will always satisfy $\gamma_w\lesssim  10 $. As mentioned before, we are interested in the expansions with much larger $\gamma_w$ factors, so that we do not discuss one step phase transition further.

\subsection{Two step FOPT with relativistic bubbles}
\label{sec:qualitative}
Two-steps realisations of the EWPT have already been studied in many works, see for example\cite{Profumo:2007wc, Espinosa:2011ax,  Patel:2012pi,Curtin:2014jma, Huang:2014ifa,Jiang:2015cwa, Kurup:2017dzf, Chiang:2017nmu}. The novelty of our study is that we will be focusing on the parameter space with relativistic bubbles which was previously ignored.  We organize the discussion as follows: In section \ref{sec:qualitative}, we show qualitative results based on approximate treatment of the potential and then in the section \ref{sec:num_results} we present the exact numerical results obtained with our code. 
The two step  phase transition 
\bea
(0,0)\to (0,\vev{s}) \to (v_{EW},0)\;,  
\eea
can happen if the $m_s^2$ parameter of Eq.(\ref{eq:scalar_pot}) is positive. In this case it is convenient to parameterize the Lagrangian in the following way
\bea
V_{\rm tree}(h, s)
= - \frac{m_h^2}{4}h^2 +\frac{m_h^2}{8v_{EW}^2}h^4  -\frac{m_s^2}{4}s^2 +\frac{\lambda_{hs}}{4} s^2 h^2 +\frac{m_s^2}{8v_s^2}s^4\;.
\label{eq:potential_1}
\eea
where $v_{EW}= \sqrt{m_h^2/2\lambda}$ GeV and $v_s=\sqrt{m_s^2/2\lambda_s}$ correspond to the local minima at $(\vev{ h}=v_{EW},{\vev{s}=0)}$ and  $(\vev{h}=0,{\vev{s}=v_s)}$ respectively. 
The origin of two-step PT can be intuitively understood from the 
following considerations. For simplicity let us ignore the Coleman-Weinberg potential and restrict the discussion  by considering only the thermal 
masses. Then the potential will be given by 
 \bea
 \label{eq:pot-app}
 V({{\cal H}, s})&\approx& V_{\rm tree}({\cal H}, s)+\frac{T^2}{24}\l[\sum_{bosons} n_i M_i^2({\cal H}, s))+\frac{1}{2} \sum_{fermions}n_F M_F^2({\cal H}, s)\r]\;,\nonumber\\
&=& V_{\rm tree}({\cal H}, s)
\nn 
&+&T^2\l[h^2\l(\frac{g'^2}{32}+\frac{3 g^2}{32}+\frac{m_h^2}{8 v_{EW}^2}+\frac{y_t^2}{8}+\frac{\lambda_{hs}}{48}\r)+s^2\l(\frac{m_s^2}{16 v_s^2}+\frac{\lambda_{hs}}{12}\r)\r].\nonumber\\
 \eea
 From this expression we can clearly see that the temperatures when the minima with non zero vevs appear for the Higgs and singlet fields can be different. Then it can happen that the $Z_2$ breaking phase transition occurs before the EW one. This means that there will be first a phase transition from $(0,0)\to (0, v_s)$. After this phase transition the Universe keeps cooling down and the minimum with $\langle h \rangle\neq 0$ will be generated. Choosing the appropriate values of  masses and  couplings we can make sure that the {minimum} with $(v_{EW},0)$ is the true {minimum} of the system. The transition $(0,v_s) \rightarrow (v_{EW},0)$ will be of the first order if there is a potential barrier in the between two minima. One of the necessary condition in this case will be $\d^2 V/\d h^2|_{s=v_s,h\to 0}>0$, which using Eq.\eqref{eq:pot-app} we get
\bea
\label{eq:cond-min}
\frac{-m_h^2}{4}+\frac{\lambda_{hs} v_s^2}{4}+T^2\l(\frac{g'^2}{32}+\frac{3 g^2}{32}+\frac{m_h^2}{8 v^2}+\frac{y_t^2}{8}+\frac{\lambda_{hs}}{48}\r)>0 \;.
\eea
From this discussion  we can see that there are qualitatively two different cases depending on whether the potential barrier between two minima remains or disappears at zero temperature, \textit{i.e.} when $m_h^2 \gtrless \lambda_{hs} v_s^2$. In the first case the phase transition will necessarily happen before the ``No Barrier"(NB) temperature
\bea
T^{\rm NB}=\sqrt{\frac{m_h^2-\lambda_{hs} v_s^2}{\frac{g'^2}{8}+\frac{3 g^2}{8}+\frac{m_h^2}{2 v^2}+\frac{y_t^2}{2}+\frac{\lambda_{hs}}{12}}} \sim  \frac{v_s\sqrt{-\lambda_{hs}+\big(\frac{m_h}{v_s}\big)^2}}{0.8}\;,
\eea
since the bounce action drops  once the potential barrier between two minima reduces.

In the other case, when the barrier remains even for zero temperature potential, the phase  transition is obviously of the first order. However  it might happen that the tunneling rate is too slow  and the system  remains stuck in the false vacuum.

From  Eq.\eqref{eq:cond-min}
we can see that size of the potential  barrier between two minima will be controlled by the coupling $\lambda_{hs}$. Increasing this parameter will enhance the potential barrier and will lead to the reduction of $T^{\rm NB}$ so that $(v_s,0)$ remains a local {minimum} even at zero temperature. At some point we expect the potential barrier to become so large that the system will remain trapped in the false vacuum forever. From this discussion we can expect  that the lowest temperatures will be achieved at the boundary of  the region where no PT can occur. The lowest temperature will correspond to the (global) minimum of  $S_3/T$ controlling the tunneling rate, as explained later in section \ref{sec:num_results}. If this minimum happens at temperatures much lower than $T_{\rm crit}$, then there will be super-cooling ($T_{\rm nuc}\ll T_{\rm crit}$).
Even though this discussion was made by considering only the thermal masses in the potential, numerically we find that  for the effective potential with truncated full dressing \cite{Curtin:2016urg} the qualitative behaviour does not change, and only explicit conditions for $m_h^2$ and $T^{\rm nuc}_{\rm min}$ are modified.

At last we always check whether the condition for the EWSB minimum to be the global minimum at zero temperature is satisfied:
\bea 
m_s^2 v_s^2 < m_h^2 v_{EW}^2 \qquad \text{(EWSB is global minimum)}\;,
\eea 
where above equation is valid up to small loop-level corrections.

\section{Numerical results}
\label{sec:num_results}

\begin{figure}
\centering
  \includegraphics[scale=.555]{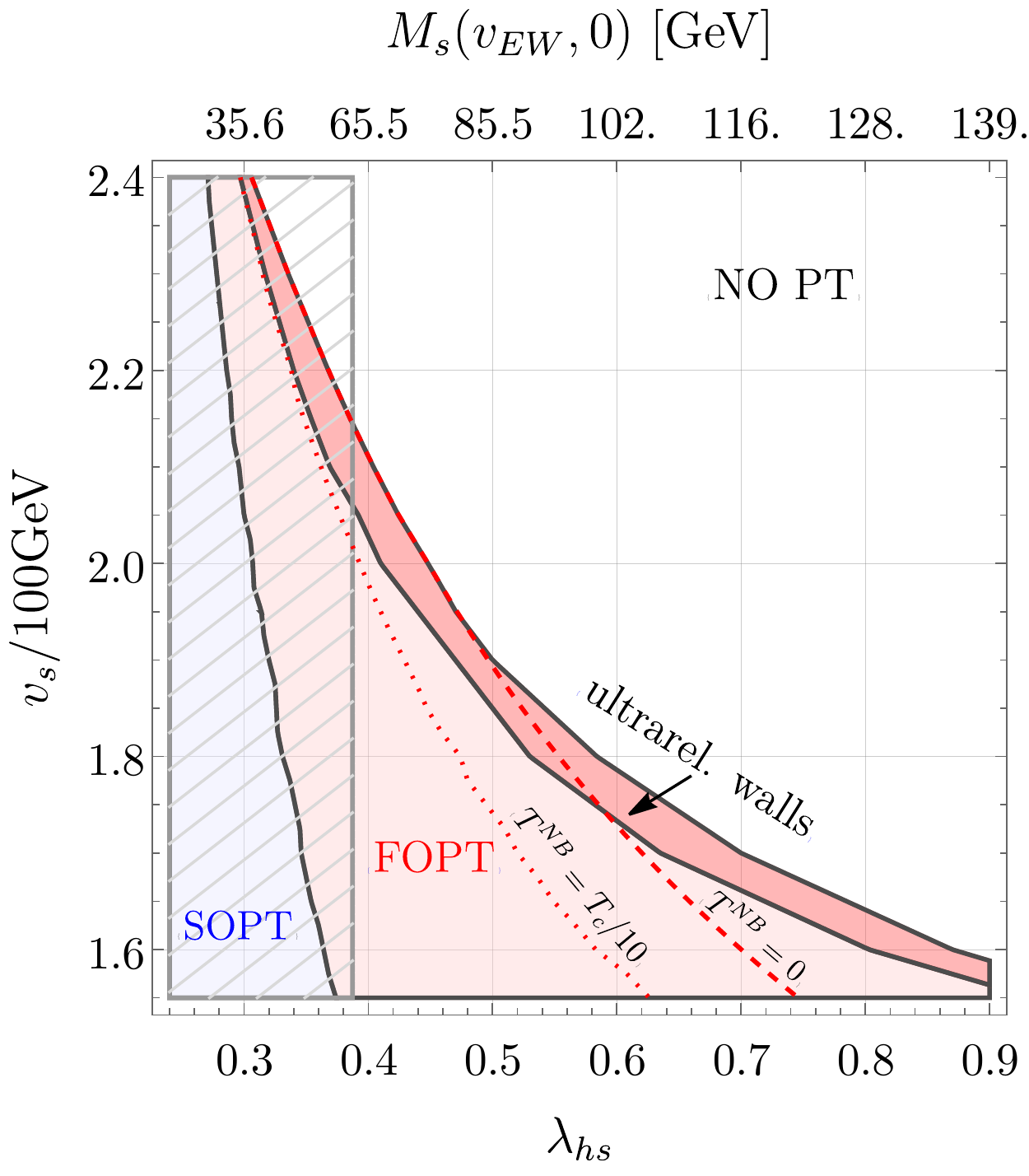}
  \includegraphics[scale=0.8]{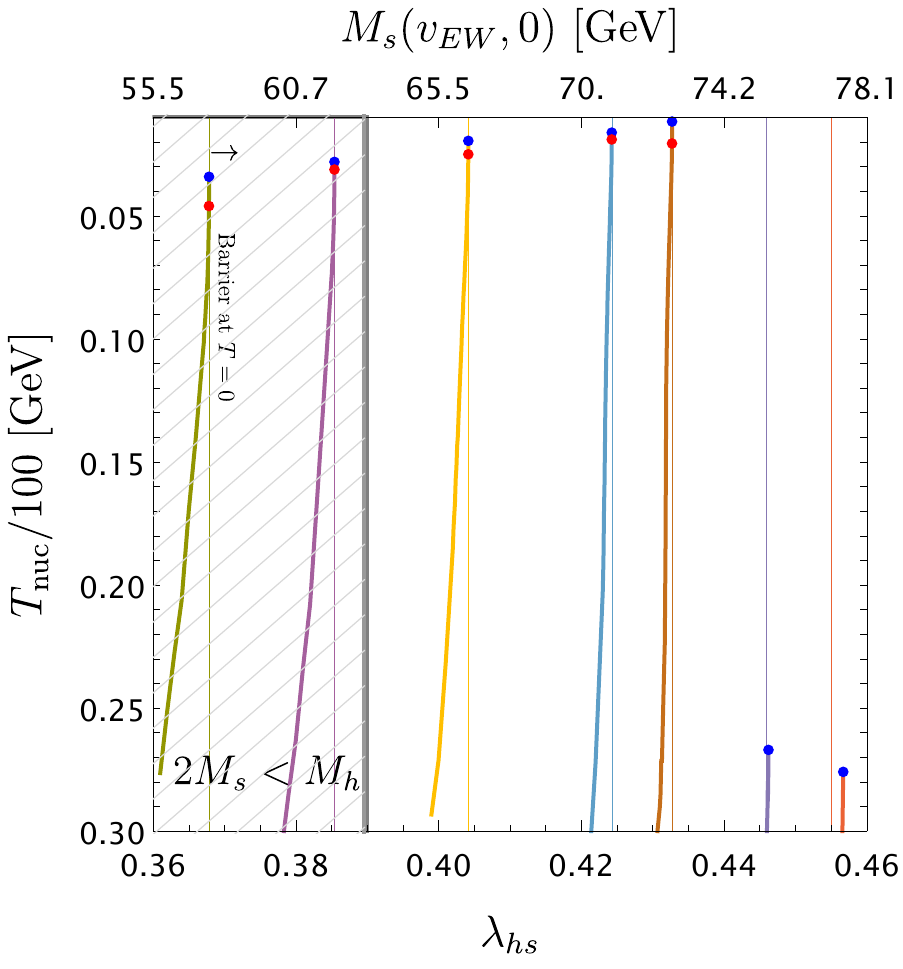}\\
  \includegraphics[scale=.58]{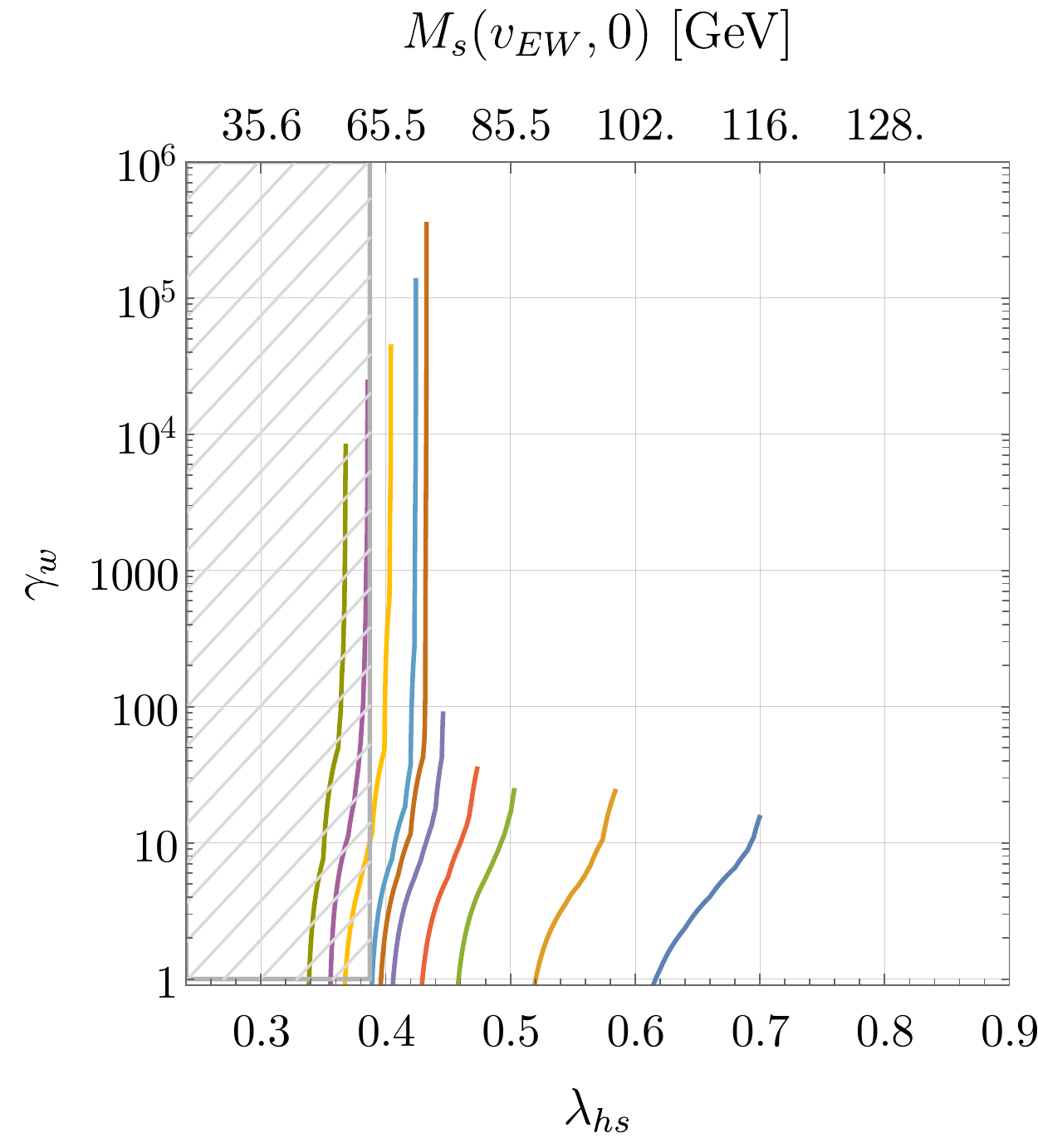}$\ \ $
  \includegraphics[scale=.56]{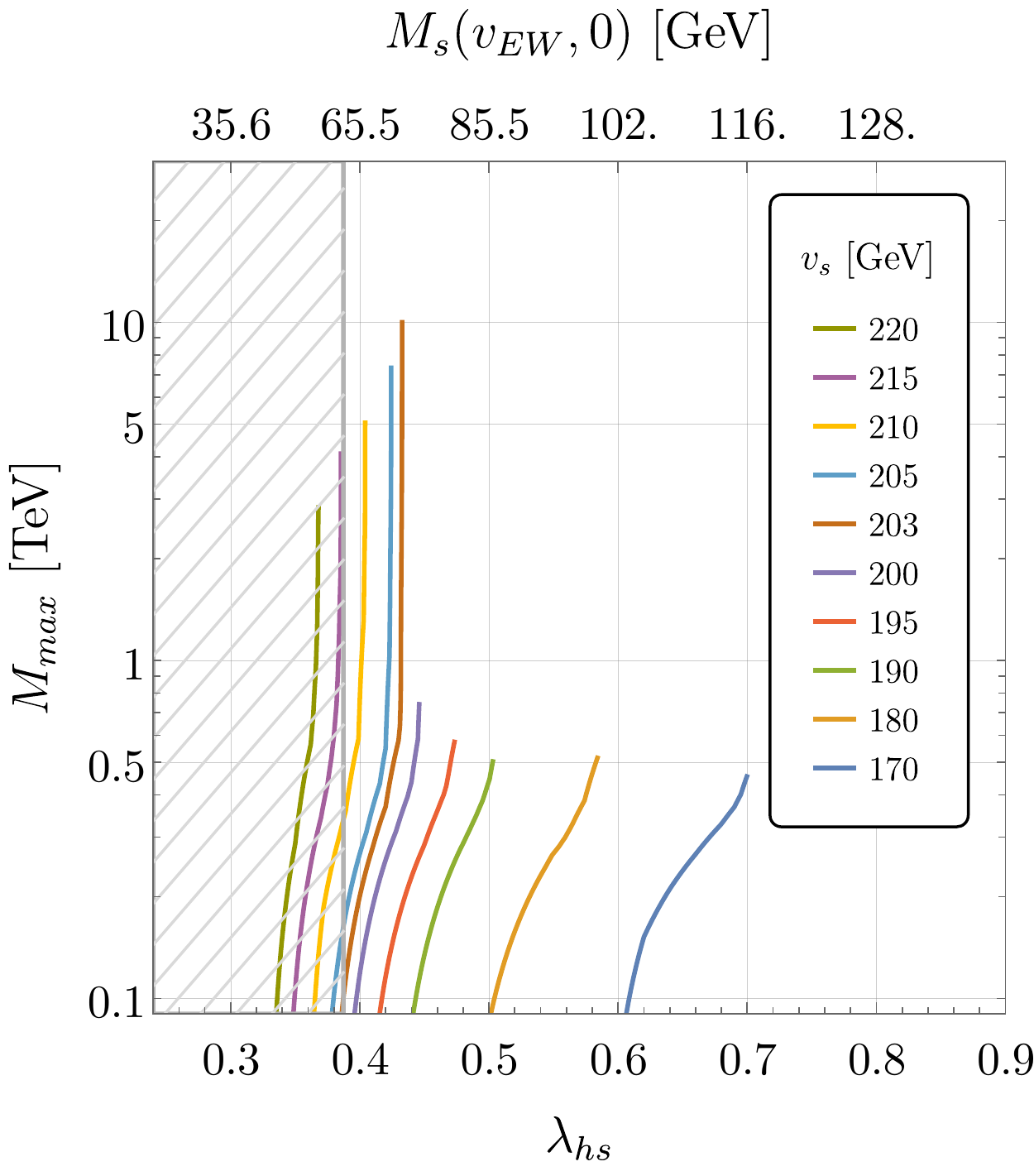}
 \caption{ {\bf Top-Left}:
 Scan of the parameter space in the plane $\lambda_{hs}-v_s$ for $m_s = 125$ GeV. The four regions are as follows: i)  white - NO PT, the region where the transition never completes because the barrier remains at zero temperature and the function $S_3/T$ never passes below the nucleation condition, ii) light and dark red are the regions where the FOPT happens. Dark (light) red corresponds to the region with relativistic (non-relativistic) bubble expansion. The boundary between two regions is given by Eq.\eqref{eq:relat_cond}
 iv) blue - the phase transition is of the second order. 
 The gray meshed region is the one in which $M_s(v_{EW},0)<m_h/2$, that is constrained from collider experiments.  {\bf Top-Right and Bottom}: the dependencies of $T_{\rm nuc},\gamma_w,\ M_{max}$ on the $\lambda_{hs}$ coupling for $m_s=125$ GeV. The blue dot on the Top-Right plot designates the end of the curve, when the tuning becomes $10^{-6}$ and the red dot signals the appearance of a barrier at zero temperature (all the points above the red dot have a barrier at $T = 0$). For the last three plots we varied the $\lambda_{hs}$ parameter with the steps of $10^{-6}$. The value of $v_s$ is encoded in color according to the bottom-right plot.  }
  \label{fig:scan}

 \end{figure}

After the qualitative discussion, let us proceed to the numerical calculations. The bounce action for $(0, v_s) \to (v_{EW},0)$  was computed using our own dedicated code (cross checked against \texttt{FindBounce}\cite{Guada:2020xnz}) and we relegate the details and methods of this calculation to Appendix \ref{app:codefor2D}. We parameterize our model in terms of  ($m_s,\lambda_{hs},v_s$) parameters (see Eq.\eqref{eq:potential_1}). Instead of analyzing all the possible values of $m_s$, in this section we report  the results (see Fig.\ref{fig:scan} left panel) by fixing $m_s=125 $ GeV (there is no particular reason for this value of $m_s$ and the results for the other values of $m_s$ are similar and are shown in the appendix \ref{sec:numeric-extra}).  In the plane  (see Fig.\ref{fig:scan}) $v_s-\lambda_{hs}$, we identify  four regions with different behaviours under the phase transition. The blue region shows second order transition when there is no barrier between the two separated vacuum. Next to it there is a light red region where the transition is of the first order. We indicate separately (dark red) the region,  where the transition is of the first order and the bubbles are relativistic. In particular the boundary between the regions with relativistic and non-relativistic bubbles is defined by the criteria of Eq.\eqref{eq:relat_cond}, \textit{i.e.} when the LO pressure for relativistically expanding bubbles is less than the driving force. At last, there is NO PT region, where the system remains stuck in the false vacuum since the tunnelling rate is too small.

The structure of the  diagram  (on the Fig.\ref{fig:scan} top$-$left panel)   can be  easily understood  from the qualitative discussion  in the previous section. Indeed, keeping $v_s$ fixed, the size the potential barrier is controlled by the coupling $\lambda_{hs}$. Moving from left to right the size of the potential barrier increases and we are gradually moving from the region of second order phase transition $\to$ FOPT $\to$ FOPT with relativistic bubbles $\to$ no PT region. Similarly moving up (increasing $v_s$ for fixed values of $\lambda_{hs}$) also corresponds to the increase of the potential barrier as we pass though the regions with different PT in the same order. On the upper axis we report  physical  mass of the singlet in the true vacuum $M_s(v_{EW}, 0)$ and exclude the constrained region where $h \to ss$ is kinematically allowed (the gray meshed region).

Next we plot the values of  $T_{\rm nuc}$ and $\gamma^{\rm terminal}_w$ as a function of $(\lambda_{hs}, v_s)$  (Fig.\ref{fig:scan}). As discussed in the previous section, the region with the smallest values of the nucleation temperature (thus the fastest bubbles) is located near the ``NO PT" region, \textit{i.e.} where the system remains trapped in the false vacuum. The blue dot on the $T_{\rm nuc}$ plot indicates the last point we have found before the system enters the regime of no phase transition (NO PT). In the bottom right we plot $M_{\rm max}=\sqrt{\gamma_w T_{\rm nuc} v_{EW}}$ quantity which indicates the maximal energy in the c.o.m frame for the plasma particle$-$bubble wall collision, which  corresponds to the largest mass  of the heavy particles we can produce  (see discussion in sec \ref{sec:heavypart-prod}).

In order to better understand the dependencies of $T_{\rm nuc}$ and $\gamma_w$ on the parameters of the model, it is useful to separate further the parameter space depending on whether the potential barrier disappears at zero temperature or not.  This is important since the bounce action in such cases has very different dependencies on the temperature. 
  \begin{figure}
    \centering
    \includegraphics[width=.46\textwidth]{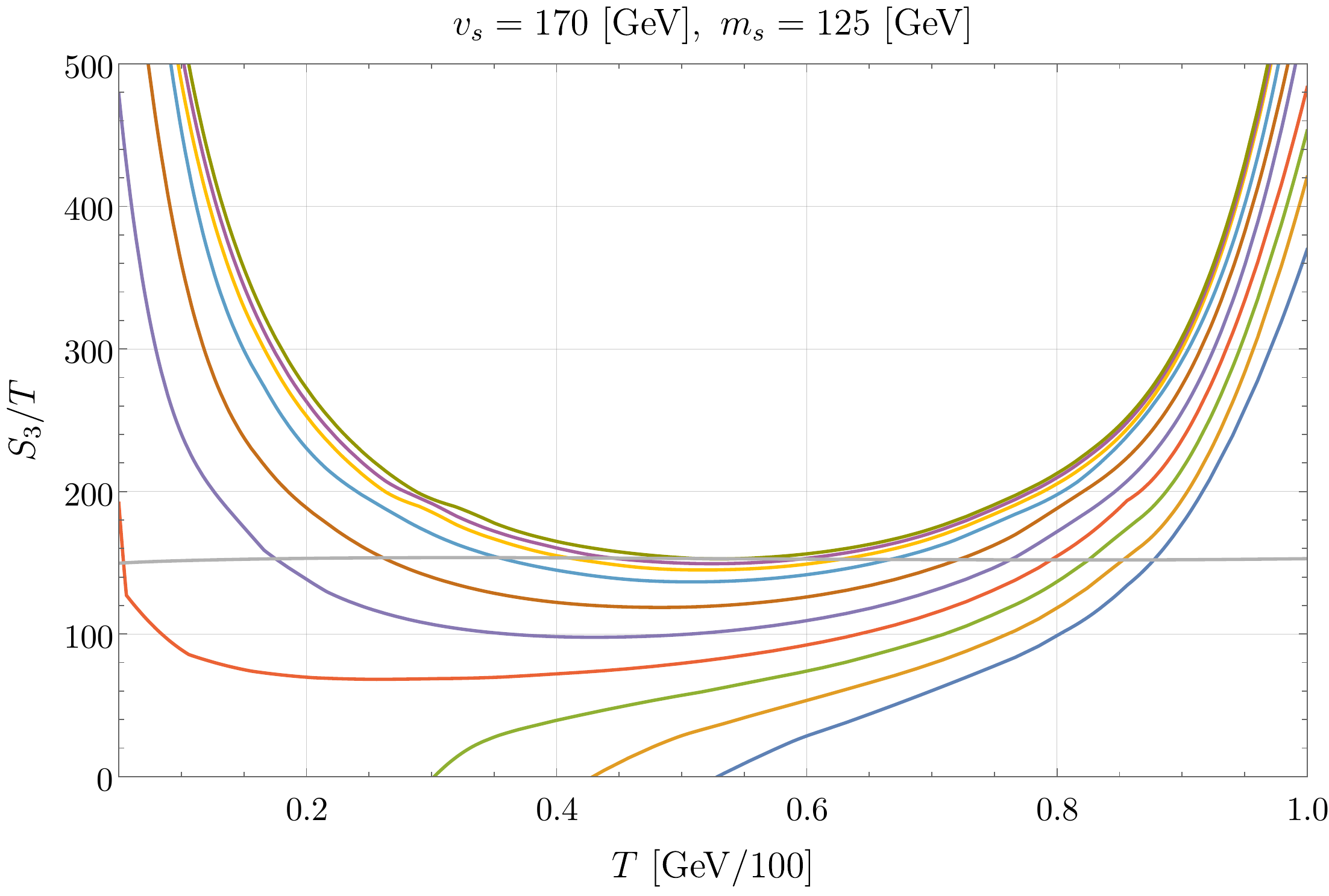}
    \includegraphics[width=.53\textwidth]{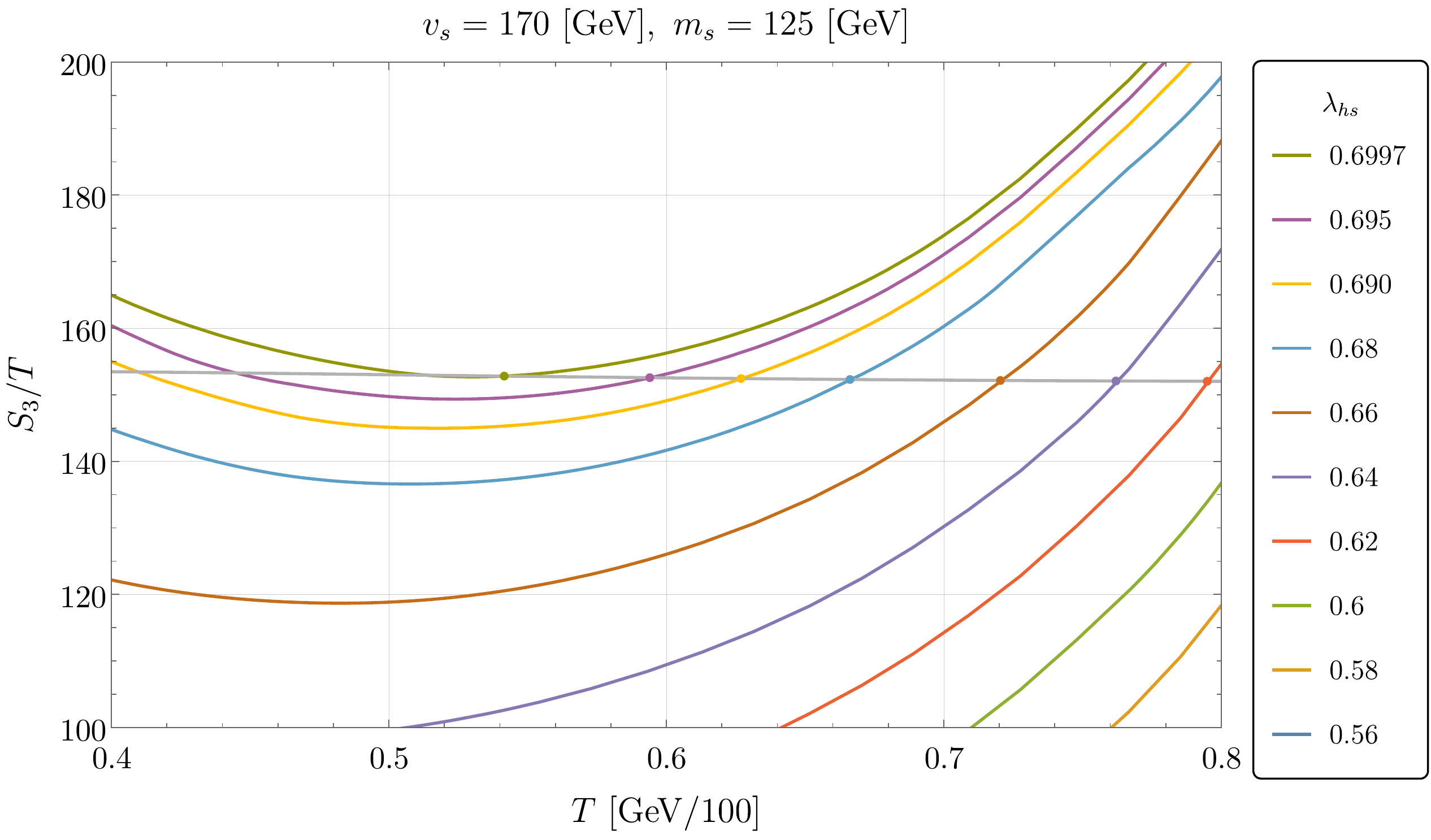}
    \caption{{\bf Left}: $S_3/T$ function with $v_s=170$ GeV and $m_s = 125$ GeV. We observe that the nucleation temperature saturates around $T_{\rm nuc} \approx 52$ GeV. The horizontal gray line satisfy the nucleation condition $S_3/T_{\rm nuc} = {3 \over 2} \log \left(S_3/ 2 \pi T_{\rm nuc}\right) + 4 \log\left(T_{\rm nuc}/ H\right)$. {\bf Right}: Zooming in on the region with lower nucleation temperatures. \label{fig:vs170}}
\end{figure}
See for example Fig.\ref{fig:vs170}, where we have plotted the $S_3/T$ for $v_s=170,m_s=125$ GeV for the various couplings  $\lambda_{hs}$. In the case when the potential barrier disappears at some temperature $T^{\rm NB}$ the function $S_3/T$ drops to zero, but if the barrier remains even at zero temperature $S_3/T$ has a  global minimum for $T\neq 0$ which will be controlling the lowest nucleation temperature possible.

\subsection{No potential barrier at zero temperature}
\label{sec:nobarrier}
Let us start by defining the region  where there is no potential barrier at zero temperature. On Fig.\ref{fig:scan}, we demonstrate the curves where the barrier disappears for various values of the temperatures ($T^{\rm NB}$). For $T^{\rm NB}=0$ case, approximate curve can be obtained analytically by looking at the leading terms in the zero temperature CW potential 
  \bea 
 \lambda_{hs} \lesssim \frac{m_h^2}{v_s^2} - \frac{n_t y_t^4}{32\pi^2} \frac{v_{EW}^2}{v_s^2}\;.
 \label{eq_lambda_ls}
 \eea 
The agreement between this  equation and exact $T^{\rm NB}=0$ curve is at the level of a  few permille discrepancies. To the right  of $T^{\rm NB}=0$ curve, the  potential barrier between the two minima remains even at zero temperatures. For the values  $v_s\gtrsim 200$ GeV, we find that the line $T^{\rm NB}=0$ approximately coincides with the boundary of no phase transition region (where system remains stuck in the false vacuum) but obviously the boundary of ``NO PT" is always to the right of $T^{\rm NB}=0$ curve. The size of this narrow strip is of the order $10^{-4}$ in $\lambda_{hs}$ values. One can see it from the $T_{\rm nuc}$ panel of Fig.\ref{fig:scan} where we have indicated the value of $\lambda_{hs}$ when $T_{\rm NB}=0$ by vertical thin line and red dot (for intersection)  and the position of the blue dot which is the last point where the transition is of the first order before we enter NO PT region. The boundaries of this region were obtained by  numerical calculations where we  have scanned $\lambda_{hs}$ parameter with a step $10^{-6}$. We postpone the discussion of the FOPT in this narrow region to the next section \ref{sec:tunnelbarrier}.

In this section we  restrict our discussion only on the region to the left of $T^{\rm NB}=0$ curve. Then the phase transition will be always completed before the universe cools down to $T^{\rm NB}$, \textit{i.e.} $T^{\rm NB}< T_{\rm nuc}$, which provides a lower bound for the nucleation temperature. At the same time  the velocity of the bubbles  become largest for the smallest possible values for the nucleation temperature. So that the fastest bubbles will be near $T^{\rm NB}=0$  curve. Looking at Fig.\ref{fig:scan}  we can see that the largest $\gamma_w$ (Lorentz boost factor) and lowest nucleation temperatures happen for $v_s\gtrsim 200$ GeV, where the $T^{\rm NB}=0$ curve passes very close to the NO PT boundary. The shape of the lines in Fig.\ref{fig:scan} clearly indicate the necessity of tuning in order to obtain low nucleation temperatures (large $\gamma_w$). In particular for the values of $v_s\gtrsim 200$ GeV we can see that the nucleation temperature  drops by choosing $\lambda_{hs}$ close to the NB value (similarly $\gamma_w$  becomes maximal see Fig.\ref{fig:scan}).
 \begin{figure}
 \centering
 \includegraphics[scale=1]{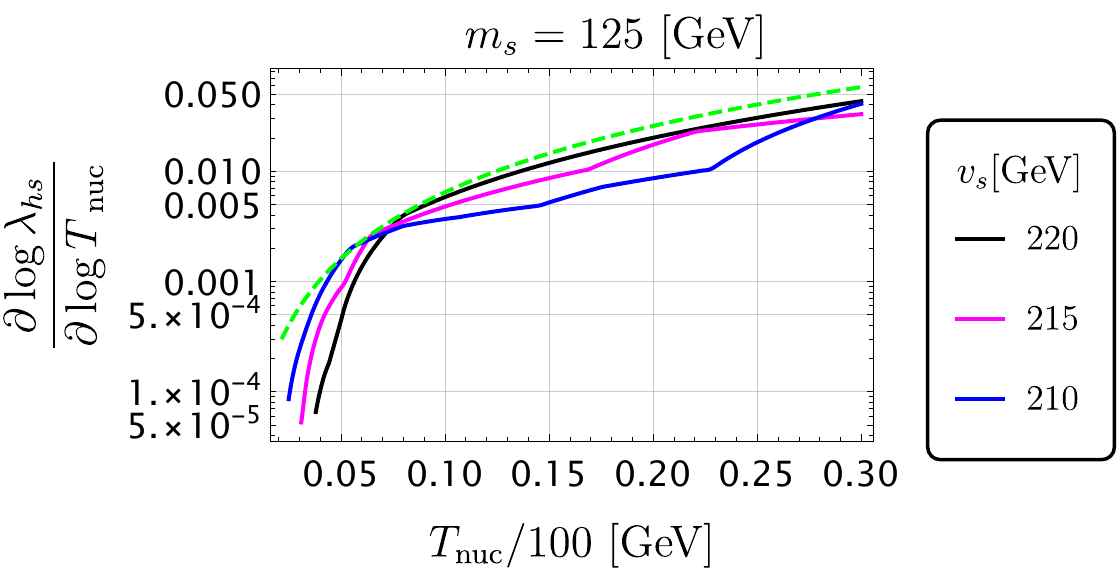}
 \caption{Tuning of the coupling $\lambda_{hs}$ as a function of the nucleation temperature. The dashed green line represent the naive tuning $\sim (T_{\rm nuc}/m_h)^2$. We observe that this naive estimation for the tuning is rather precise at large nucleation temperature but can underestimate the tuning by one order of magnitude for very low nucleation temperature.}
 \label{fig:tuning_supercooled}
 \end{figure}
We can estimate this tuning  by looking at $\d \log \lambda_{hs}/\d \log T_{\rm nuc}$ quantity (analogue of Giudice-Barbieri \cite{Barbieri:1987fn} measure of the tuning) as a function of $T_{\rm nuc}$.  This result agrees with our expectation from the steepness of the curves in the Fig.\ref{fig:tuning_supercooled} and with the naive tuning expectation which scales as $\sim \big(T_{\rm nuc}/m_h\big)^2$\footnote{This expression follows Eq.(\ref{eq:cond-min}) if we require the cancellation between the terms independent of temperature.}.

At last we would like to remind that the discussion in this section always  assumed that the phase transition completes before the potential barrier disappears. We have checked numerically that this is always the case. Indeed the  time of the phase transition is given  approximately by the bubble radius at the moment of percolation \cite{Ellis:2018mja,Enqvist:1991xw}
\bea
\label{eq:R*}
R_\star\equiv\frac{(8\pi)^{1/3}}{\tilde \beta}= \l(\int_{T_{\rm per}}^{T_{\rm crit}}\frac{d T}{T}\frac{\Gamma(T)}{H(T)}\l(\frac{T_{\rm per}}{T}\r)^3\r)^{-1/3}.
\label{eq:radius}
\eea
This radius is related 
to the   $\beta \equiv -\frac{d}{dt} \frac{S_3}{T} = H T \frac{d}{dT}\frac{S_3}{T}$ 
parameter by an approximate relation \cite{Enqvist:1991xw}
\begin{align}
  &{\beta \over H}\biggl|_{T_{\rm per}}\simeq\frac{\tilde \beta}{H}= \frac{(8\pi)^{1/3}}{R_\star H}\;,
 \label{eq:beta2}
 \end{align}
where we find $ R_\star^{-1}\sim 
 \beta_{\rm typical} \sim (10-10^4) H.
$ 
At this point  the temperature drops during the bubble expansion will scale as
\bea
\label{eqn: DeltaT}
\Delta T\sim T_{\rm nuc} (H \Delta t)\sim \frac{ T_{\rm nuc} H}{\beta}\;.
\eea
Due to the large value of $\beta/H$  we find numerically that this drop of the temperature is not enough for the barrier to disappear or in other words
\bea
T_{\rm nuc}-\Delta T > T^{\rm NB}\;.
\label{eq:criterion}
\eea
Such behaviour can be understood from the following consideration near $T_{\rm NB}$ the bounce action drops very quickly  and  so that the tunneling  becomes very efficient almost instantaneous and typical bubble radiuses are much smaller than the Hubble scale. This leads to another prediction that GW signal will be suppressed as well since it is controlled by the ($\beta/H$) quantity Eq.\eqref{eq:beta2}.
As we will see, even with this suppression the GW signal is efficient enough to be detected in the future.

\subsection{Tunneling with potential barrier at zero temperature}
\label{sec:tunnelbarrier}
Let us proceed to the analysis of the case when the potential barrier does not disappear at zero 
temperature. The parameter space with the lowest nucleation temperatures (fastest bubbles) will be 
located again near the ``NO PT" boundary. However in this case the nucleation temperature will be controlled by the local  minima of the $S_3/T$ function.
(see Fig.\ref{fig:vs170}). At least a minimum is expected since the potential at low temperature 
becomes fully temperature independent and
  $S_3(T\to 0) \to \text{const.}$, so that  $S_3/T$
necessarily starts to grow {for $T\to0$}.

Numerically (see Fig.\ref{fig:scan}) for the value of $m_s=125$ GeV we find that  for  $v_s\lesssim 180$ GeV  entire region with the fast bubbles has a potential barrier at zero temperature.
 On Fig.\ref{fig:vs170}, we present the euclidean action for $v_s = 170$ GeV and $m_s=125$ GeV. Going back to Fig.\ref{fig:scan}, we see that for those values the ``NO PT" curve and the ``$T_{\rm NB} =0$" curve are largely separated. This is not a surprise since in this region of parameter space the bounce action $S_3/T \sim O(10^2)$  is small enough to guarantee the successful tunneling even when the barrier remains at zero temperature.
 In the range of $\lambda_{hs}$ from $0.544$ to $0.58$, numerically we find that nucleation temperatures are between $65-44$ GeV, with a clear saturation at $T^{\rm sat}_{\rm nuc} \approx 44$ GeV, and the corresponding Lorentz factor for the velocities of the bubble expansion in the ranges of $\lesssim 10$.  Interestingly we find  that in this case the bubble radius $R_\star \sim \frac{(8\pi)^{1/3}v_w}{\beta}$ are a little bit larger than the ones discussed in the section \ref{sec:nobarrier}, corresponding to a bit smaller values of $\beta/H$ parameter. 

\begin{figure}
    \centering\hspace{-2cm}
   \begin{minipage}{.52\textwidth}
   
 \includegraphics[scale=0.44]{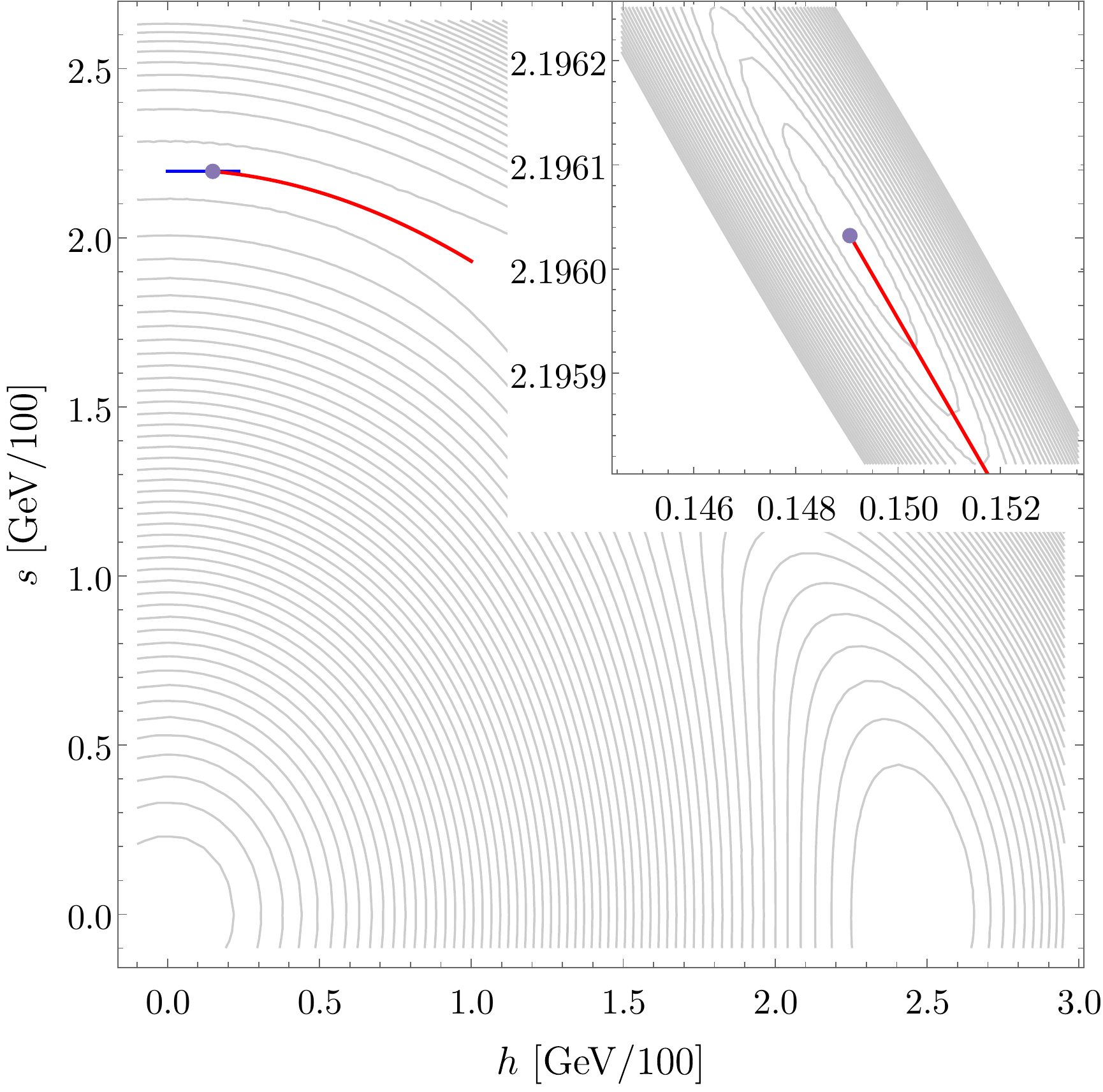}
 \end{minipage}
 \begin{minipage}{.45\textwidth}
  \includegraphics[scale=.65]{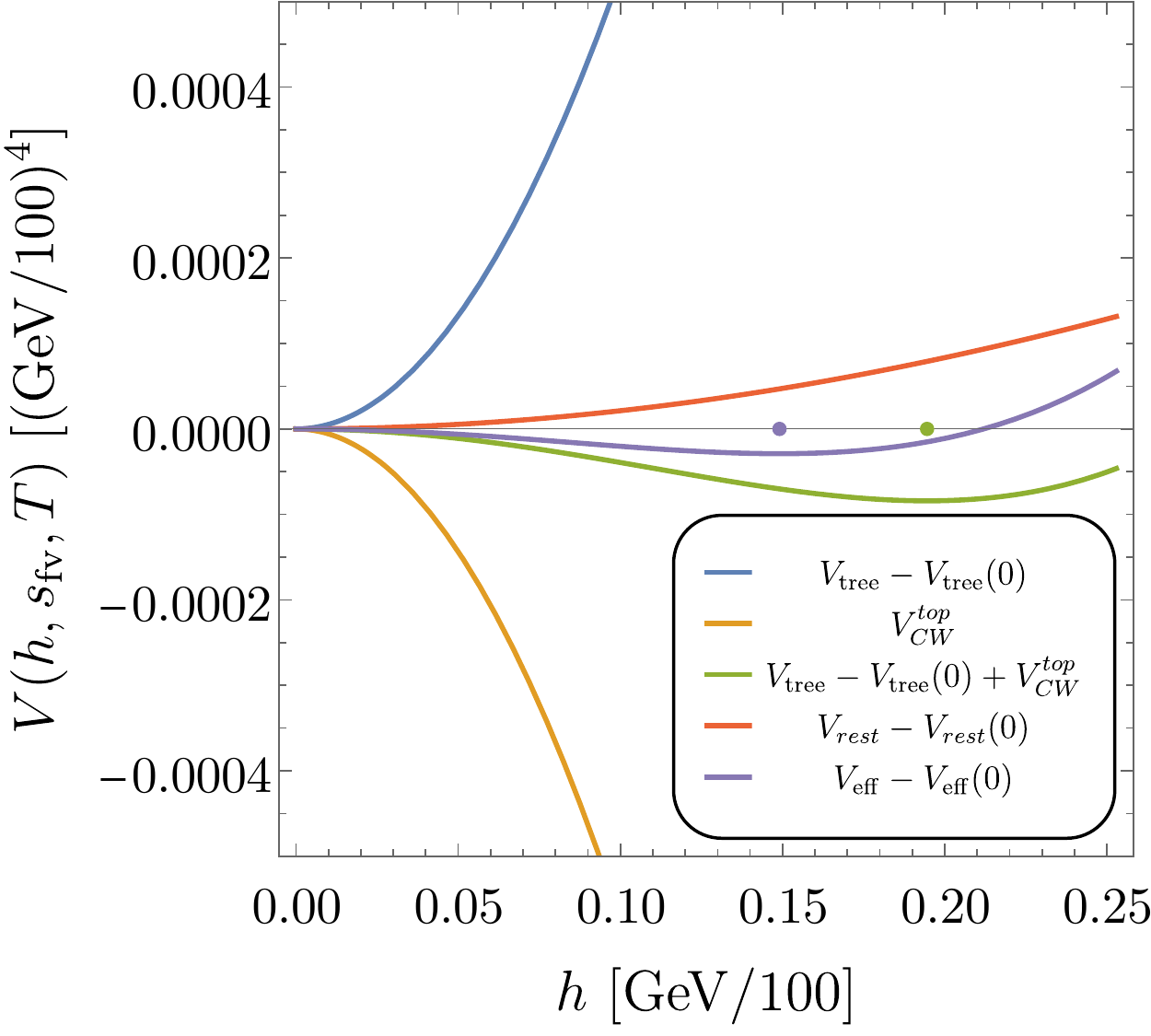}
 \end{minipage}
\caption{Here is presented an explicit example of displacement of the false vacuum for $\{\lambda_{hs}, v_s, T\}=\{0.36784, 220\ {\rm GeV}, 3\ {\rm GeV}\}$. {\bf Left}: we show the 2D potential where the blue line corresponds to the part of the potential plotted in the right panel and the (purple) dot is the position of the displaced false minimum, in both the plots. Red line indicates the bounce trajectory. {\bf Right}: plot of the different contributions to the potential. We see that a displaced minimum can be generated balancing the tree level and the CW potential of the top quark, for low enough temperature, in such a way all the other particles, that are massless in the false vacuum, have a negligible contribution. It can be shown that they cause, as the temperature increases, the 
shift of the local minimum towards $h=0$.
      \label{fig:flat dir}
    }
  \end{figure}

\paragraph{Super fine-tuned region}
We finally comment on the parameter space with $v_s\gtrsim 200$ GeV (again we are fixing $m_s = 125$ GeV), where the curves  ``NO PT" and ``$T_{\rm NB} = 0$" almost superimpose (the region between red and blue dots on the $T_{\rm nuc}$ panel of the Fig.\ref{fig:scan}). There will be a very narrow strip between the curves ``NO PT" and ``$T_{\rm NB} = 0$" regions, where the tunnelling will happen even though the barrier remains at zero temperature. We find (see Fig.\ref{fig:S3curves}) that the region corresponds to the  variations of the $\lambda_{hs}$ parameter of the order $\delta \lambda_{hs}\sim O(10^{-4})$, \textit{i.e.} two order of magnitude smaller than the full region with relativistic bubbles.
In this very small region  various additional effects can start playing a role. For example let us look at the Fig.\ref{fig:S3curves} we can see that the bounce action $S_3/T$ has a local maximum and a deeper (global) minimum with respect to the standard scenario. Such a behaviour of the action is coming from the cancellations of various terms in the effective potential. For simplicity let us look at $T=0$ case. Then there is a region of parameter space where purely polynomial potential has no local minimum at $(0, v_s)$,  but the effects of the $\frac{-3 M_t^4(h)}{8\pi^2}\big(\log \frac{M_t(h)}{M_t(v_{EW})}\big) $ terms in Coleman-Weinberg contribution lead to the appearance of the local minimum at $(\delta v_h, v_s)$.  On Fig.\ref{fig:flat dir} we plot  the contributions of the various terms in the effective potential leading to the appearance of this local minimum and the trajectory of the typical bounce solution in this case. As a result the distance in the fields space between the two minima decreases and the tunneling becomes faster, which leads to the appearance of 
the second (global) minimum in $S_3/T $.

 \subsubsection{Benchmark points}


  \begin{figure}
 \centering
 \begin{minipage}{.48\textwidth}
 \includegraphics[width=1\textwidth]{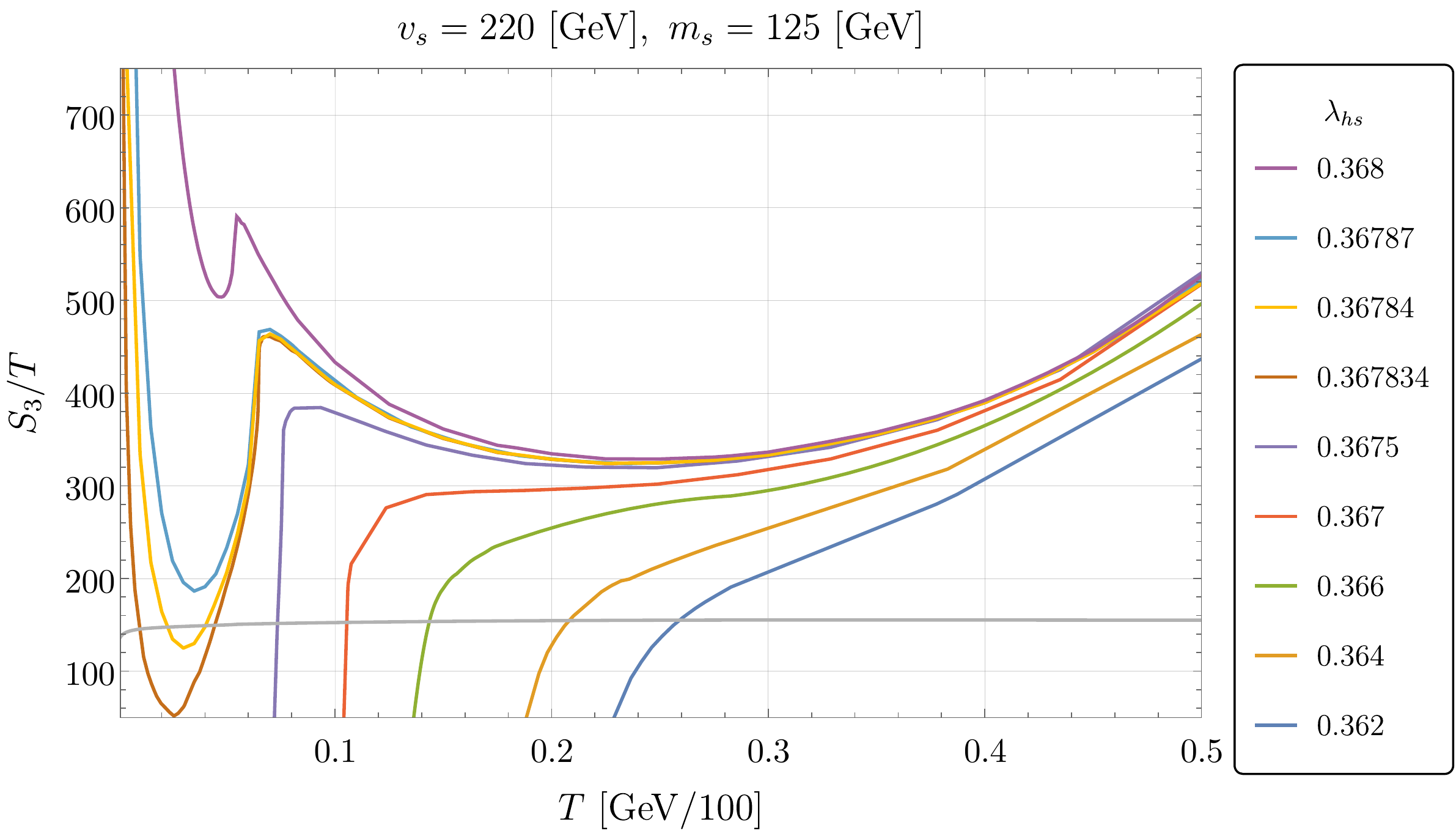}
 \end{minipage}
 \begin{minipage}{.48\textwidth}
\includegraphics[width=1.02\textwidth]{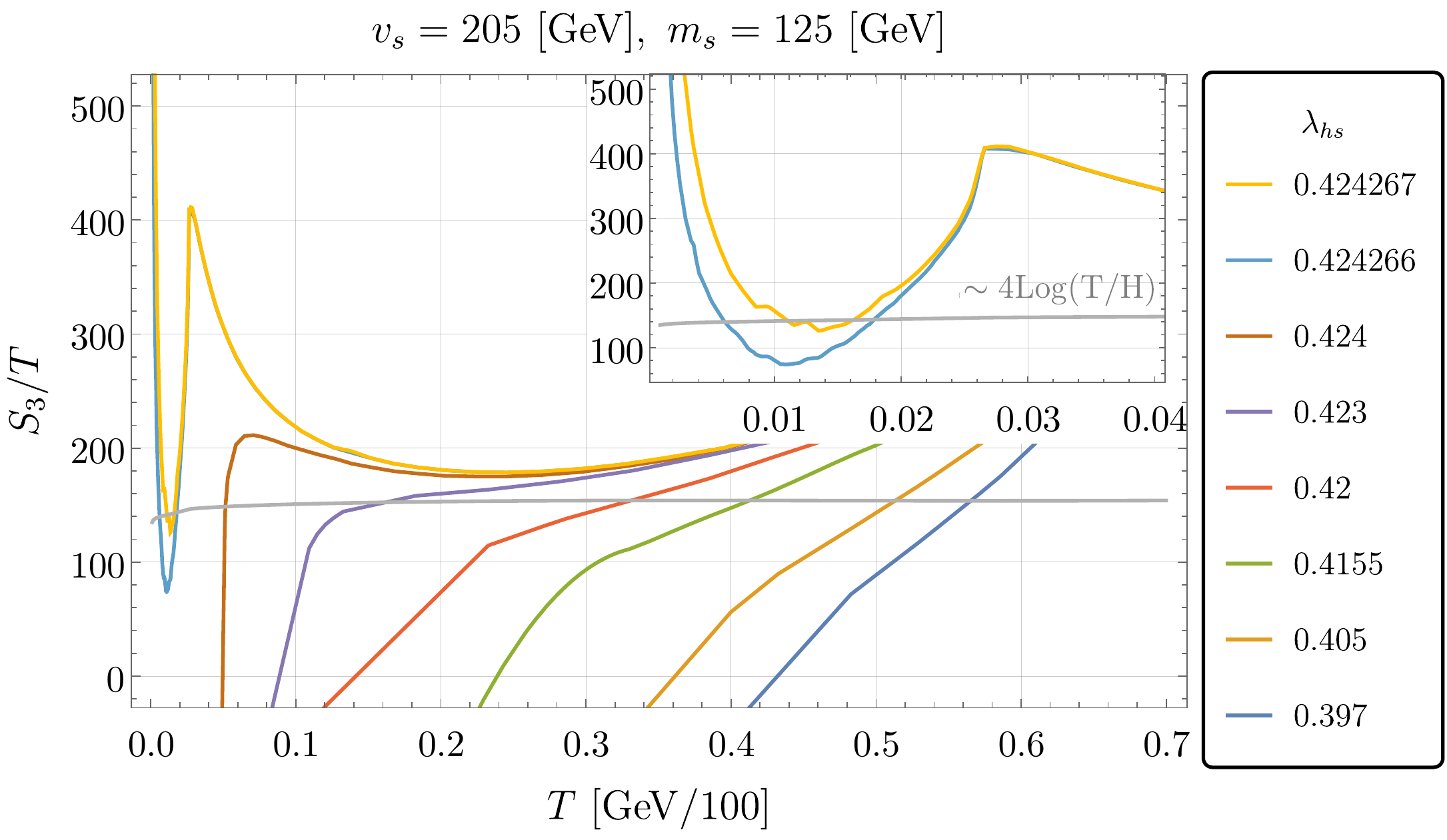}
 \end{minipage}
 \caption{\textbf{Left}: Plot of $S_3(T)/T$ as a function of the temperature, for different values of $\lambda_{hs}$ in the case where $m_s = 125$ GeV and $v_s=$ 220 GeV. As we increase the value of the coupling $\lambda_{hs}$, the disappearance of the potential barrier happens later, allowing for longer supercooling, until it is large enough to remain even at zero temperature. For the first four curves from the top, we observe a second drop in the function at very low temperature. This second drop corresponds to the displacement on the false minimum that we describe in this section. \textbf{Right}: Same plot as in the left panel, but with a lower value for $v_s$. The pattern we found is the same, but lowering $v_s$ causes a lowering of the curves and the displacement of the false minimum is less pronounced.
 }
 \label{fig:S3curves}
 \end{figure}
 
In Table \ref{benchmarkpoint} and \ref{benchmarkpointbis}, we give typical values of the nucleation temperature, the Lorentz factor $\gamma_w$, the $\beta/H$ factor, 
 and we indicate if the barrier remains at zero temperature, applying the criterion in Eq.\eqref{eq:criterion}.  
We can see that the largest bubble radius at the collision (smallest $\beta$) correspond to the case when $S_3/T$ is monotonic and very flat near the tunneling temperature (c.f. the right panel of Fig.~\ref{fig:S3curves}).

  \begin{table}[]
\footnotesize
\begin{center}
\begin{tabular}{|l|l|l|l|l|l|l|l|l|l|l|l|}
\hline
\multicolumn{7}{|c|}{$m_s=125$ GeV, \quad $v_s=170$ GeV}  
 &
 \multicolumn{1}{l|}{} 
\\
\hline
\multicolumn{1}{|l|}{$\lambda_{hs}$}  &
\multicolumn{1}{l|}{$\frac{T_{\rm reh}}{100 \text{GeV}}$}  &
\multicolumn{1}{l|}{$\frac{T_{\rm nuc}}{100 \text{GeV}}$}  &
\multicolumn{1}{l|}{$\frac{T_{\rm per}}{100 \text{GeV}}$} &
\multicolumn{1}{l|}{$\gamma_w$} &
\multicolumn{1}{l|}{${\tilde\beta\over H}={(8\pi)^{1/3} \over R_\star H}$}&
\multicolumn{1}{l|}{$m_H^{\rm False}/\text{GeV}$} 
&
 \multicolumn{1}{l|}{FM$_0$} 
\\
\hline
\multicolumn{1}{|l|}{$0.56$}  &
\multicolumn{1}{l|}{$0.880$}  &
\multicolumn{1}{l|}{$0.877$}  &
\multicolumn{1}{l|}{$0.850$} &
\multicolumn{1}{l|}{$-$}  &
\multicolumn{1}{l|}{$434$}  &
\multicolumn{1}{l|}{$35.4$} 
&
\multicolumn{1}{l|}{No}   
\\
\hline
\multicolumn{1}{|l|}{$0.58$}  & 
\multicolumn{1}{l|}{$0.855$}  &
\multicolumn{1}{l|}{$0.851$}  &
\multicolumn{1}{l|}{$0.822$} &
\multicolumn{1}{l|}{$-$}  &
\multicolumn{1}{l|}{$355$}  &
\multicolumn{1}{l|}{$37.3$} 
&
\multicolumn{1}{l|}{No}   
\\
\hline
\multicolumn{1}{|l|}{$0.6$}  & 
\multicolumn{1}{l|}{$0.829$}  &
\multicolumn{1}{l|}{$0.824$}  &
\multicolumn{1}{l|}{$0.790$} &
\multicolumn{1}{l|}{$-$}  &
\multicolumn{1}{l|}{$296$}  &
\multicolumn{1}{l|}{$39.2$} 
&
\multicolumn{1}{l|}{No}   
\\
\hline
\multicolumn{1}{|l|}{$0.62$}  &
\multicolumn{1}{l|}{$0.800$}  &
\multicolumn{1}{l|}{$0.795$}  &
\multicolumn{1}{l|}{$0.762$} &
\multicolumn{1}{l|}{$1.2$}  &
\multicolumn{1}{l|}{$209$}  &
\multicolumn{1}{l|}{$40.1$} 
&
\multicolumn{1}{l|}{No}   
\\
\hline
\multicolumn{1}{|l|}{$0.64$}  &
\multicolumn{1}{l|}{$0.769$}  &
\multicolumn{1}{l|}{$0.762$}  &
\multicolumn{1}{l|}{$0.714$} &
\multicolumn{1}{l|}{$2.4$}  &
\multicolumn{1}{l|}{$158$}  &
\multicolumn{1}{l|}{$42.5$} 
&
\multicolumn{1}{l|}{No}   
\\
\hline
\multicolumn{1}{|l|}{$0.66$}  &
\multicolumn{1}{l|}{$0.729$}  &
\multicolumn{1}{l|}{$0.720$}  &
\multicolumn{1}{l|}{$0.661$} &
\multicolumn{1}{l|}{$4$}  &
\multicolumn{1}{l|}{$108$}  &
\multicolumn{1}{l|}{$43.4$} 
&
\multicolumn{1}{l|}{No}   
\\
\hline
\multicolumn{1}{|l|}{$0.68$}  &
\multicolumn{1}{l|}{$0.678$}  &
\multicolumn{1}{l|}{$0.666$}  &
\multicolumn{1}{l|}{$0.582$} &
\multicolumn{1}{l|}{$6.6$}  &
\multicolumn{1}{l|}{$51$}  &
\multicolumn{1}{l|}{$44.8$}
&
\multicolumn{1}{l|}{No}  
\\
\hline
\multicolumn{1}{|l|}{$0.69$}  &
\multicolumn{1}{l|}{$0.642$}  &
\multicolumn{1}{l|}{$0.627$}  &
\multicolumn{1}{l|}{$0.506$} &
\multicolumn{1}{l|}{$8.8$}  &
\multicolumn{1}{l|}{$18$}  &
\multicolumn{1}{l|}{$45.0$}
&
\multicolumn{1}{l|}{No}   
\\
\hline
\multicolumn{1}{|l|}{$0.695$}  &
\multicolumn{1}{l|}{$0.612$}  &
\multicolumn{1}{l|}{$0.594$}  &
\multicolumn{1}{l|}{$0.412$} &
\multicolumn{1}{l|}{$11$}  &
\multicolumn{1}{l|}{$5$}  &
\multicolumn{1}{l|}{$44.7$}
&
\multicolumn{1}{l|}{No}  
\\
\hline
\multicolumn{1}{|l|}{$0.6997$}  &
\multicolumn{1}{l|}{$0.566$}  &
\multicolumn{1}{l|}{$0.542$}  &
\multicolumn{1}{l|}{$0.237$} &
\multicolumn{1}{l|}{$15$}  &
\multicolumn{1}{l|}{$1.4$}  &
\multicolumn{1}{l|}{$43.8$} 
&
\multicolumn{1}{l|}{No}   
\\
\hline
\end{tabular}
\end{center}
\caption{We report for Fig.\ref{fig:vs170}, $m_s = 125$ GeV and $v_s =170$ GeV, reheating, nucleation and percolation temperatures, respectively, for different values of $\lambda_{hs}$ and $\gamma_w$ reached by the expanding walls as well as the parameter $\tilde{\beta}/H$ computed using Eq.(\ref{eq:beta2}). We also show the effective Higgs mass in the false vacuum at the nucleation temperature defined as $(m_H^{\rm False})^2={\partial^2 V \over \partial h^2}\big|_{T=T_{\rm nuc}}$, relevant for DM production explained in section \ref{sec:DM-production}. 
In the last column,  FM$_0$ concerns {the displacement of the false minimum. No if it is at the (Higgs) origin, yes if it moved. In this case, the minimum is always at the origin.}
}
\label{benchmarkpoint}
\end{table}
 
 \begin{table}[]
\footnotesize
\begin{center}
\begin{tabular}{|l|l|l|l|l|l|l|l|l|l|l|l|}
\hline
\multicolumn{7}{|c|}{$m_s=125$ GeV, \quad $v_s=205$ GeV}  
&
\multicolumn{1}{l|}{} 
\\
\hline
\multicolumn{1}{|l|}{$\lambda_{hs}$}  &
\multicolumn{1}{l|}{$\frac{T_{\rm reh}}{100 \text{GeV}}$}  &
\multicolumn{1}{l|}{$\frac{T_{\rm nuc}}{100 \text{GeV}}$}  &
\multicolumn{1}{l|}{$\frac{T_{\rm per}}{100 \text{GeV}}$} &
\multicolumn{1}{l|}{$\gamma_w$} &
\multicolumn{1}{l|}{${\tilde\beta\over H}={(8\pi)^{1/3} \over R_\star H}$}&
\multicolumn{1}{l|}{$m_H^{\rm False}/\text{GeV}$} 
&
\multicolumn{1}{l|}{FM$_0$} 
\\
\hline
\multicolumn{1}{|l|}{$0.397$}  &
\multicolumn{1}{l|}{$0.577$}  &
\multicolumn{1}{l|}{$0.564$}  &
\multicolumn{1}{l|}{$0.544$} &
\multicolumn{1}{l|}{$4$}  &
\multicolumn{1}{l|}{$371$}  &
\multicolumn{1}{l|}{$19.1 $} 
&
\multicolumn{1}{l|}{No} 
\\
\hline
\multicolumn{1}{|l|}{$0.405$}  & 
\multicolumn{1}{l|}{$0.530$}  &
\multicolumn{1}{l|}{$0.512$}  &
\multicolumn{1}{l|}{$0.488$} &
\multicolumn{1}{l|}{$8$}  &
\multicolumn{1}{l|}{$268$}  &
\multicolumn{1}{l|}{$19.1 $} 
&
\multicolumn{1}{l|}{No} 
\\
\hline
\multicolumn{1}{|l|}{$0.4155$}  & 
\multicolumn{1}{l|}{$0.448$}  &
\multicolumn{1}{l|}{$0.412$}  &
\multicolumn{1}{l|}{$0.379$} &
\multicolumn{1}{l|}{$18$}  &
\multicolumn{1}{l|}{$130$}  &
\multicolumn{1}{l|}{$17.7$}
 &
 \multicolumn{1}{l|}{No} 
\\
\hline
\multicolumn{1}{|l|}{$0.42$}  &
\multicolumn{1}{l|}{$0.393$}  &
\multicolumn{1}{l|}{$0.330$}  &
\multicolumn{1}{l|}{$0.290$} &
\multicolumn{1}{l|}{$37$}  &
\multicolumn{1}{l|}{$72$}  &
\multicolumn{1}{l|}{$15.2 $}
&
\multicolumn{1}{l|}{No} 
\\
\hline
\multicolumn{1}{|l|}{$0.423$}  &
\multicolumn{1}{l|}{$0.339$}  &
\multicolumn{1}{l|}{$0.161$}  &
\multicolumn{1}{l|}{$0.124$} &
\multicolumn{1}{l|}{$270$}  &
\multicolumn{1}{l|}{$66$}  &
\multicolumn{1}{l|}{$7.1$}
&
\multicolumn{1}{l|}{No} 
\\
\hline
\multicolumn{1}{|l|}{$0.4234$}  &
\multicolumn{1}{l|}{$0.335$}  &
\multicolumn{1}{l|}{$0.107$}  &
\multicolumn{1}{l|}{$0.095$} &
\multicolumn{1}{l|}{$805$}  &
\multicolumn{1}{l|}{$109$}  &
\multicolumn{1}{l|}{$3.9$} 
&
\multicolumn{1}{l|}{No} 
\\
\hline
\multicolumn{1}{|l|}{$0.424$}  &
\multicolumn{1}{l|}{$0.335$}  &
\multicolumn{1}{l|}{$0.051$}  &
\multicolumn{1}{l|}{$0.051$} &
\multicolumn{1}{l|}{$5.7 \cdot 10^3$}  &
\multicolumn{1}{l|}{$3.3 \cdot 10^3$}  &
\multicolumn{1}{l|}{$0.7$} 
&
\multicolumn{1}{l|}{No} 
\\
\hline
\multicolumn{1}{|l|}{$0.4242$}  &
\multicolumn{1}{l|}{$0.335$}  &
\multicolumn{1}{l|}{$0.0337$}  &
\multicolumn{1}{l|}{$0.0337$} &
\multicolumn{1}{l|}{$ 1.8 \cdot 10^4$}  &
\multicolumn{1}{l|}{$3.2\cdot 10^4$}  &
\multicolumn{1}{l|}{$0.25$}
&
\multicolumn{1}{l|}{No} 
\\
\hline
\multicolumn{1}{|l|}{$0.42424$}  &
\multicolumn{1}{l|}{$0.335$}  &
\multicolumn{1}{l|}{$0.028$}  &
\multicolumn{1}{l|}{$0.0279$} &
\multicolumn{1}{l|}{$3.0\cdot 10^4$}  &
\multicolumn{1}{l|}{$1.8\cdot 10^3$}  &
\multicolumn{1}{l|}{$4.4$}
&
\multicolumn{1}{l|}{No} 
\\
\hline
\multicolumn{1}{|l|}{$0.424266$}  &
\multicolumn{1}{l|}{$0.335$}  &
\multicolumn{1}{l|}{$0.018$}  &
\multicolumn{1}{l|}{$0.017$} &
\multicolumn{1}{l|}{$1.0\cdot 10^5$}  &
\multicolumn{1}{l|}{$99$}  &
\multicolumn{1}{l|}{$6.2 $} 
&
\multicolumn{1}{l|}{Yes} 
\\
\hline
\multicolumn{1}{|l|}{$0.424267$}  &
\multicolumn{1}{l|}{$0.335$}  &
\multicolumn{1}{l|}{$0.016$}  &
\multicolumn{1}{l|}{$0.014$} &
\multicolumn{1}{l|}{$1.3\cdot 10^5  $}  &
\multicolumn{1}{l|}{$44$}  &
 \multicolumn{1}{l|}{$6.3$} 
&
\multicolumn{1}{l|}{Yes} 
\\
\hline
\end{tabular}
\end{center}
\caption{Same as Table \ref{benchmarkpoint}, but for Fig.\ref{fig:S3curves} and with $v_s=205$ GeV. {We observe that the last two points display a displacement of the false minimum.}
}
\label{benchmarkpointbis}
\end{table}

 \section{Consequences for production of dark matter and Baryogenesis}
 \label{sec:prod}

    One of the motivation for the study of a model of EWPT with relativistic bubbles is the relation between relativistic expansion and the out-of-equilibrium production of heavy states presented for the first time in \cite{Vanvlasselaer:2020niz}, when the field undergoing the PT (here the Higgs) is coupled to some heavy dark sector at typical mass $M_N$. In this section, we remind the principle of the production mechanism and we study the scenario of the production of Dark Matter\cite{Azatov:2021ifm} and Baryogenesis\cite{Azatov:2021irb}, that were previously agnostic about the EWPT realisation.

    First of all, the strong FOPT involves a supercooling represented by a dilution factor, \bea D\equiv \frac{g^{\rm sym}_{\star s}(T_{\rm nuc})}{g_{\star s}(T_{\rm reh})} \left(\frac{T_{\rm nuc}}{T_{\rm reh}}\right)^3,\eea 
 with $g_{\star s}(T)$ ($g^{\rm sym}_{\star s}(T)$) being the number of relativistic degrees of freedom of the entropy in the broken (symmetric) phase. 
 This means that with $D\ll 1$, any type of dark matter production or Baryogenesis mechanism that happens much earlier than the PT should provide values denser than the conventional estimation by a factor of $1/D$ (see for example \cite{Hambye:2018qjv, Baldes:2021aph}). For instance, the WIMP cross section should be  $\sigma \sim \frac{D}{10^{-3}} 10^{-29} \rm cm^3/s$ to produce a correct dark matter abundance. This is the case when freeze-out happens at temperatures much higher than the reheating.

\subsection{Production of heavy states during ultra-relativistic expansion}
\label{sec:heavypart-prod}
There are few mechanisms which can lead to heavy particle production during FOPT. This can happen  if the incoming massless particle in the unbroken phase gets a very large mass from the Higgs vev \cite{Baldes:2021vyz,Bodeker:2009qy} (mass gain), or due to the bubble- bubble collision \cite{Falkowski:2012fb,Katz:2016adq} or due to the plasma particle$-$bubble collision \cite{Vanvlasselaer:2020niz}. Our study will be focused on the later one. 

Let us assume that FOPT happens and the bubble expansion is indeed relativistic with $\gamma_w^{\rm terminal}\gg 1$. The simplest model where the production of heavy particles during plasma$-$bubble wall collision can be realized, is described by the following Lagrangian \cite{Vanvlasselaer:2020niz}:
\bea 
\mathcal{L} =
{1 \over 2}(\d_\mu h)^2+i \bar q \sl\d q + i \bar N \sl\d N- M_N \bar N N- Y h \bar N q-V(h)
\label{eq:Toy_model_1}\;,
\eea
where $q$ is a massless particle in the symmetric phase and $N$ is a heavy field with large vev$-$independent mass $M_N\gg v_{EW}$. $h$ is the Higgs field undergoing a FOPT. With no loss of generality, we go to the basis where fermion masses are real. Before the strong phase transition starts, the abundance of heavy states $N$ in the plasma is strongly Boltzmann suppressed and they would naively seem irrelevant for the dynamics of the transition. In an homogeneous vacuum, the transition from light to heavy state $q \to N$ is obviously forbidden by the conservation of momentum. However, in the presence of the bubble wall, the conservation of momentum along the $z$ direction is broken (assuming a planar wall expanding in the $x-y$ plane) and a computation using WKB phases for the $q$ and $N$ fields demonstrates that the probability $\mathcal{P}(q\to N)$ is non-vanishing \cite{Vanvlasselaer:2020niz} and is given by
 \bea
 \mathcal{P}(q \to N) \approx  \frac{Y_{}^2 v_{EW}^2}{M_N^2}\Theta(\gamma_w T_\text{nuc} - M_N^2 L_w)\;,
 \label{eq:prod_tree}
 \eea
with $L_w \sim 1/v_{EW}$ the
width of the wall. Behind the bubble wall a large abundance of $N$ and $\bar{N}$, $n_N^{BE}$ is produced. Let us emphasize that this abundance is much larger than its equilibrium value. 

Another possibility of the heavy particle
production can be realized for the following interaction
    \bea 
    \Delta \mathcal{L} \supset \frac{\lambda_{h\phi}}{2} \phi^2 h^2 + \frac{1}{2}M^2_\phi \phi^2
    \label{eq:portal}.
    \eea 
In this case $\phi$ is a heavy scalar field with mass $M_\phi \gg v_{EW}$, then in the vicinity of the wall, the process $h\to \phi \phi$ has the probability \cite{Vanvlasselaer:2020niz}
\bea
\qquad \mathcal{P}(h \to \phi^2) \approx  \bigg(\frac{\lambda_{h\phi} v_{EW}}{M_\phi}\bigg)^2\frac{1}{24\pi^2} \Theta(\gamma_w T_\text{nuc} - M_\phi^2 L^h_w).
\eea

The results in Sec. \ref{app:dynamics} on the terminal velocity in the singlet extension of SM allow us to compute the maximal mass of the particles which can be produced  during the electroweak FOPT in the singlet extension. Indeed saturating the step function in the above equation and assuming  the  $L_w \sim 1/v_{EW}$  we get approximately:
\bea 
M^{MAX} \approx \frac{{400}{\rm~ GeV}}{\log^{1/2} \frac{M_z}{g T_{\rm nuc}}}\l(\frac{\Delta V-\Delta P_{\rm LO}}{(100 {\rm~ GeV})^4}\r)^{1/2}\l(\frac{100 {\rm GeV}}{T_{\rm nuc}}\r).
\label{eq:mass_prod}
\eea
Numerical results for the maximal mass $M^{\rm MAX}$ are reported in Fig.\ref{fig:scan}. We can see that the maximal mass we can produce  is roughly
$\sim 10$ TeV scale.

We would like to note that our results can be easily applied for the mass gain mechanism of the heavy state production \cite{Baldes:2021vyz}. Indeed in this case the maximal mass will be $M_{\rm mass ~gain}\simeq \gamma_w T$, and can be read off from the bottom right plot of the Fig.\ref{fig:scan} by noting that it will scale as $M_{\rm mass ~gain}\sim M_{\rm MAX}^2/v_{EW}$. Since the mass of the heavy field comes from the vev of the Higgs, it will additionally be bounded by the unitarity considerations to be below $\lesssim 2$ TeV.

\subsection{Dark Matter production}
\label{sec:DM-production}
In this section we will apply the results for the velocity of the bubble expansion for DM model building.
\subsubsection{Scalar DM coupled to the Higgs portal}

We assume a heavy scalar $\phi$ coupled to the SM via the traditional Higgs portal 
\bea
\label{eq:lag_DM}
{\cal L}_{DM}=\frac{1}{2}(\d_\mu \phi)^2-\frac{M_\phi^2 \phi^2}{2}-{\frac{\lambda_{\phi h}}{2}}h^2\phi^2\;.
\eea
The DM ($\phi$) field is stabilized by some additional $Z_2^{\phi}$ (we use this subscript to differentiate it from $Z_2$ of the singlet potential). After the Higgs transition, the abundance of massive $\phi$, $n^{\rm BE}_\phi$, behind the wall is given by 
\bea
n_\phi^{\text{BE}} &\approx & \frac{2}{\gamma_w v_w}  \int \frac{d^3p}{(2\pi)^3}\frac{p_z}{p_0} \mathcal{P}(h \to \phi^2) \times f_h (p, T_{\text{nuc}})\;. 
\eea
We can see that
DM production  during the bubble expansion
is strongly dependent on the density  of the Higgs field available at the nucleation temperature $f_h (p, T_{\text{nuc}})$. The relevant parameter for the discussion is the ratio \bea 
\sqrt{\frac{d^2 V}{dh^2}\bigg|_{\rm fv}} \frac{1}{T} \equiv \frac{m_H^{\rm False}}{T}\;,
\eea 
where fv denotes the position of the false vacuum and $\sqrt{d^2 V/dh^2\big|_{\rm fv}}$ is the mass of the Higgs $m_H^{\rm False}$ in the false vacuum. As soon as this quantity becomes larger than 1, we expect exponential suppression of the Higgs abundance
\bea 
 C_{\rm eff} \frac{\zeta (3)T_{\text{nuc}}^3}{\pi^2} \equiv\int\frac{d^3p}{(2\pi)^3}  f_h (p, T_{\text{nuc}}) \approx  \begin{cases}
\frac{\zeta (3)T_{\text{nuc}}^3}{\pi^2}\quad \text{if} \quad m_H^{\rm False} <T\;,
\\
\bigg(\frac{m_H^{\rm False} T_{\text{nuc}}}{2\pi}\bigg)^{3/2}e^{-m_H^{\rm False}/T_{\text{nuc}}}\quad \text{if} \quad   m_H^{\rm False}>T\;.
\end{cases}  
\label{eq:ceff}
\eea 
Here we define $C_{\rm eff}$ to take into account the Boltzmann suppression. After redshifting to today, the stable produced abundance takes the form
\begin{align}
\Omega^{\text{today}}_{\phi,\text{BE}}h^2 & = \frac{M_\phi n_\phi^{\text{BE}} }{\rho_c/h^2} \frac{g_{\star S0}T_0^3}{g_{\star S}(T_{\text{reh}})T_{\text{reh}}^3} \approx 6.3\times 10^8\; \frac{M_\phi n_\phi^{\text{BE}} }{\text{GeV}}\frac{1}{g_{\star S}(T_{\text{reh}})T_{\text{reh}}^3}\;,
\nn
&\approx {5.4}\times 10^5  \times  \bigg(\frac{C_{\rm eff}\lambda_{h\phi}^2 v_{EW}}{ M_\phi g_{\star S}(T_{\text{reh}})}\bigg)\bigg(\frac{v_{EW}}{\text{GeV}}\bigg)\bigg(\frac{T_\text{nuc}}{T_{\text{reh}}}\bigg)^3 \times e^{-\frac{M_\phi^2}{2\gamma_w v_{EW} T_{\rm nuc}}}\;.
\label{eq:scaled_ab}
\end{align}
This expression has to be supplemented with the \emph{freeze-out}(FO) contribution which is produced \emph{before} the phase transition
\bea 
\label{eq:fo}
\Omega^{\text{today}}_{\phi,\text{FO}}h^2 &\approx & 0.1\times \bigg(\frac{T_\text{nuc}}{T_{\text{reh}}}\bigg)^3 \times \bigg(\frac{0.03}{\lambda_{\phi h}}\bigg)^2 \bigg(\frac{M_\phi}{100 \text{ GeV}}\bigg)^2\;,
\nn 
\Omega^{\text{today}}_{\phi,\text{tot}}h^2 &=& \Omega^{\text{today}}_{\phi,\text{BE}}h^2+ \Omega^{\text{today}}_{\phi,\text{FO}}h^2\;.\eea 
Note that the FO contribution is suppressed by the factor $\big(T_\text{nuc}/T_{\text{reh}}\big)^3$ due the brief stage of inflation during the phase transition. Obviously  the prediction for relic density must match the experimental observations: $\Omega^{\text{today}}_{\phi,\text{tot}}h^2 \approx 0.1$.  We can see from Eqs.(\ref{eq:scaled_ab}),(\ref{eq:fo}) that for small values of the portal coupling $\lambda_{h\phi}$, DM production will be dominated by the freeze out mechanism while bubble expansion takes over for larger values of $\lambda_{h\phi}$. 

Next we can check whether this mechanism for DM production can lead to viable phenomenology, given the results on bubble dynamics in section \ref{sec:num_results}. Instead of making a scan of the parameter space, we will just focus on a few representative benchmark points.
\begin{figure}
    \centering
   \begin{minipage}{.52\textwidth}
 \includegraphics[scale=0.58]{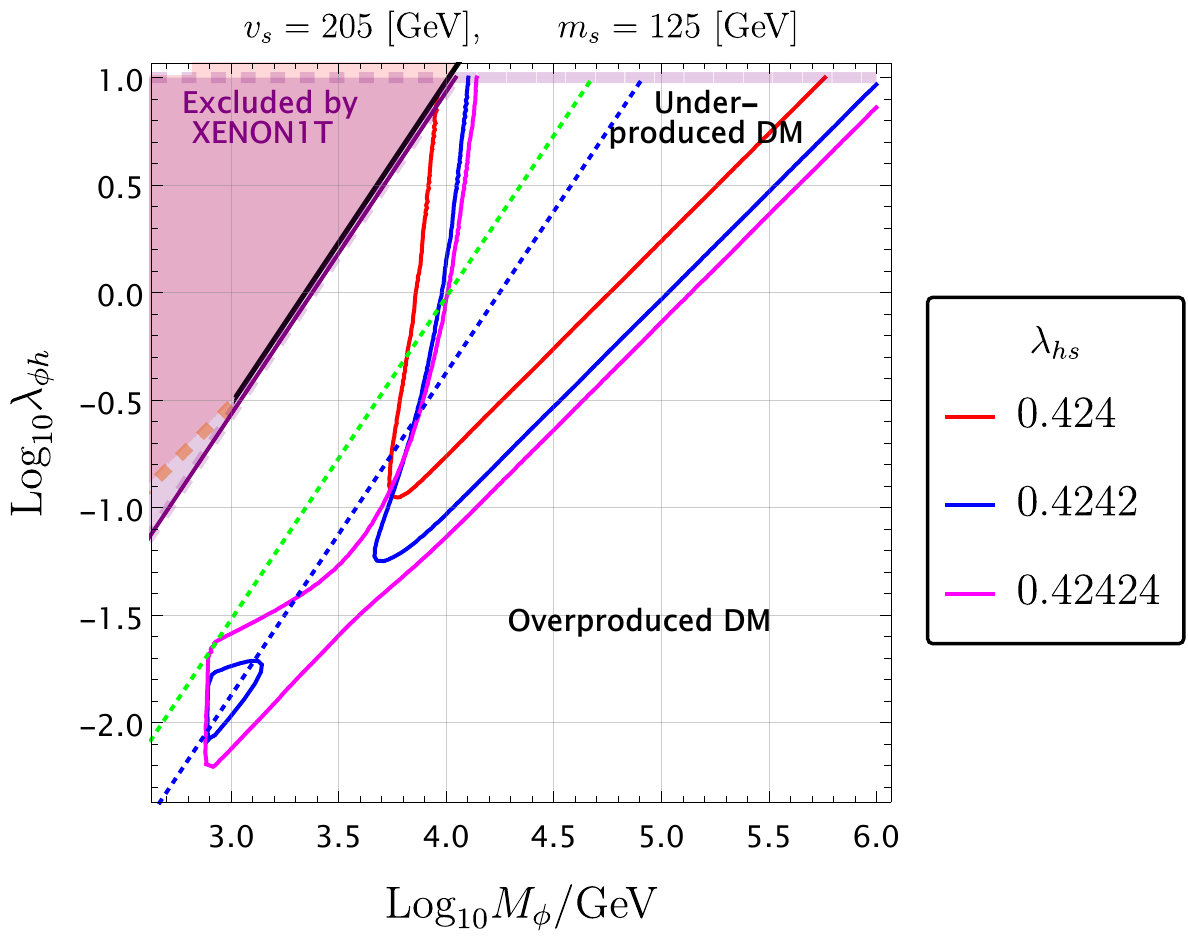}
 \end{minipage}
 \begin{minipage}{.45\textwidth}
  \includegraphics[scale=.58]{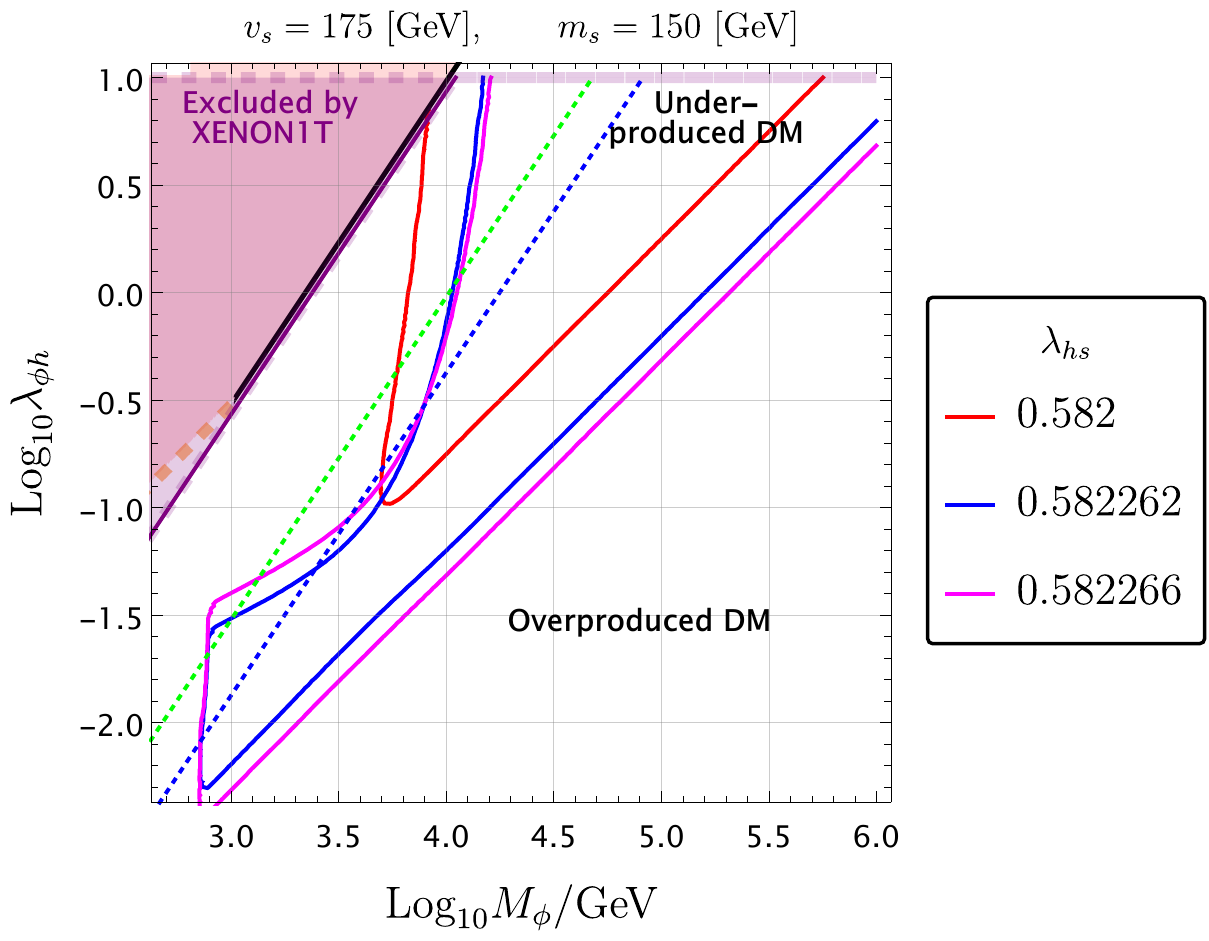}
 \end{minipage}
    \caption{{\bf Left}: DM abundance in the parameter space $\lambda_{\phi h}-M_\phi$ for different values of $\lambda_{hs}$ by fixing $v_s= 205$ GeV and $m_s=125$ GeV (that fixes the values of $T_{\rm nuc}, T_{\rm reh}$ and $\gamma_w^{\rm terminal}$). The solid lines represent correct  DM abundance, while underproduced inside and overproduced outside. The lower part of each contour is dominated by freeze-out and the upper part via bubble expansion. The connecting vertical line (independent of the portal) comes from thermal production after the reheating of the transition. The magenta shaded region is excluded by XENON1T while the dotted green and blue lines are projected limits from XENONnT and DARWIN respectively.  {\bf Right}: same plot for $v_s= 175$ GeV and $m_s=150$ GeV. As expected, increasing the tuning from red to magenta increases the amplitude of the curve. The values used are extracted from Table \ref{benchmarkpointthird}.  
    \label{fig:DM_plot}
}
\end{figure}

For $m_s=125, v_s=205$ GeV, we show in Fig.\ref{fig:DM_plot} the isocontours reproducing the correct relic density for three reference values of $\lambda_{hs}$ (corresponding nucleation temperatures can be found in the Table \ref{benchmarkpoint}). Firstly, $C_{\rm eff}\simeq 1$ for all three reference points. For $\lambda_{hs}=0.424$ the upper red curve corresponds to the case when DM production is dominated by the BE (bubble expansion) and the lower curve by FO. The steepness of the upper red curve (BE) comes from the fact that we are always in the region of parameter space where $\exp\big[{-M_\phi^2/(2\gamma_w v_{EW} T_{\rm nuc})}\big] \ll 1$, leading to a very strong sensitivity  on $M_\phi$ mass. Physically this means that the model generically predicts large overproduction of DM in BE process unless the Boltzmann suppression $\exp\big[{-M_\phi^2/(2\gamma_w v_{EW} T_{\rm nuc})}\big]$ is playing a role. For the other two reference points $\lambda_{hs}=0.4242,0.42424$ we can see that there is an additional part of parameter space for the DM masses $M_\phi \sim 1-4$ TeV, which corresponds to the region without the Boltzmann suppression $\exp\big[{-M_\phi^2/(2\gamma_w v_{EW} T_{\rm nuc})}\big] \sim 1$. This is related to larger values of $M^{\rm MAX}\sim \sqrt{\gamma_w v_{EW} T_{\rm nuc}}$ and smaller values of the nucleation temperature, reducing the excess of the DM abundance. On the right panel of Fig.\ref{fig:DM_plot}, we report similar plots for  $v_s=175, m_s=150$ GeV.

Finally, before closing this section, we comment about the possibility of considering the singlet $s$ itself, in the limit of very precise $Z_2$, as DM. After the phase transition $T\sim 40$ GeV, the singlet is in thermal equilibrium and we can apply straightforwardly the freeze-out expression:
    \bea
\Omega^{\text{today}}_{s,\text{ FO}}h^2 \approx 0.1 \bigg(\frac{0.06}{\lambda_{hs}}\bigg)^2 \bigg(\frac{M_s(v_{EW}, 0)}{100 \text{ GeV}}\bigg)^2\;.
\label{eq:relic_FO}
\eea
From this estimate of the FO abundance for $s$ and recalling that we considered $\lambda_{hs}\sim 0.3-0.6$ and $M_s(v_{EW}, 0) \sim 100$ GeV, we conclude that the abundance of $s$  produced in this fashion, today, is underproduced by one or two orders of magnitude to fit the observed amount of DM $\Omega^{\text{today}}_{s,\text{ FO}}h^2 \approx 0.1$. {Even in this underproduced case, there are severe bounds from the direct detection experiments except for the resonant region, where \eqref{eq:relic_FO} is over-estimated. However, as we will discuss in the Appendix.~\ref{app:DW}, we will have a $Z_2$ explicit breaking which makes $s$ decay much before today.}

\subsubsection{Singlet portal DM}
\label{sec:singletportalDM}

\begin{figure}
    \centering
  \includegraphics[scale=0.65]{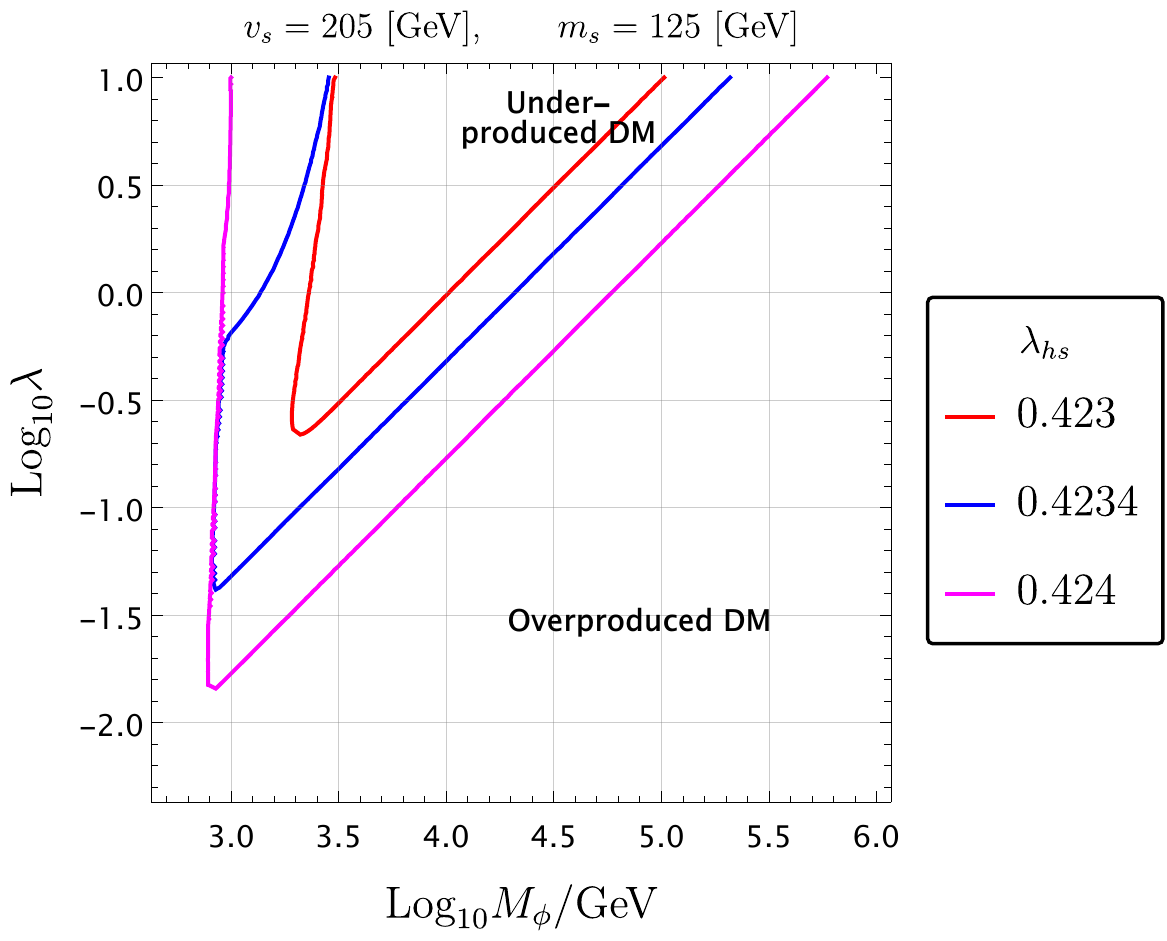}
    \caption{DM production for the singlet portal model. In this context, the Boltzmann suppression $\sim e^{-m_s/T_{\rm nuc}}$ plays a strong role and allows for bubble expansion produced DM with much higher nucleation temperature $T_{\rm nuc} \sim 15$ GeV.
    \label{fig:singlet_portal}}
\end{figure}

In this section we mention an alternate possibility of coupling the DM ($\phi$) to the singlet field $s$ via ``singlet portal"
      \bea 
    \Delta \mathcal{L} \supset \frac{\lambda_{s\phi}}{2} \phi^2 s^2 + \frac{1}{2}M^2_\phi \phi^2\;, \quad \mathcal{ P}(s \to \phi^2) \approx \ \bigg(\frac{\lambda_{s\phi} v_{s}}{M_\phi}\bigg)^2\frac{1}{24\pi^2} \Theta(\gamma_w T_\text{nuc} - M_\phi^2 L^s_w)\;,
    \label{eq:portal_2}
    \eea 
    where the width of the singlet wall is similar to the length of the Higgs wall $L_w$. Interestingly even though FOPT is from $(0, v_s)\to (v_{EW}, 0)$, the singlet scattering of the wall can lead to the production of the $\phi$ field. Phenomenology of DM production is very similar to the Higgs portal case discussed in the previous section, with one main difference: in the false vacuum, the mass of the singlet is not small and the factor $C_{\rm eff}$ introduced in the Eq.(\ref{eq:ceff}) plays an important role. The results are shown in Fig.\ref{fig:singlet_portal}. For example, if we compare the curves for $\lambda_{hs}=0.424,m_s=125,v_s=205$ in Fig.\ref{fig:singlet_portal} and in Fig.\ref{fig:DM_plot}, we can see that, for the Higgs portal DM, the isocontour has the same shape, but with the larger values of DM $M_\phi$ masses. As shown in Eq.(\ref{eq:scaled_ab}), this is due to the proportionality between the DM relic abundance produced during the bubble expansion with $\propto C_{\rm eff}/M_\phi $.

\paragraph{Singlet portal with additional field}

A slight modification of this scenario is to further introduce a light scalar $\tilde{s}$. Then we can have, \bea {\cal L}={\cal L}_{SM}- \tilde{\lambda} \tilde{s} s \phi^2 -\frac{M_\phi^2}{2}\phi^2,\eea where again $\phi$ is the DM and we did not write down the mass terms for simplicity of notation. 
We assume that $\tilde s$ is in the thermal bath before the PT. Then due to the field change of $s$ in the bubble wall, the momentum conservation violating process $\tilde{s}\to \phi \phi$ can occur ($h\to \phi\phi$ may also occur if there is the $h^2 \phi^2$ term.) In this model $s\to \tilde{s}+SM$ particles happen via the DM loop. 

\subsubsection{Fermion-mediated Dark Matter}

In the previous section we noticed that the DM production during bubble expansion strongly depends on the mass of the incoming particle in the symmetric phase, due to the Boltzmann suppression factor. In case of an incoming scalar, generically this effect is relevant and crucially modifies the phenomenology, as we have seen in the section \ref{sec:singletportalDM}. In this section we construct a model where the incoming particle is a massless fermion in the symmetric phase, so that $C_{\rm eff}\equiv 1$ by definition. The  model consists of a vector-like neutral fermion $N$ which is a singlet under SM and a couple of $Z_2$ odd fields $\phi$ and $\chi$:
\bea
{\cal L}={\cal L}_{SM}+ Y_* \bar L H N+ M_N \bar N N + Y_{DM} \bar N \chi \phi\;.
\eea
Here, $L,H$ are SM lepton and Higgs doublets, respectively. The production mechanism works as follows: the heavy field $N$ is produced during the phase transition $L\to N$ and it will subsequently decay into $N\to \chi \phi, N\to LH$. The field $N$ can be Majorana or Dirac (in the former case there  will be a relation to neutrino masses and in the later it will be completely independent from  neutrinos). 

In this model, heavy $N$ are produced via $L \to N$ with a probability 
\bea
 \mathcal{P}^{\rm tree}(L \to N) \approx  \frac{Y_{\star}^2 v_{EW}^2}{M_N^2}\Theta(\gamma_w T_\text{nuc} - M_{N}^2 L_w)\;.
 \label{eq:prod_tree}
 \eea
 As a consequence, unstable heavy $N$ accumulate behind the wall with initial density given by
\bea
n_N^{\text{BE}} &\approx&
 \frac{Y_\star^2 v_{EW}^2}{M_{N}^2 \gamma_w v_w}  \int \frac{d^3p}{(2\pi)^3} \frac{p_z}{p_0} \times f_{L} (p,T_\text{nuc})\Theta ( p_z- M_N^2/v_{EW}) \;,
 \nn 
&\approx& \frac{Y_\star^2 v_{EW}^2  T_\text{nuc}^3 }{2\pi^2 M_N^2}  e^{-  \frac{M_N^2}{2v_{EW}T_\text{nuc} \gamma_w }}  + \mathcal{O}(1/\gamma_w)\;,
\label{eq:density_alt}
\eea 
where $v_w = \sqrt{1- 1/\gamma_w^2}$, we expanded for large $\gamma_w$ and approximated the Fermi-Dirac distribution as a Boltzmann distribution. Compared to the original proposal in Ref.\cite{Azatov:2021ifm}, the density of the heavy fields inside the  bubble will be additionally enhanced by $\sim 16\pi^2$ factor since $1\to 1$ transitions are more effective than $1\to 2$.  Let us assume that $M_\phi < M_\chi$ so that $\phi$ is the DM candidate, then DM production will happen  via the following chain of processes:
\bea 
\chi \underset{\text{via PT}}{\rightarrow} N \underset{\text{via decay}}{\rightarrow} \phi \chi \to \phi\phi + \text{SM}\;.
\eea
However, the heavy $N$ has two channels of decay: toward the heavy dark sector $\phi, \chi$ and back to the light $L$. The abundance of $\phi, \chi$ after the transition is thus suppressed and given by 
\bea 
n_\chi \approx n_\phi \approx \frac{Y_{DM}^2Y_\star^2}{Y^2_{DM}+Y^2_{\star}}\frac{ v_{EW}^2  T_\text{nuc}^3 }{2\pi^2 M_N^2}  e^{-  \frac{M_N^2}{2v_{EW}T_\text{nuc} \gamma_w }}  + \mathcal{O}(1/\gamma_w)\;,
\eea
and the final relic abundance redshifted to today thus reads 
\bea
\Omega^{\text{today}}_{\phi,\text{BE}}h^2&\approx &   {1.5}\times 10^8 \times \frac{Y^2_\star Y_{DM}^2}{Y_{\star}^2+Y^2_{DM}}\frac{2M_\phi}{M_N} \bigg(\frac{v_{EW}}{ M_N}\bigg)\bigg(\frac{v_{EW}}{ 246 {\rm GeV}}\bigg) \bigg(\frac{T_\text{nuc}}{T_{\text{reh}}}\bigg)^3 e^{-  \frac{M_N^2}{2v_{EW}T_\text{nuc} \gamma_w }}\;.\nn
\label{eq:relic_ab_alt}
\eea
For the freeze-out process in the symmetric phase, we have: $\phi \phi \to L H L H$ by neglecting co-annihilation. The cross-section is highly phase space suppressed (closing a loop for a 2 to 2 annihilation gives a similar scaling): $\sigma_{\phi \phi \to (LH)^* LH}\sim \frac{M_\chi^2 (Y_{\rm DM} Y_*)^2 }{(16\pi^2)^2 4\pi M_N^4}.$ The abundance by taking account the supercooling is 
\bea
\Omega^{\text{today}}_{\phi,\text{FO}}h^2= 10^3 \l( \frac{T_\text{nuc}}{T_{\text{reh}}}\r)^3 \frac{M_N^4/M_\chi^2}{(6{\rm TeV})^2}\frac{10}{(Y_{\rm DM} Y_*)^4}\;.
\label{eq:FOloop}
\eea 
The total DM density today will be given by the sum of Eq.(\ref{eq:relic_ab_alt})-(\ref{eq:FOloop}). Therefore, this scenario leads to the over-production of DM unless  $M_\phi, M_\chi\lesssim 10$ GeV. However, note that these equations are valid only  for {the} heavy DM candidates which do not go back to equilibrium after the phase transition. Otherwise, the final density will be given by Eq.(\ref{eq:FOloop}) only without $\big(T_\text{nuc}/T_{\text{reh}}\big)^3$ and we are going back to the normal freeze-out scenario.

\begin{figure}
    \centering
  \includegraphics[scale=0.65]{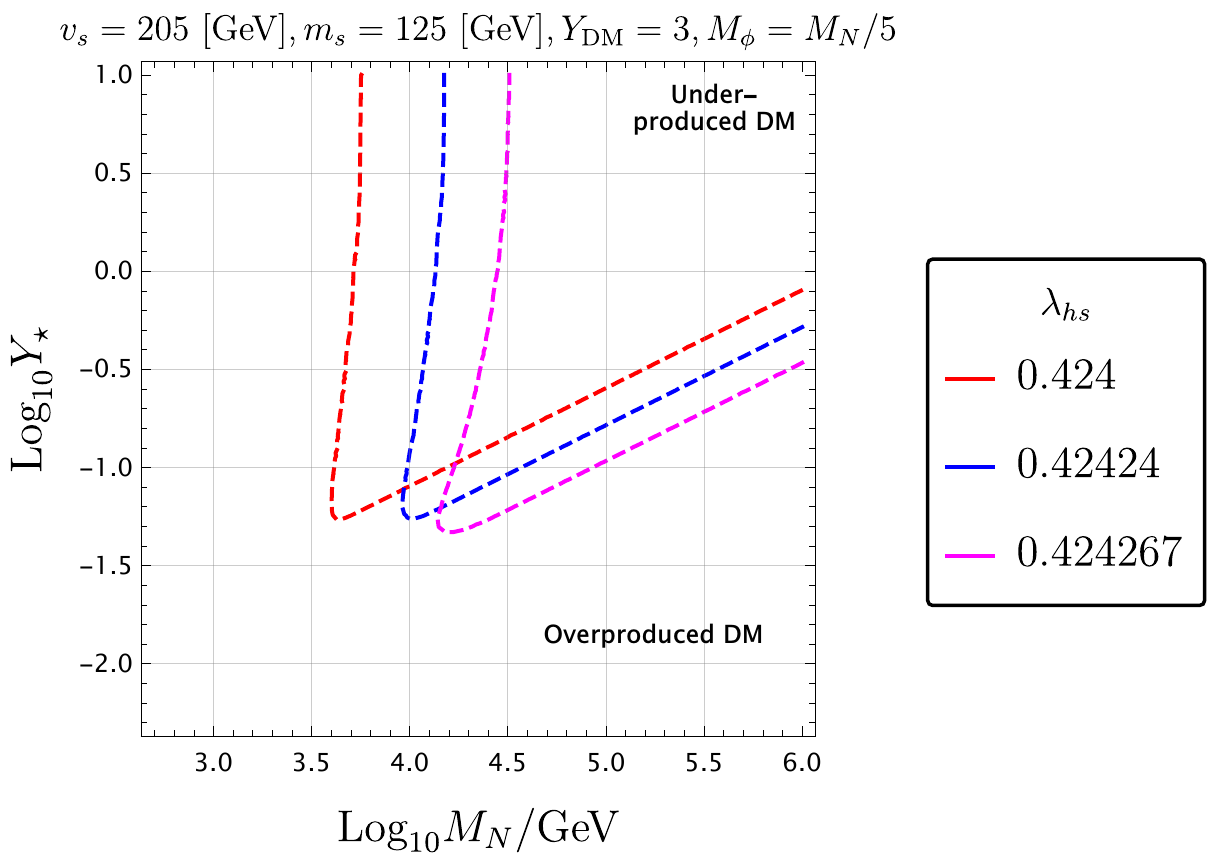}
    \caption{DM production in the fermionic portal model. We fix $M_{\rm DM}=M_N/5$, and $Y_{\rm DM}=3$. Only the co-annihilation regime where $|M_\phi-M_\chi| \lesssim M_\phi/20$ appears to be viable and shown by dashed lines.}
    \label{fig:singlet_fer}
\end{figure}

Let us now investigate the regime $M_\phi\simeq M_\chi$, precisely  $|M_\phi-M_\chi| \lesssim M_\phi/20$, where  the co-annihilation takes place. In this case we have the channel $\phi \chi \to H {\bar L}$ to decrease the abundance of $\phi$ as well as $\chi$. The cross-section is $\sigma_{\phi \chi \to H {\bar L}}\sim \frac{(Y_{\rm DM} Y_*)^2}{4\pi M_N^2}.$ Therefore, we have \bea\Omega^{\text{today}}_{\phi,\text{FO,Co}}h^2\sim  0.1  \l( \frac{T_\text{nuc}}{T_{\text{reh}}}\r)^3\times \frac{M_N^2}{(10{\rm TeV})^2}\frac{1}{(Y_{\rm DM} Y_*)^2}\;.
\label{eq:FOCO}
\eea 
Summing this estimate with the $\Omega^{\text{today}}_{\phi,\text{BE}}h^2$ in Eq.(\ref{eq:relic_ab_alt}) we find that it becomes possible to reproduce the observed DM abundance. However we see that bubble expansion tends to overproduce the DM and the relic abundance in BE can be reproduced if only the factor $\exp[-M_N^2/(2 v_{EW} T_{\rm nuc}\gamma_w)]$ starts playing a role in suppressing  DM relic density (left boundary of  Fig.\ref{fig:singlet_fer} is almost vertical).
\label{sec:singletportal}

\subsection{Baryogenesis mechanism}
\label{sec:baryogenesis}
Now we remind the scenario of Baryogenesis with relativistic bubble walls that was proposed in \cite{Azatov:2021irb}. As a prototype, we worked with the following model, reminiscent of the toy model of Eq.\eqref{eq:Toy_model_1}, (omitting the kinetic terms)
\bea
\label{eq:modelB}
{\cal L} =&&{\cal L}_{SM} +  m_\eta^2 |\eta |^2+\sum_{I=1,2} M_{I} \bar B_I B_I \nn
+&&\l(\sum_{I=1,2}Y_I (\bar B_I H)P_L Q+
{y_I \eta^*\bar B_I P_R\chi}
+ \kappa \eta^c d u
 +\frac{1}{2} m_\chi \bar{\chi^c} \chi + h.c. \r) .
\eea
Thus, additionally to the SM and the singlet sector, the model contains a Majorana field $\chi$ and two  vector-like $B$ quarks with the masses $M_{1,2}\sim m_\chi$.
$\eta$ is a scalar field in the fundamental representation of QCD  and with electric charge $Q(\eta)=1/3$. We defined $Q,u,d$  as  the  SM quark doublet and singlets respectively, and ignored the flavour indices. $H$ is the SM Higgs.

We first proved that the production mechanism, when computed up to one loop-level, indeed transforms a CP-violating phase into a chiral asymmetry in the abundances produced, in a fashion very similar to usual leptogenesis decay. Let us now sketch the mechanism in itself. First, $b-$quarks collide with the relativistic wall and produce $B_I, B_I^c$ via the mechanism explained above. Thus inside the bubble we have
\bea
n_{B_I}-n_{B^c_{I}}=-\bigg(\frac{Y_I v_{EW}}{M_I}\bigg)^2 \epsilon_I n_b^{\rm inc}\;,\qquad 
n_{b}-n_{b^c}=\sum_I\bigg(\frac{Y_I v_{EW}}{M_I}\bigg)^2 \epsilon_I n_b^{\rm inc}\;,
\label{eq:asymmB}
\eea
where $n_b^{\rm inc}$ is the number density of the bottom-type quark colliding with the wall from outside, $n_b$ and $n_B$ are the abundances inside of the bubble and  $\epsilon$ is a loop suppressed coefficient which parametrizes the $CP$ violating phase and the resonance between one-loop and tree-level diagrams. After the passage through the wall, the following asymmetric abundances are then generated
\bea
\sum_I  \l(n_{B_I}-n_{ B_I^c}\r)
= -( n_{b}- n_{b^c})\;.
\eea
From this expression we can read an apparent asymmetry in the bottom quark abundances. However, if the heavy $B$ freshly produced were to decay back into $b$, the asymmetry would be washed out. This is however not the case if they decay in a dark sector containing $\chi, \eta$, where the asymmetry is enhanced by the presence of a Majorana mass for $\chi$. The final unsuppressed produced asymmetry is given by
\bea
 \frac{\Delta n_{\rm Baryon}}{s}  
\approx&\frac{135 \zeta(3)}{8 \pi^4} \sum_{I,{J}}\bigg(\frac{Y_I \vev{H}}{M_I}\bigg)^2 \frac{|y_I|^2}{|y_I|^2+|Y_I|^2}\times \frac{g_b}{ g_\star}\bigg(\frac{T_{\rm nuc}}{T_{reh}}\bigg)^3\nn
&
\times\; {\rm Im}  (Y_I Y^*_J y_I^*y_J )
\l(-\frac{2 {\rm Im} [f_B^{IJ}]}{|Y_I|^2}+\frac{4 
{\rm Im} [f_B^{IJ}]|_{m_{\chi,\eta}\to 0}
}{|y_I|^2}
\r)\;,
 \eea
 where $s$ is the entropy at the moment of the production, $g_b$ is number of degrees of freedom of the bottom and $g_\star$ the number of relativistic degrees of freedom. The loop functions $f_B^{IJ}$  have been computed in \cite{Azatov:2021irb} and are controlled by the CP violating sector. 
 Absence of strong wash-out conditions 
 \bea
M_{B,\chi,\eta}\gtrsim 30 T_{\rm reh} \sim 10^3 \text{ GeV}\;,\qquad  \text{(suppressed wash-out)}
\eea
as well as experimental signatures (direct production in colliders, flavor violation, neutron oscillations) pushed the heavy particles to be 
\bea 
M_{B,\chi,\eta} \gtrsim 2\times 10^3 \text{ GeV}\;.
\eea 
In the context of singlet extension with $Z_2$ that we studied, this opens up the range 
\bea 
M_{B,\chi,\eta} \in [2, 10] \text{ TeV}\;,
\eea
where the Baryogenesis mechanism proposed above is operative.

\subsection{Impact of the heavy sector on the phase transition}
The models we are considering by construction have new heavy fields coupled to the Higgs boson. These will lead to the  finite  corrections to the scalar parameters of the form (assuming a Yukawa type connection $yBHb$)
\bea 
\label{eq:tuningmh}
\delta m_h^2 \sim  \frac{-8 g_N M_N^2y^2}{64\pi^2} \bigg(\log \frac{M_N^2}{v_{EW}^2} - \frac{3}{2}\bigg), \qquad
  \delta\lambda \sim  4g_N \frac{y^4}{64\pi^2} \bigg(\log \frac{M_N^2}{v_{EW}^2}-\frac{3}{2}\bigg),
  \eea 
 where $g_N$ is the number of heavy degrees of freedom and 
 $M_N$ is the typical mass of the heavy sector. One can
 wonder how these corrections can effect the tuning of the 
 Higgs potential. However note that in our model the Higgs mass hierarchy problem is not 
 addressed and generically we expect the size of $m_h^2$ to 
 be of the order of the cut off scale  ($M_{\rm pl}$ in SM).  So the corrections in 
 Eq.(\ref{eq:tuningmh}) do not make the tuning worse. 
 
 In case the Higgs hierarchy problem is solved at the scale of the heavy fields in Eq.(\ref{eq:lag_DM}),(\ref{eq:modelB}) the 
 tuning in the Higgs potential will be roughly,
 \bea
 {\rm tuning}\sim \frac{m_h^2}{\Delta_{\rm Heavy~physics} m_h^2}\sim \frac{8 \pi^2 m_h^2}{y^2  M_N^2}.
 \eea
We can combine this estimate with a tuning  for low nucleation temperatures (see discussion in  section \ref{sec:nobarrier}) which are necessary for the heavy field production and the tuning estimate becomes:
\bea
{\rm tuning}\sim \frac{T^2_{\rm nuc}}{\Delta_{\rm Heavy~physics} m_h^2}\sim \frac{8 \pi^2 T_{\rm nuc}^2}{y^2 M_N^2}.
\eea
Using the estimates of the maximal values of $\gamma_w$ and the maximal mass of  heavy particles which can be produced during the bubble-plasma collisions  
(see Eq.(\ref{eq:terminal_velo}) and Eq.(\ref{eq:prod_tree})) 
we get the following estimate for the maximal tuning in the model
\bea
{\rm  tuning}^{MAX}\sim \l(\frac{T_{\rm nuc}}{20  {\rm GeV}}\r)^4 ,
\eea
where we remind the reader that this estimate is valid only if the Higgs hierarchy problem is solved at the heavy fields scale.

\section{Gravitational waves emitted}
\label{sec:GW}
It is well known that strong gravitational waves background will be emitted, with peak frequency around the mHz if the EWPT happens to be strongly first order. This is the optimal range of sensitivity of the forthcoming LISA detector~\cite{amaroseoane2017laser,Caprini:2019egz}
and also GW detectors such as eLISA\cite{Caprini:2015zlo}, LIGO\cite{vonHarling:2019gme, Brdar:2018num}, BBO\cite{Corbin:2005ny, Crowder:2005nr}, DECIGO\cite{Seto:2001qf, Yagi:2011wg, Isoyama:2018rjb}, ET\cite{Hild:2010id,Sathyaprakash:2012jk,Maggiore:2019uih}, AION\cite{Badurina:2019hst}, AEDGE\cite{Bertoldi:2019tck}.

The signal produced at the moment of the transition can be separated into different contributions: the \emph{bubble collision}\cite{Cutting:2018tjt} contribution, the \emph{plasma sound waves}\cite{Caprini:2019egz} and finally the \emph{turbulence}. Only the two first sources of GW are well understood. Another nice feature of those two sources is that they are expected to dominate in different physical situations; the bubble collision would dominate in case of runaway wall and the sound waves if the wall reaches a terminal velocity. We have already mentioned that the EWPT, if first order, will always happen in the regime of terminal velocity, because of the large number of strongly coupled vector bosons\footnote{A possible exception would be the case of extreme cooling, as hinted in \cite{Gouttenoire:2021kjv} where even the pressure from gauge bosons cannot stop the acceleration of the wall. However, in our study, we do not find such a situation.}. For GW produced by plasma sound wave, the peak frequency and amplitude are given by 
\begin{align}
\Omega_{\text{plasma}}^{\text{peak}}h^2 &\approx 0.7\times 10^{-5} \bigg(\frac{100}{g_\star}\bigg)^{1/3}\bigg(\frac{\kappa_{sw}\alpha}{1+\alpha}\bigg)^2 (H_{\text{reh}}R_\star),
\nn
 f_{\text{peak}} &\approx 2.6\times 10^{-5}\bigg(\frac{1}{H_{\text{reh}} R_\star}\bigg)\bigg(\frac{z_p}{10}\bigg)\bigg(\frac{T_{\text{reh}}}{100 \text{ GeV}}\bigg)\bigg(\frac{g_\star}{100}\bigg)^{1/6} \text{ Hz} 
 \label{eq:GWsignal}
\end{align}
with $z_p \sim 10$, $\kappa_{sw}$ is the efficiency factor for the production of sound waves in the plasma\cite{Espinosa:2010hh},
\bea
\kappa_{sw} \approx  \frac{\alpha}{0.73+ 0.083\sqrt{\alpha} + \alpha},
\eea
$\alpha$ and $R_{\star}$ have been defined in Eqs.\eqref{eq:alpha} and \eqref{eq:R*} respectively, $R_\star \sim \mathcal{O}(10^{-1}-10^{-4})H^{-1}$ is the approximate size of the bubble at collision and all quantities $(T,H,g_\star)$ have to be evaluated at \emph{reheating}. 

As we have seen in sections \ref{sec:baryogenesis}-\ref{sec:DM-production},  for the baryogenesis and DM production  we need relativistic walls with relatively low 
 nucleation temperature $\lesssim 10$ GeV. In this context, $\alpha\gg 1$ and $\kappa_{sw} \to 1$. The peak frequency and the signal amplitude are only function of the size of the bubbles at collision, which are reported in Table \ref{benchmarkpointbis} and \ref{benchmarkpointthird}. We can observe that in this range $\beta/H$ spans the value between $[50, 10^4]$, with a preference for lower values. Going back to Eq.\eqref{eq:GWsignal}, emitted amplitude and frequencies will be of the order
\bea 
\Omega_{\text{plasma}}^{\text{peak}}h^2 \in [5\times 10^{-7}, 2 \times 10^{-9} ], \qquad f_{\text{peak}} \in [10^{-4}, 0.03] \quad \text{Hz}
\eea 
where we set $z_p = 10, g_\star = 100$. This range of frequencies and amplitude are largely in the expected sensitivity of the coming observer LISA\cite{Caprini:2015zlo,Caprini:2019egz}, as expected for this class of models\cite{Ellis:2019oqb}. We thus conclude that strong GW signal in the LISA with spectrum controlled by the plasma sound waves is a generic prediction of Baryogenesis with relativistic bubble walls. This is in sharp opposition with the general expectation that usual EWBG demands slow walls, and thus suppressed signals. 

As a final comment, it should however be noticed that the current simulations do not directly provide a solutions for the regime of large $\alpha$, and we only have an extrapolation of the numerical result. Thus, the conclusion above should be taken with a grain of salt.

\section{Conclusion}
\label{sec:conclusion}
In this study we have presented the first explicit realization of  the baryogenesis and DM production during electroweak phase transition  for ultra-relativistic bubble expansion. The work is based on the proposals in \cite{Vanvlasselaer:2020niz,Azatov:2021ifm,Azatov:2021irb,Baldes:2021vyz} where new heavy particles are produced in plasma$-$bubble wall collisions.  We have shown that the model with SM extended by a real singlet with a $Z_2$ symmetry can indeed lead to ultra-relativistic bubbles, where the Lorentz factor $\gamma_w$ can reach the values $\sim 10^{5-6}$. Such fast bubbles can appear if the symmetry breaking occurs in two steps: first  discrete $Z_2$ is spontaneously broken and in the second step  electroweak symmetry breaking is accompanied by $Z_2$ restoration $(0,0) \xrightarrow{SOPT} (0,v_s) \xrightarrow{FOPT} (v_{EW},0) $. We find that there exists a region of parameter space where the nucleation temperature can become as low as $1-2$ GeV and the collision of the bubble wall with the plasma particles can  lead to the non-thermal  heavy particle production with the masses  up to $\sim 10$ TeV. Interestingly we find that the mechanism is most efficient for relatively low masses of the singlet field ${M_s(v_{EW}, 0)}\sim 70-100\ {\rm GeV}$, close to the region excluded by the Higgs invisible decays.  Subsequently, this region of parameter space will be probed by HL-LHC (\cite{Curtin:2014jma,Argyropoulos:2021sav}) in the singlet production mediated by off-shell Higgs boson. By noting the slight $Z_2$ breaking, $s$, if produced, can decay into $b\bar{b}$ in collider experiments. Depending on the size of the breaking displaced vertices of $b \bar{b}$ may be probed. We find the  typical bubble radius parameter of the order of $R_{\star}\sim (10^{-4}-1)H^{-1}$ so that stochastic gravitational background signal becomes observable at GW experiments like LISA\cite{Caprini:2015zlo,Caprini:2019egz}. 

The model necessarily requires tuning $\propto (T_{\rm nuc}/m_h)^2$ which numerically turns out to be of the order of $10^{-4}-10^{-2}$ (using Giudice-Barbieri measure) for successful baryogenesis and DM production mechanism. In spite of this we believe it  can provide a useful guidance for more appealing models where these hierarchies can appear naturally.

\section*{Acknowledgements}
AA and SC were supported by the MIUR contract 2017L5W2PT. WY was supported by JSPS KAKENHI Grant Nos.20H05851, 21K20364, 22K14029, and 22H01215. MV thanks the DESY institute for hospitality during the writing of this manuscript as well as Alberto Mariotti and Simone Blasi for fruitful discussions in the late stages of the completion of the paper. 

\appendix

\section{The bounce in two dimensions}
\label{app:codefor2D}
In this paper we  studied  numerically the phase transition from the minimum $(0, v_s)$ (or in the vicinity of it) to $(v_{EW},0)$.
The bounce computation can be done using existing codes  for example
\texttt{FindBounce} or  \texttt{CosmoTransition}. However we  have found that in the regime  of long supercooling where the potential around the false vacuum is very flat and, the existing codes are often not stable and lead to numerical errors.
Thus we have  developed our own code (more stable for the flat potentials), following the procedure described in \cite{Bruggisser:2018mrt}, while cross-checking the available values with \texttt{FindBounce}. 
\subsection{Computation of the bounce profile}
In this Appendix, we briefly review the standard computation of the bounce action with only one field before going to describe the algorithm we used for the same computation but for the case of two fields PT. In order to compute the vacuum tunneling probability from the false vacuum to the true one in $d$ dimensions, we need to minimize the Euclidean action given by 
\bea 
S_E = \int d^d x\bigg[\frac{1}{2}(\partial_\mu \phi)^2+V[\phi]\bigg].
\label{eq:bounce_action}
\eea 
It is known that the field configurations leading to the minimal action are the ones that exhibit an $O(d)$ spherical symmetry, then the so-called bounce solution is the solution of the following Cauchy problem 
\bea 
\frac{d^2\phi}{dr^2} + \frac{d-1}{r}\frac{d\phi}{dr} = \frac{d V[\phi]}{d\phi}, \qquad
\lim_{r \to \infty} \phi(r) = 0, \qquad \frac{d\phi}{dr}\bigg|_{r= 0} = 0,
\label{eq:problem}
\eea 
where we have chosen the false minimum to be at $\phi = 0$. If we interpret the parameter $r$ as a time and $\phi$ as a position, this problem becomes formally equivalent to the evolution of a mechanical ball in a potential $-V[\phi]$ undergoing a friction $\frac{d-1}{r}\frac{d\phi}{dr}$, released with vanishing velocity and stopping its evolution for $r \to \infty$ at $\phi = 0$. It is well known that this problem can be solved by applying numerically an overshoot/undershoot method on the position of the released point. Releasing the ball too close to the {true} vacuum would induce an overshoot configuration (the ball would continue after crossing $\phi = 0$), we would thus shift the release point toward the {false} vacuum, while releasing it too close would end up in an undershoot configuration (the ball would never reach $\phi = 0$ and starts oscillate around the minimum of $-V[\phi]$) and we correct it by shifting the release point farther from the false vacuum. Iterating between those two situations, we are able to find the correct release point and obtain the \emph{bounce solution}. 

It is well known that the case of a PT triggered by temperature fluctuation at temperature $T$ is formally equivalent to imposing a periodicity $T^{-1}$ in the imaginary time $t_E$, which imposes the following constraint on the field
\bea 
\phi [t_E, \vec{x}] = \phi [t_E+T^{-1}, \vec{x}]
\eea 
and the computation of thermally induced phase transition thus amounts to take $d=3$ in the above equations. 
\subsection{Bounce action in two dimensions and path deformation}
The problem complexifies when the transition involves many fields. Here there is no straightforward intuition for the path followed by the fields in field space during the tunneling. One can think that a straight line, connecting the two minima, could be a reasonable guess, but it turns out that it cannot be considered as a good approximation of the euclidean action\footnote{Let us emphasize that in the region of the parameter space we studied, the straight line between the false and the true vacuum gives an Euclidean action which is often wrong by \emph{orders of magnitudes}, as the path is often very far from the straight line, as a consequence, we cannot dispense from the effort of studying the exact 2D path in field space.}. Here we thus describe the algorithm \cite{Bruggisser:2018mrt} to find the right path in field space. In a multi-field case, the Eq.\eqref{eq:problem} becomes 
\begin{figure}
    \centering
    \includegraphics[width=.5\textwidth]{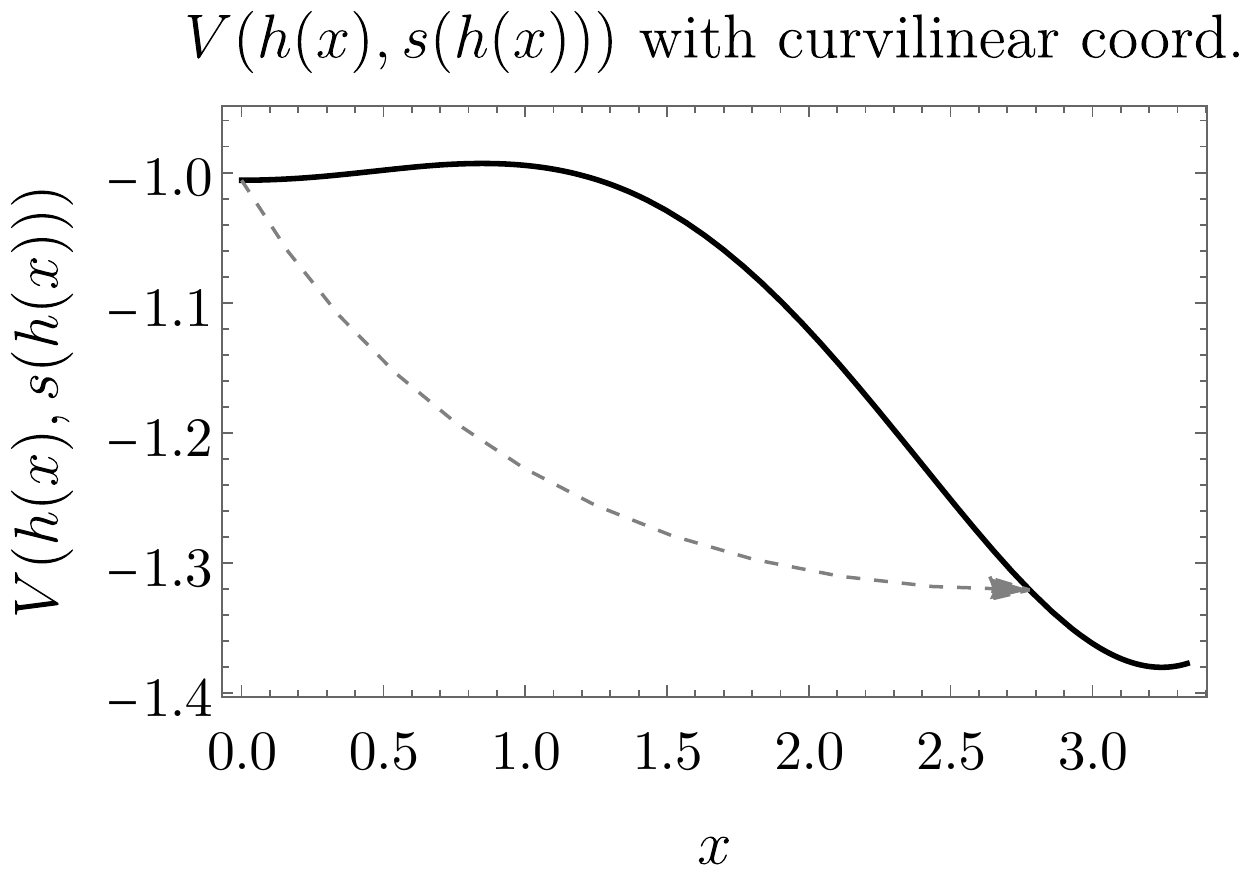}\includegraphics[width=.5\textwidth]{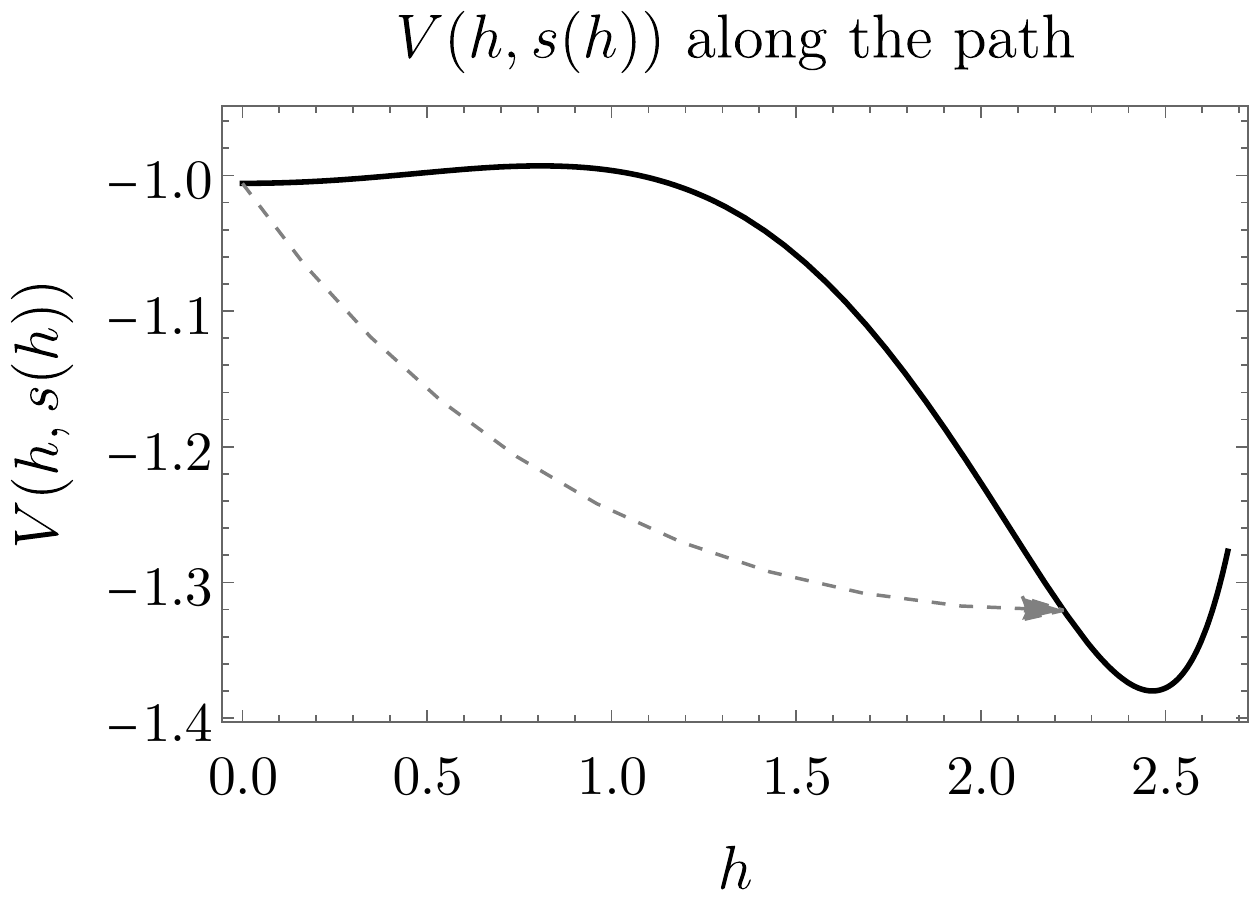}
    \caption{On the left, the potential, along the path, experienced by the field $x$ and the corresponding escape point $x_\star$. On the right the potential, along the path, projected on the $h$ direction.}
    \label{fig:2bounceplot1}
\end{figure}
\bea 
\frac{d^2\vec\phi}{dr^2} + \frac{d-1}{r}\frac{d\vec\phi}{dr} = \vec \nabla V[\vec\phi], \qquad
\lim_{r \to \infty}\vec \phi(r) = 0, \qquad \frac{d\vec\phi}{dr}\bigg|_{r= 0} = 0.
\label{eq:problem_2field}
\eea 
Since in this case an overshoot/undershoot procedure cannot be easily applied, the idea is to reduce the problem to one dimensional tunneling. In order to do so we start guessing the path, $\vec{\phi}_g(x)$, where $x$ is now to be understood as the parameter that measure the distance along the path, \textit{i.e.} the so-called curvilinear abscissa. For the present case, if we parametrize the path in the field space as $(h(t), s(t))=(t, f(t))\equiv(h, s(h))$ it is defined as
\bea 
x(h) = \int^h_{h_{fm}} \sqrt{1+\left(\frac{ds(h')}{dh'}\right)^2} dh',
\eea 
where $h_{fm}$ is the value of the Higgs field in the false minimum. With this choice of field coordinate to parametrize the path, the condition $\big|\frac{d\vec\phi(x)}{dx}\big|^2=1$ is satisfied, and the Euclidean equation of motion in Eq.~(\ref{eq:problem_2field}) can be rewritten along the parallel and the perpendicular direction 
\bea 
\frac{d^2x}{dr^2} + \frac{d-1}{r}\frac{dx}{dr} = \partial_x V[\vec\phi_g(x)],
\nn 
\frac{d^2\vec\phi_g(x)}{dx^2}\bigg(\frac{dx}{dr}\bigg)^2=\vec\nabla_\perp V[\vec\phi_g(x)].
\eea 
Here, we have been able to separate the dynamics along the parallel and perpendicular direction in such a way the first equation defines a new undershoot/overshoot problem, that we solve to obtain the value of the \emph{escape point}, $\vec\phi_0(x_\star)$, and the Euclidean action corresponding to the potential along the path considered $\vec\phi_g$, as in Fig.\ref{fig:2bounceplot1}. On the other hand, the second equation can be seen as a condition that the bounce solution has to satisfy and can be thought as a force field acting on the path, defined as following
\bea 
\vec{N} \equiv \frac{d^2\vec\phi_g(x)}{dx^2}\bigg(\frac{dx}{dr}\bigg)^2-\vec\nabla_\perp V[\vec\phi_g(x)]
\label{eq:force}.
\eea 
The right path will be the one where $\vec{N}$ is vanishing. The algorithm proceeds iteratively: first we guess a path, the straight line connecting the two minima, then we find the bounce solution along this path, we compute the normal force and deform the guessed path according to it. In practice, to define the path at the step $n$, $\vec\phi_n$, we need to solve for the bounce profile for the path at $\vec\phi_{n-1}$, extract the escape point $x_{\star, n-1}$, that is $(h(x_{\star, n-1}), s(x_{\star, n-1}))$ in field space, we then discretize the path in the interval $x \in [0, x_{\star, n-1}]$, creating a series $(\vec\phi_{n-1})_j$, for $j = 1, ... , N$ and a series of values for the normal force $(\vec N_{n-1})_j$. We then shift each point of the discretized path by 
\bea 
(\vec\phi_{n})_j = (\vec\phi_{n-1})_j + \rho(\vec N_{n-1})_j \qquad j = 1, \dots, N.
\eea 
In the end, we fit a path $\vec\phi_{n}$ along the shifted points from $(\vec\phi_{n-1})_j$. The procedure of deformation of the path will produce a series of paths $\vec\phi_i[x]$, over which we compute the Euclidean action according to Eq.\eqref{eq:bounce_action} at each step of the deformation, like in Fig.\ref{fig:2bounceplot2}. The algorithm stops when the difference in the bounce action, $S_3$, between two successive iterations is below some imposed precision. At a definite temperature $T$, we start by identifying the two minima, the false and the true ones
\begin{figure}
    \centering\hspace{-1cm}
    \begin{minipage}{.55\textwidth}
     \includegraphics[scale=.65]{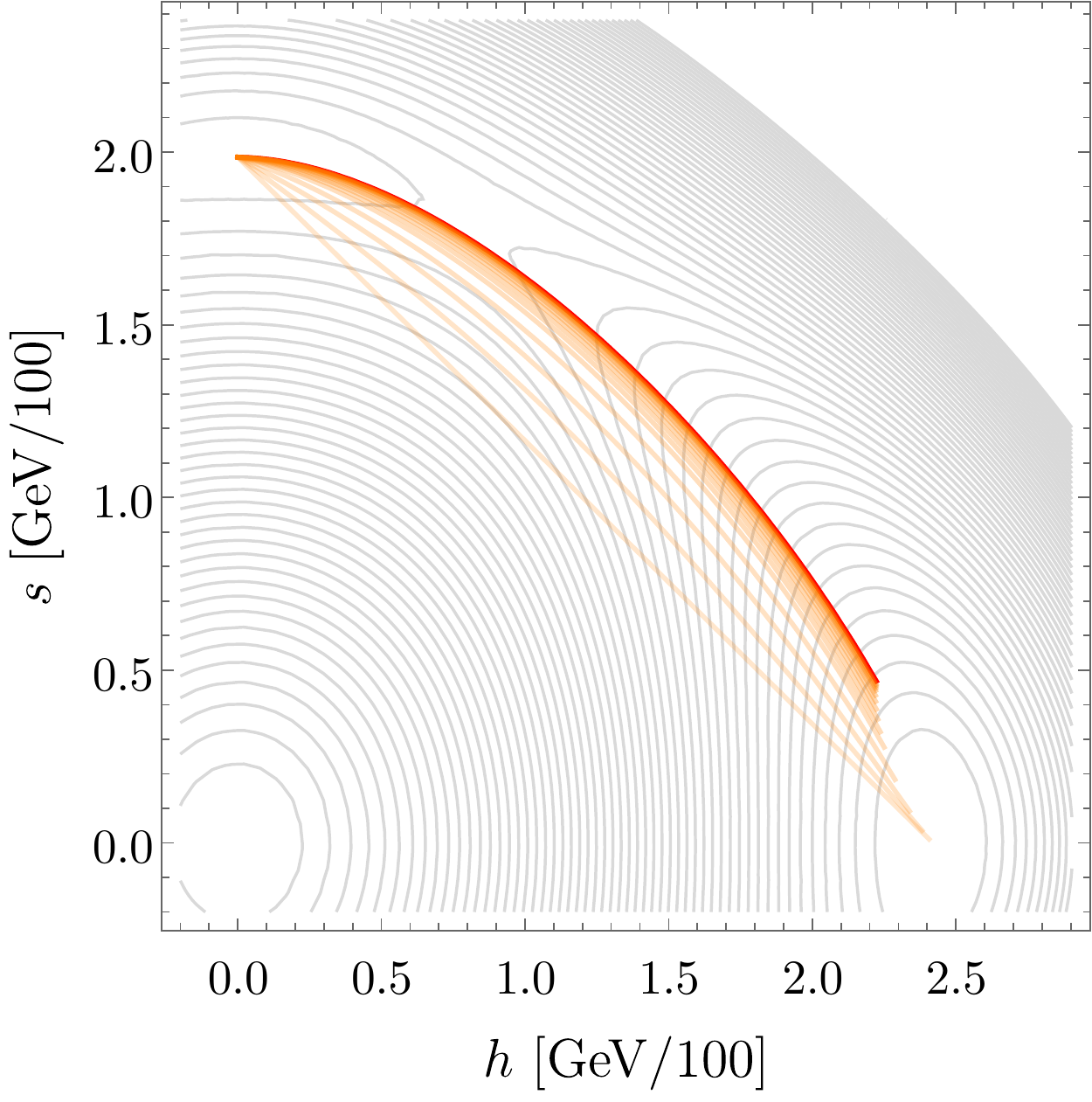}
     \end{minipage}
      \begin{minipage}{.43\textwidth}
     \includegraphics[scale=.55]{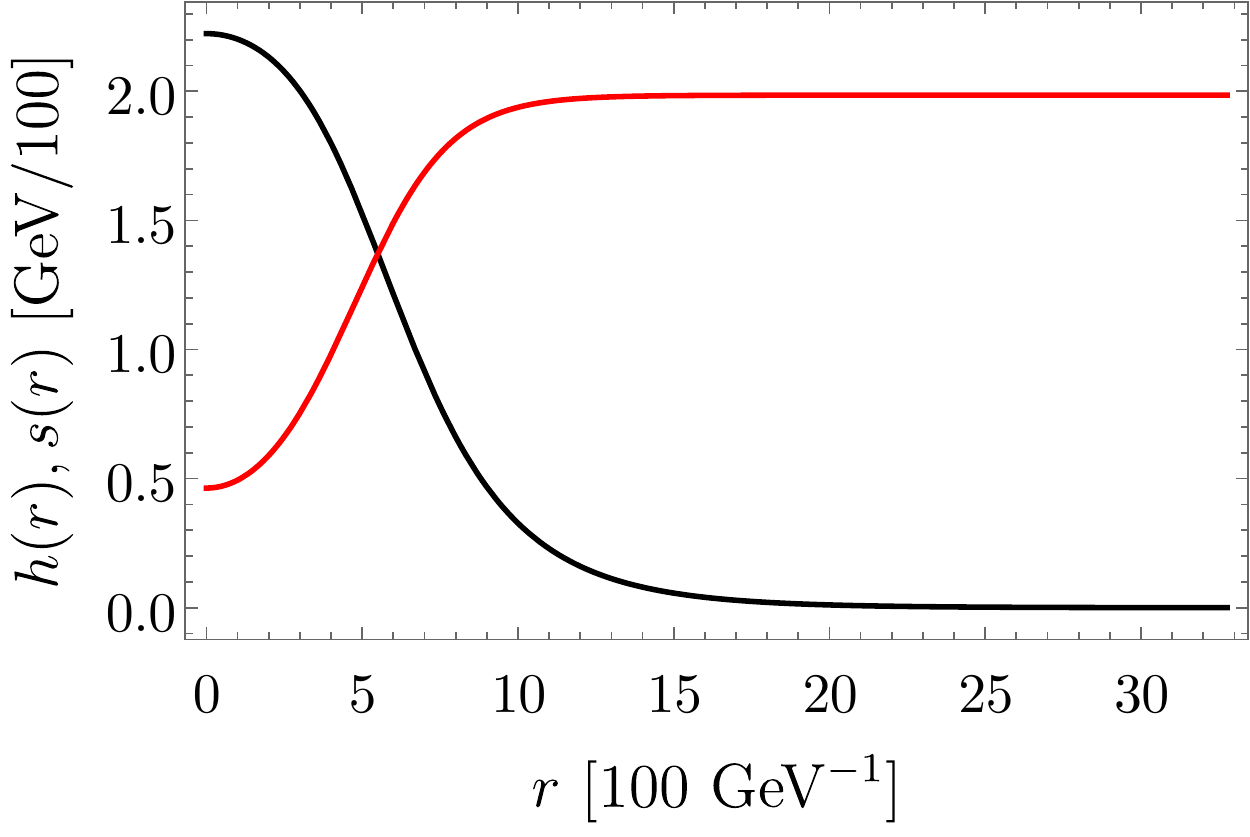}
    \end{minipage}
    
    \caption{\textbf{Left}: iterative procedure for the correct path, starting from the straight line connecting the two minima and then modified according to the field $\vec{N}$. \textbf{Right}: bounce profile of the fields (black for $h(r)$ and red for $s(r)$) on the correct path.}
    \label{fig:2bounceplot2}
\end{figure}
\bea 
(\langle h\rangle,\langle  s\rangle)_{\rm fm} = (v(T), v_s(T)) \to (v_{EW}, 0), 
\eea 
and will keep the false minimum fixed during the whole procedure of deformation. Generally, especially when we have a sizable amount of supercooling, the escape point is just behind the barrier, so the escape point $(v_\star (T_n), v_{s, \star}(T_n))$ will be different from the, zero-temperature, EWSB vacuum, but when the tunneling happens the system will classically roll down towards the global minimum, as we can see from Fig.\ref{fig:2bounceplot2}. We do not track the evolution of the fields profile after the tunneling.

\section{Supplemental numerical results  }
\label{sec:numeric-extra}

\begin{figure}
    \centering
    \includegraphics[width=.65\textwidth]{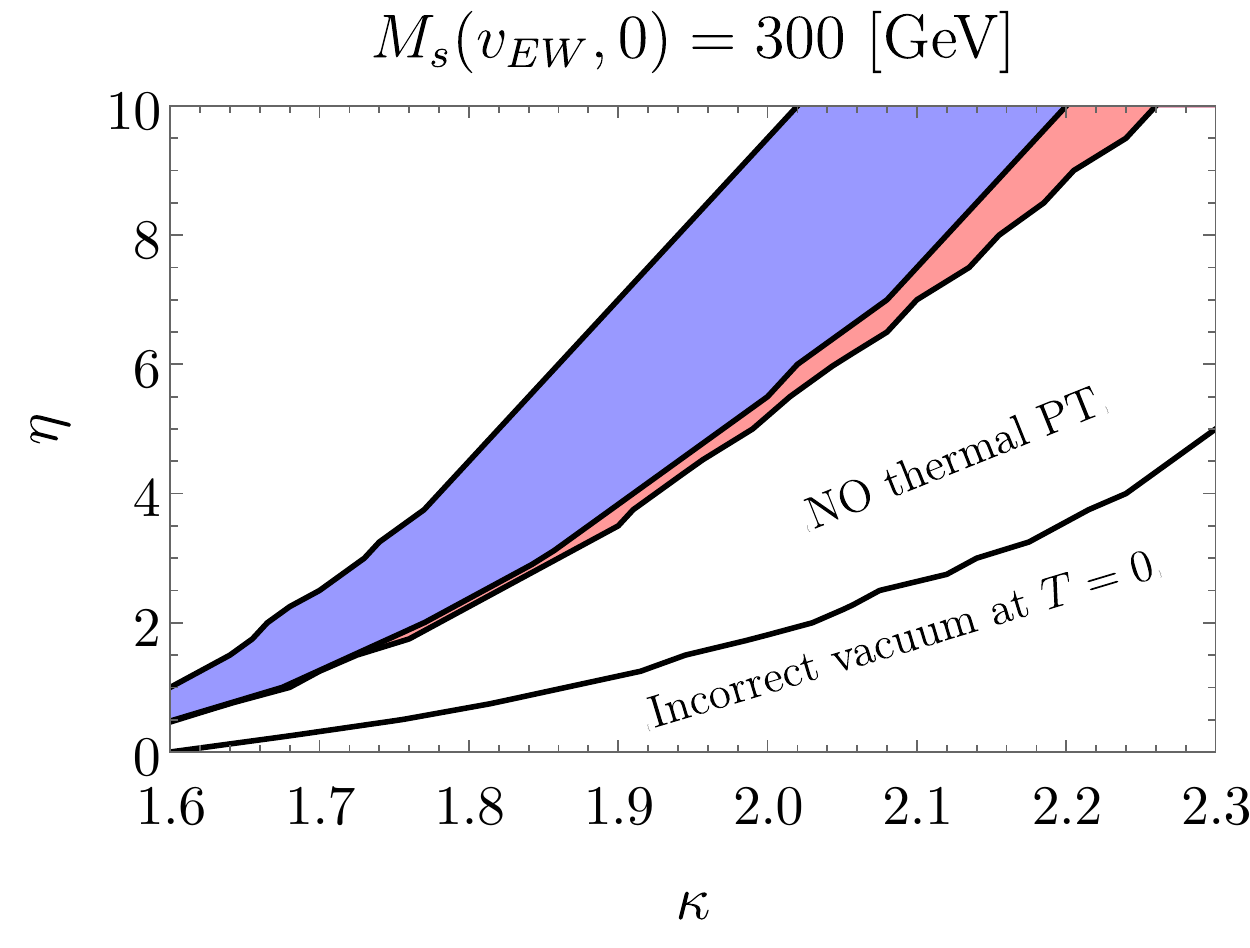}
    \caption{Here is presented the same results found in Fig. 6 of Ref. \cite{Kurup:2017dzf}. It has to be noted that these results are obtained with only the thermal potential and without Daisy resummation, \textit{i.e.} without thermal masses. The relation with our parameters is $(\lambda_s,\ \lambda_{hs})=(\eta,\ 2 \kappa)$ and $M_s(v_{EW}, 0)= 300$ GeV.}
    \label{fig:KPplot}
\end{figure}

 \begin{figure}
 \centering
  \includegraphics[scale=0.54]{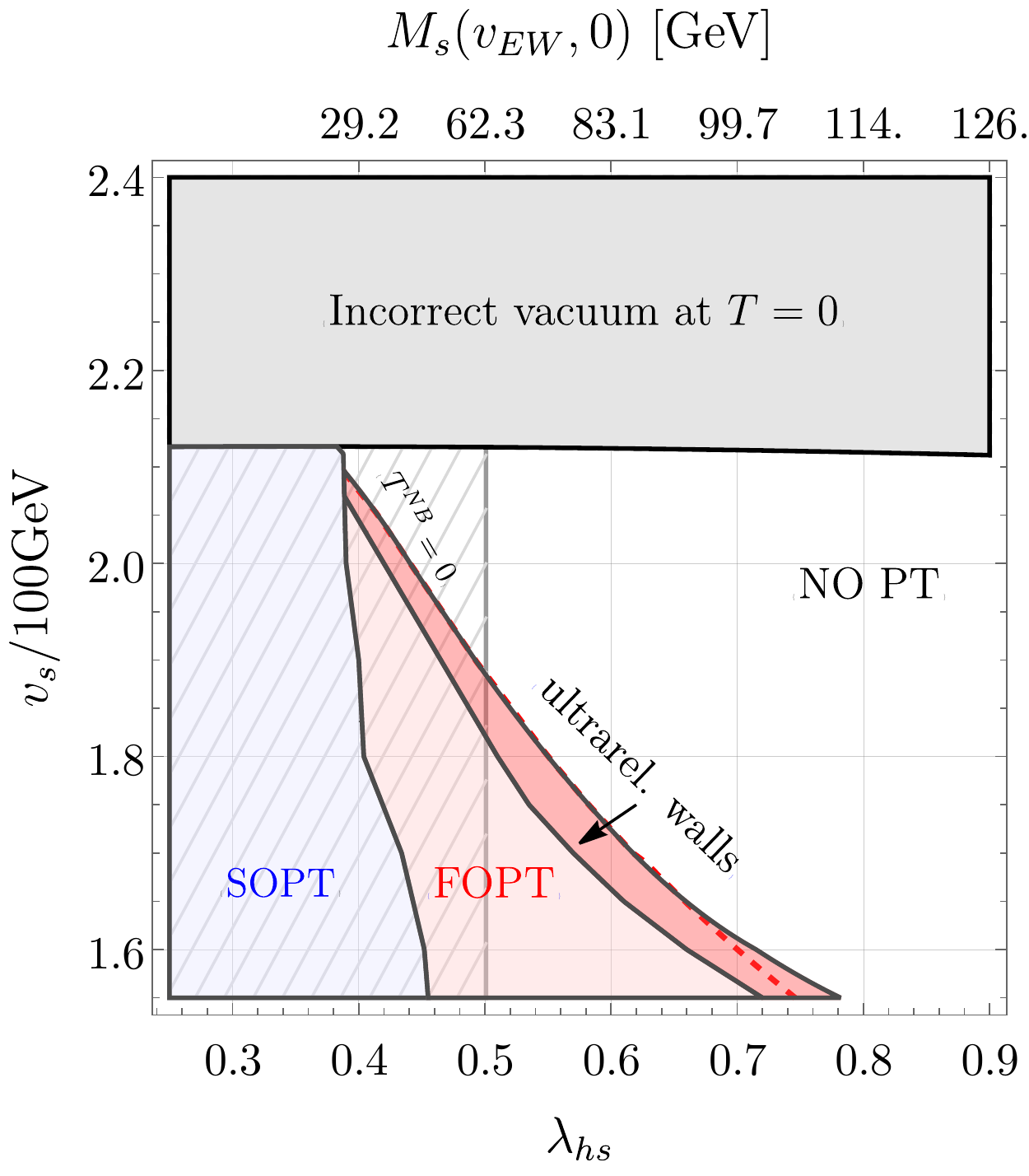} \includegraphics[scale=0.77]{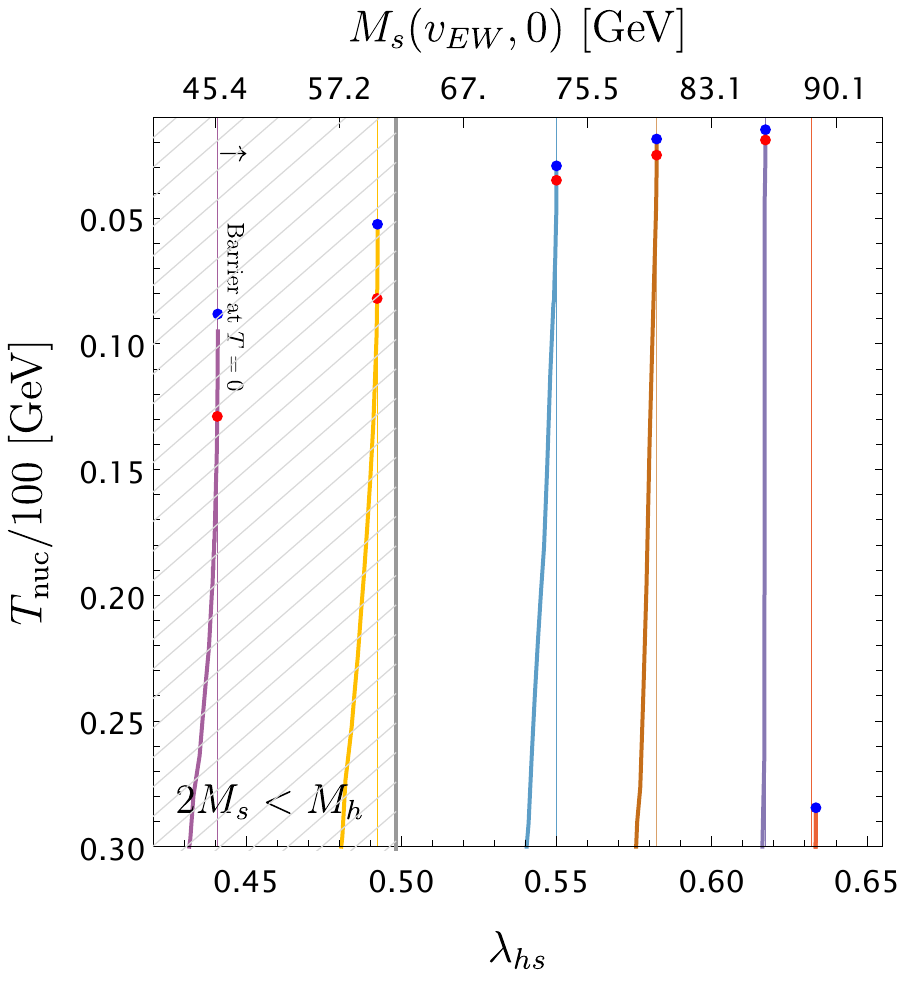}
  \\
     \includegraphics[scale=0.57]{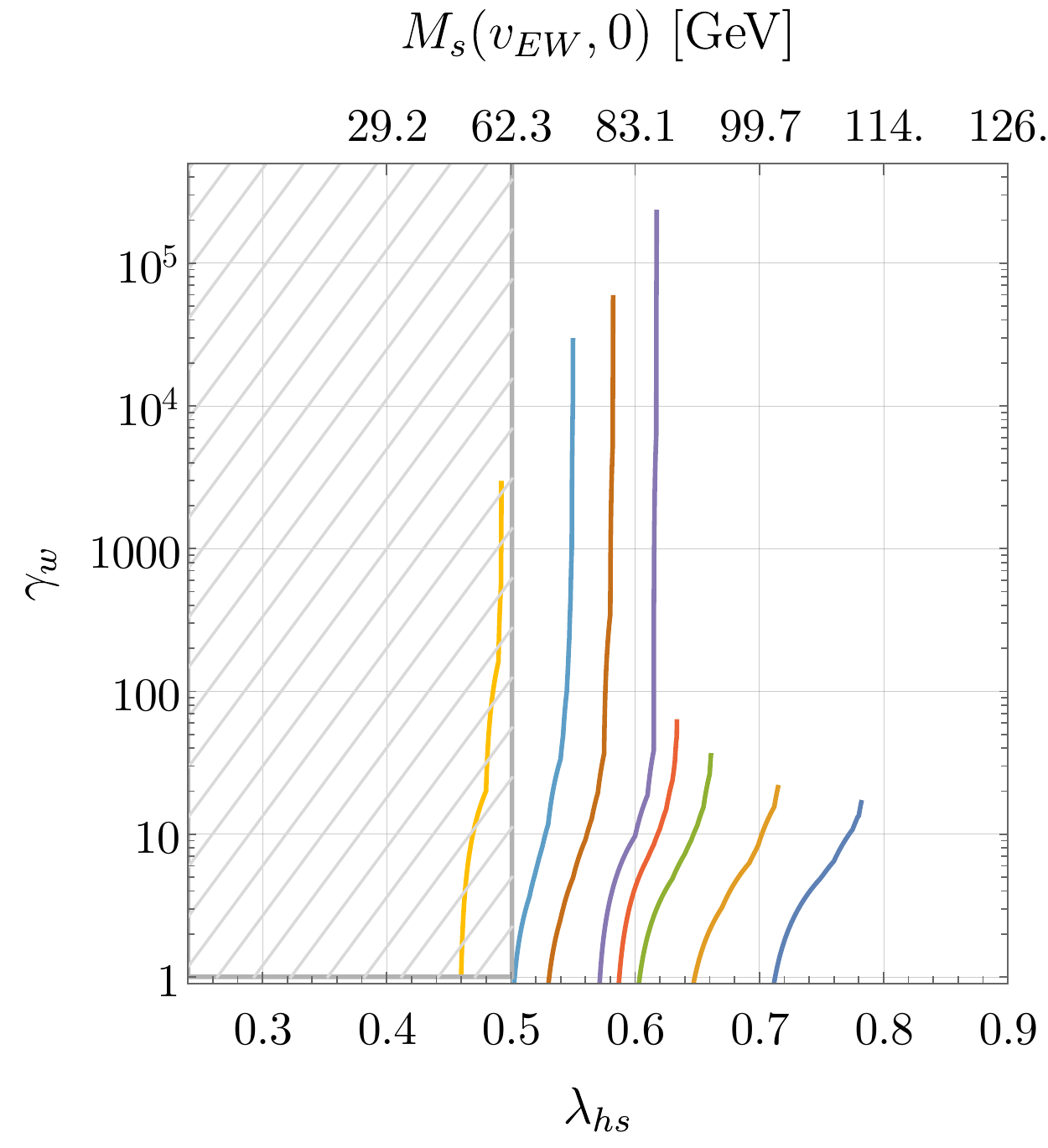}  \includegraphics[scale=0.55]{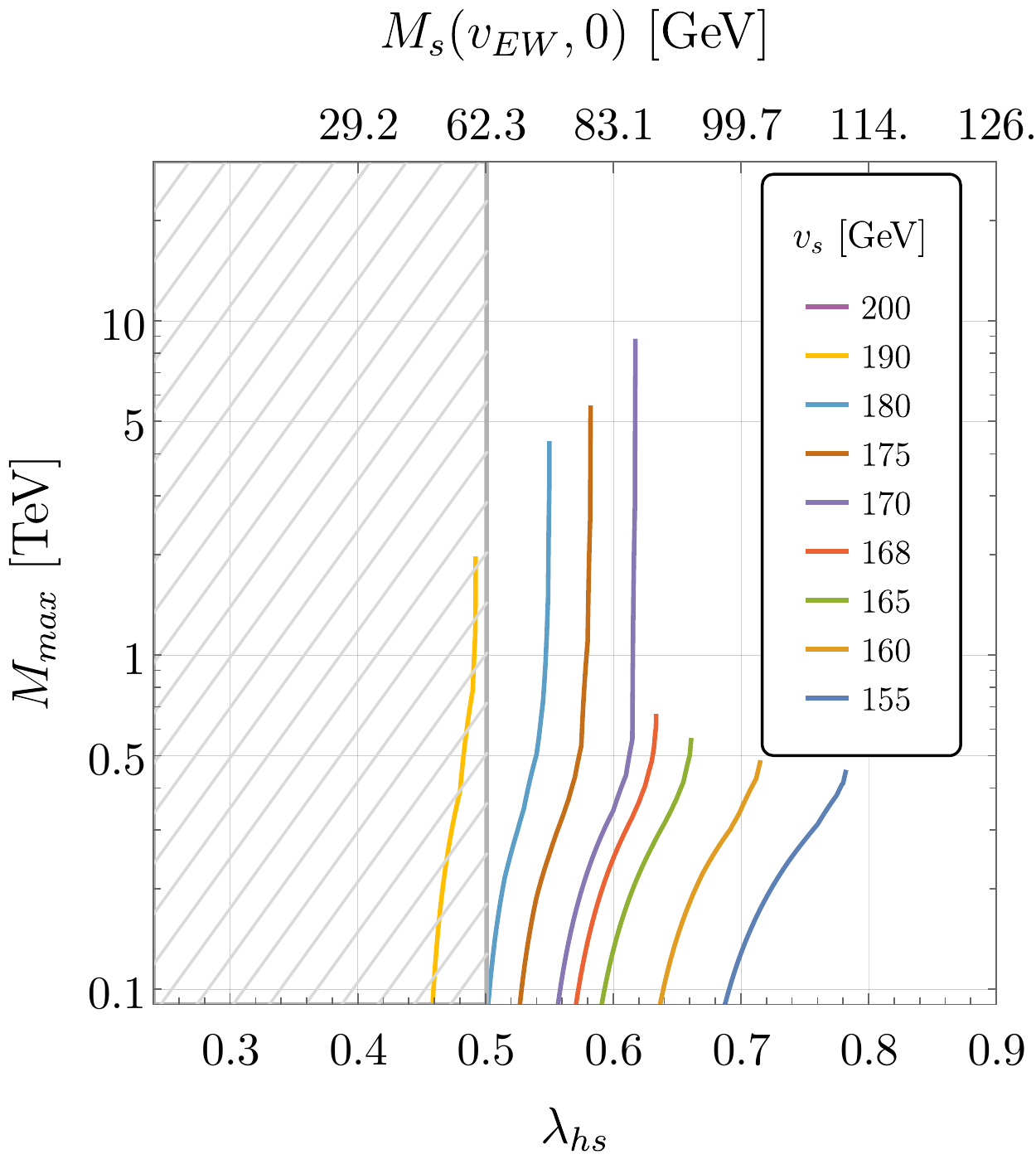}
     \\
     \includegraphics[scale=0.95]{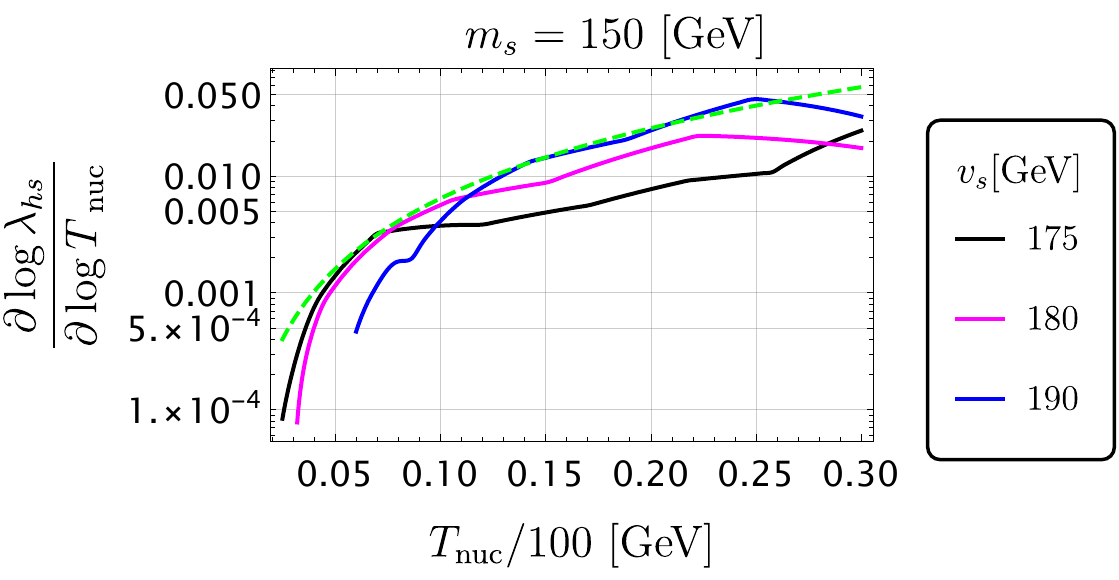}
 \caption{Similar plots than in Fig.\ref{fig:scan} and Fig.\ref{fig:tuning_supercooled} for the value of $m_s = 150$ GeV.}
 \label{fig:ms_150}
 \end{figure}

\begin{figure}
    \centering
     \includegraphics[width=.75\textwidth]{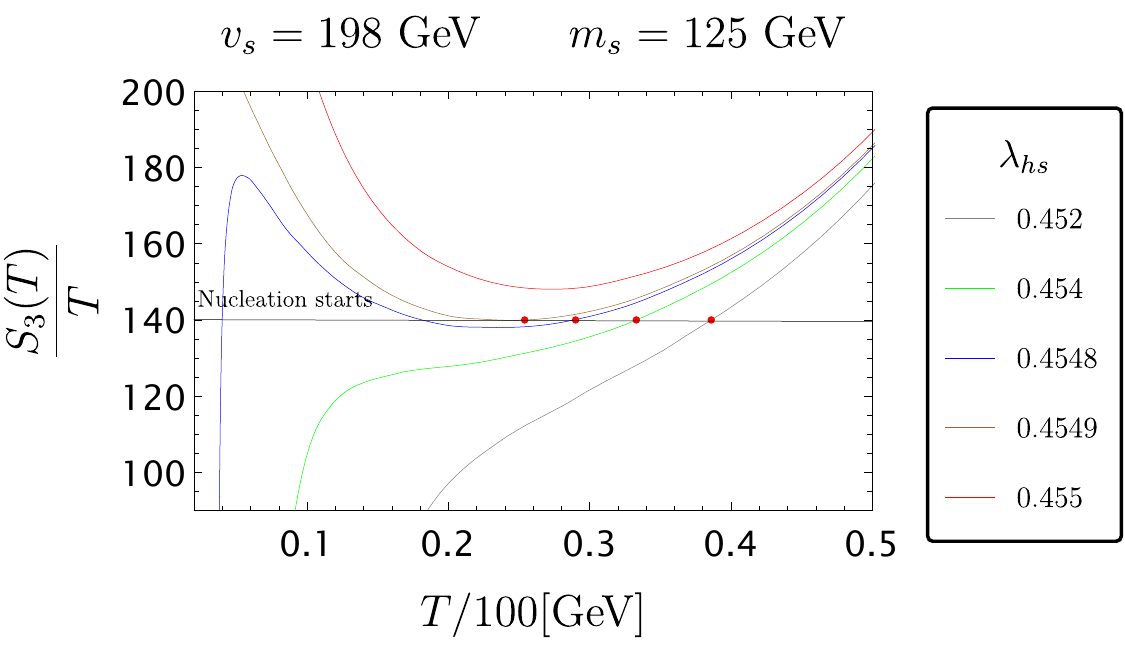}
    \caption{On this plot we show the regime of transition. We observe that several points display a disappearance of the barrier which is typical of the regime of no barrier at $T=0$. However, the nucleation temperature is controlled by the first minimum of $S_3/T$, which is typical of the regime with a barrier at $T=0$.}
    \label{fig:S3Tvs198}
\end{figure}

\begin{table}[]
\footnotesize
\begin{center}
\begin{tabular}{|l|l|l|l|l|l|l|l|l|l|l|l|}
\hline
\multicolumn{7}{|c|}{$m_s=150$ GeV, \quad $v_s=175$ GeV}  
&
 \multicolumn{1}{l|}{}

\\
\hline
\multicolumn{1}{|l|}{$\lambda_{hs}$}  &
\multicolumn{1}{l|}{$\frac{T_{\rm reh}}{100 \text{GeV}}$}  &
\multicolumn{1}{l|}{$\frac{T_{\rm nuc}}{100 \text{GeV}}$}  &
\multicolumn{1}{l|}{$\frac{T_{\rm per}}{100 \text{GeV}}$} &
\multicolumn{1}{l|}{$\gamma_w$} &
\multicolumn{1}{l|}{${\tilde\beta\over H}={(8\pi)^{1/3} \over R_\star H}$}&
\multicolumn{1}{l|}{$m_H^{\rm False}/\text{GeV}$} 
&
\multicolumn{1}{l|}{FM$_0$} 
\\
\hline
\multicolumn{1}{|l|}{$0.53$}  &
\multicolumn{1}{l|}{$0.616$}  &
\multicolumn{1}{l|}{$0.608$}  &
\multicolumn{1}{l|}{$0.592$} &
\multicolumn{1}{l|}{$-$}  &
\multicolumn{1}{l|}{$580$}  &
\multicolumn{1}{l|}{$17.4$} 
&
\multicolumn{1}{l|}{No} 
\\
\hline
\multicolumn{1}{|l|}{$0.54$}  & 
\multicolumn{1}{l|}{$0.581$}  &
\multicolumn{1}{l|}{$0.570$}  &
\multicolumn{1}{l|}{$0.552$} &
\multicolumn{1}{l|}{$3$}  &
\multicolumn{1}{l|}{$420$}  &
\multicolumn{1}{l|}{$17.8$} 
&
\multicolumn{1}{l|}{No} 
\\
\hline
\multicolumn{1}{|l|}{$0.56$}  & 
\multicolumn{1}{l|}{$0.492$}  &
\multicolumn{1}{l|}{$0.470$}  &
\multicolumn{1}{l|}{$0.444$} &
\multicolumn{1}{l|}{$9$}  &
\multicolumn{1}{l|}{$228$}  &
\multicolumn{1}{l|}{$17.7$}
&
\multicolumn{1}{l|}{No} 
\\
\hline
\multicolumn{1}{|l|}{$0.58$}  &
\multicolumn{1}{l|}{$0.329$}  &
\multicolumn{1}{l|}{$0.141$}  &
\multicolumn{1}{l|}{$0.130$} &
\multicolumn{1}{l|}{$348$}  &
\multicolumn{1}{l|}{$160$}  &
\multicolumn{1}{l|}{$4.7 $} 
&
\multicolumn{1}{l|}{No} 
\\
\hline
\multicolumn{1}{|l|}{$0.582$}  &
\multicolumn{1}{l|}{$0.327$}  &
\multicolumn{1}{l|}{$0.051$}  &
\multicolumn{1}{l|}{$0.0508$} &
\multicolumn{1}{l|}{$5.3 \cdot 10^3$}  &
\multicolumn{1}{l|}{$9.5 \cdot 10^3$}  &
\multicolumn{1}{l|}{$0.4$} 
&
\multicolumn{1}{l|}{No} 
\\
\hline
\multicolumn{1}{|l|}{$0.582262$}  &
\multicolumn{1}{l|}{$0.327$}  &
\multicolumn{1}{l|}{$0.025$}  &
\multicolumn{1}{l|}{$0.0236$} &
\multicolumn{1}{l|}{$3.7 \cdot 10^4$}  &
\multicolumn{1}{l|}{$194$}  &
\multicolumn{1}{l|}{$8.2$}
&
\multicolumn{1}{l|}{Yes} 
\\
\hline
\multicolumn{1}{|l|}{$0.582264$}  &
\multicolumn{1}{l|}{$0.327$}  &
\multicolumn{1}{l|}{$0.024$}  &
\multicolumn{1}{l|}{$0.0219$} &
\multicolumn{1}{l|}{$4.3 \cdot 10^4$}  &
\multicolumn{1}{l|}{$130$}  &
\multicolumn{1}{l|}{$8.3$} 
&
\multicolumn{1}{l|}{Yes} 
\\
\hline
\multicolumn{1}{|l|}{$0.582266$}  &
\multicolumn{1}{l|}{$0.327$}  &
\multicolumn{1}{l|}{$0.021$}  &
\multicolumn{1}{l|}{$0.017$} &
\multicolumn{1}{l|}{$5.7 \cdot 10^4$}  &
\multicolumn{1}{l|}{$24$}  &
\multicolumn{1}{l|}{$8.5$}
&
\multicolumn{1}{l|}{Yes}
\\
\hline
\end{tabular}
\end{center}
\caption{Same as Tables \ref{benchmarkpoint} and \ref{benchmarkpointbis}, but with $m_s=150 \ {\rm GeV}$ and $v_s=175$ GeV.}
\label{benchmarkpointthird}
\end{table}

In this Appendix, we present all our supplemental numerical results. First, tough we focused mostly on a more weakly coupled part of the parameter space, we would like to compare our findings with the ones in the Ref.\cite{Kurup:2017dzf} and argue that we observed only small changes, due to the inclusion of loop-corrections and Daisy resummation. On Fig.\ref{fig:KPplot} we make a reproduction of the scan of the Fig.6 of \cite{Kurup:2017dzf} using our potential and emphasize the close similarities.
The relations between the parameters $\kappa, \eta$ and the couplings in the Eq.(\ref{eq:scalar_pot})  is as follows
\bea
\kappa_{\hbox{\cite{Kurup:2017dzf}} }= \frac{\lambda_{hs}}{2},~~~
\eta_{\hbox{\cite{Kurup:2017dzf} }}=\lambda_{s}.
\eea

In the main text, we studied specifically the case where the parameter $m_s = 125$ GeV, we observed that for this 
value, the region of deep supercooling displayed small masses of the singlet in the real vacuum, being 
on the verge of detection due to $h\to ss$ at $M_s \lesssim 62$ GeV. We also concluded in section \ref{sec:num_results} that this region was closing around $M_s \approx 75$ GeV. 
We could wonder if this conclusion would change if we modify the value of the 
parameter $m_s$, and if so in which direction. On Fig.\ref{fig:ms_150} we show similar plots than in Fig.\ref{fig:scan} and \ref{fig:tuning_supercooled} for the case of $m_s = 150$ GeV. Thus, increase 
the value of $m_s$ pushes the deep supercooling region to $M_s \approx 90$ GeV, at the price of increasing the portal coupling $\lambda_{hs}$. 
However, we can observe on the last plot of Fig.\ref{fig:ms_150} that the typical tuning remains roughly the same and that we can still trust our naive $(T_{\rm nuc}/m_h)^2 $ for an order-of-magnitude estimate of the tuning. 
   
On the other hand, we also observed that decreasing the parameter $m_s$ to $\approx 100$ GeV was pushing all the deep supercooling region inside $M_s \lesssim 62$ GeV, which is thus strongly disfavored by colliders. We hope that this trend can be extrapolated to larger values of $m_s$, until we hit perturbativity bounds for $\lambda_{hs}$.

We could also wonder what happens at the \emph{upper} boundary of the deep supercooling region, as we have observed on Fig.\ref{fig:scan} a sharp decrease in the supercooling allowed around $v_s \lesssim 200$ GeV (for $m_s = 125$ GeV). This transition regime can be understood if we plot the explicit $S_3/T$ functions on Fig.\ref{fig:S3Tvs198}. Comparing the plot in Fig.\ref{fig:S3Tvs198} with the one in Fig.\ref{fig:S3curves}, we see that as we decrease $v_s$, the full pattern of $S_3/T$ is shifted toward smaller values. At some critical point around $v_s\approx 200$ GeV, the nucleation becomes controlled by the first minimum in the function $S_3/T$ and not by the disappearance of the barrier. This largely suppresses the possibility for large supercooling. 

Finally, in Table \ref{benchmarkpointthird}, we provide the value of the velocity, reheating and nucleation temperature for $m_s = 150$ GeV and $v_s = 175$ GeV that was used in Fig.\ref{fig:DM_plot}. 

\section{The coefficient of NLO pressure}
\label{sec:sum}
In this appendix we will review the calculation  of the friction coefficient for the NLO pressure for EW phase transition. We will follow closely the discussion in \cite{Gouttenoire:2021kjv} 
and report the quantity
\bea
\bigg[\sum_{abc}\nu_a g_a \beta_c C_{abc}\bigg] 
\eea
where $\nu_a= 1 (3/4)$ for $a$ a boson (fermion), $\beta_c\equiv \frac{M_c}{M_Z}$and $C_{abc}$ stands for the couplings appearing in the vertex. 
Normalization of the $C_{abc}$ coefficient is
the following:
for a chiral fermion coupled to the vector field the amplitude for the process $\psi\to \psi A_{soft}$ is equal to 
\bea
\label{eq:splitting-ferm}
g_\psi\bar \psi_L A_\mu \psi_L \Rightarrow ~~~C_{\psi \psi A}=\frac{g_\psi^2}{4\pi \alpha_{em}}.
\eea
Where in the relation Eq.(\ref{eq:splitting-ferm}) is written only for one polarization of the vector field. Similarly for the scalar field 
\bea
\label{eq:splitting-scalar}
ig_\phi (\phi^*\d_\mu \phi-\d_\mu\phi^* \phi)A^\mu  \Rightarrow ~~~C_{\phi \phi A}=\frac{g_\psi^2}{4\pi \alpha_{em}},
\eea
and the vector fields
\bea
g_{V^1 V^2 A^3}\l(V^1_{\mu\nu}V^2_\mu A_\nu+V^2_{\mu\nu}A_\mu V^1_\nu+A_{\mu\nu}V^1_\mu V^2_\nu\r)\Rightarrow C_{V_1 V_2 A}=\frac{g_{V^1 V^2A}^2}{4\pi \alpha_{em}},
\eea
where in all of these formulas $C_{abc}$ coefficients are reported only for one polarization of the vector fields both in the initial and the final states. Summing all of these contributions and taking care of the multiplicities of the initial and final states  we  find
\bea
\label{eq:sumC}
\bigg[\sum_{abc}\nu_a g_a \beta_c C_{abc}\bigg] =  {2}\l(\frac{7+14 c_w}{s_w^2}-\frac{7-15 s_w^2}{c_w^2}\r)\simeq 157.
\eea
For the interested reader we refer 
various individual contributions in the 
Table~\ref{tab:contrNLO}. 
If in the false vacuum the Higgs doublet ${\cal H}$ is too heavy its contribution must be 
subtracted and the sum in Eq.(\ref{eq:sumC}) reduces to 
\bea
\bigg[\sum_{abc}\nu_a g_a \beta_c C_{abc}\bigg]_{\rm No ~Higgs} 
\simeq 145.\eea
At last we would like to emphasize that these results  include 
only the transverse polarizations of the vector 
fields.  NLO effects of the longitudinal polarizations are not fully established and we omit them here, however these cannot qualitatively modify the results.
\begin{table}[]
\begin{center}
\begin{tabular}{lllll}
\hline
\multicolumn{1}{l|}{Process} & \multicolumn{1}{l|}{$\sum g_a C_{abc}$} & \multicolumn{1}{l|}{$\beta$} & \multicolumn{1}{l|}{$\nu$} &  
Result\\ \hline
\multicolumn{1}{l}{$\psi \to W^\pm \psi$} & \multicolumn{1}{l}{~~$\frac{24}{s_w^2}$} & \multicolumn{1}{l}{$c_w$} & \multicolumn{1}{l}{$\frac{3}{4}$} & $\frac{18 c_w}{s_w^2}$ \\ 
\\
\multicolumn{1}{l}{$\psi \to Z \psi$} & \multicolumn{1}{l}
{~~$\frac{4(3-6 s_w^2+8s_w^4)}{s_w^2 c_w^2} $} & 
\multicolumn{1}{l}{1} & \multicolumn{1}{l}{$\frac{3}{4}$} 
&$\frac{3(3-6 s_w^2+8s_w^4)}{s_w^2 c_w^2}$  \\\\ 
\multicolumn{1}{l}{${\cal H} \to W {\cal H}$} & \multicolumn{1}{l}
{~~$\frac{2}{s_w^2} $} & 
\multicolumn{1}{l}{$c_w$} & \multicolumn{1}{l}{$1$} 
&$\frac{2 c_w}{s_w^2}$  \\ \\

\multicolumn{1}{l}{${\cal H} \to Z {\cal H}$} & \multicolumn{1}{l}
{~~$\frac{1 -2s_w^2c_w^2 + c_w^4+s_w^4}{2s_w^2 c_w^2}$} & 
\multicolumn{1}{l}{$1$} & \multicolumn{1}{l}{$1$} 
&$\frac{1 -2s_w^2c_w^2 + c_w^4+s_w^4}{2s_w^2 c_w^2}$  \\ \\

\multicolumn{1}{l}{$ A\to W_{soft} W$ $\&$ $ W\to W_{soft} A$ } & \multicolumn{1}{l}
{~~$8$} & 
\multicolumn{1}{l}{$c_w$} & \multicolumn{1}{l}{$1$} 
&$8 c_w$  \\ \\

\multicolumn{1}{l}{$ Z\to W_{soft} W$ $\&$ $ W\to W_{soft} Z$ } & \multicolumn{1}{l}
{~~$\frac{8c_w^2}{s_w^2}$} & 
\multicolumn{1}{l}{$c_w$} & \multicolumn{1}{l}{$1$} 
&$\frac{8 c_w^3}{s_w^2}$  \\ \\

\multicolumn{1}{l}{$ W\to Z_{soft} W$ } & \multicolumn{1}{l}
{~~$\frac{4c_w^2}{s_w^2}$ } & 
\multicolumn{1}{l}{$1$} & \multicolumn{1}{l}{$1$} 
&$\frac{4 c_w^2}{s_w^2}$  \\ \\
\hline
\multicolumn{1}{l}{Total:} &&&&\multicolumn{1}{l}{ $ 2\l(\frac{7+14 c_w}{s_w^2}-\frac{7-15 s_w^2}{c_w^2}\r)\simeq 157 $}\\
\hline
\end{tabular}
\end{center}
\caption{Different contributions to the sum in Eq.\eqref{eq:LLpressure}. 
\label{tab:contrNLO}}
\end{table}

\section{Domain wall collapse}
\label{app:DW}
Our main discussion was focused on the two step phase transition 
$(0,0)\to (0,v_s)\to (v_h,0)$ where the first phase transition is $Z_2$ 
breaking. Obviously during such a 
phase transition domain walls will be formed which can drastically modify the cosmology of the system. We can 
avoid the stable domain walls if we assume some small $Z_2$ breaking, 
however in this case the question rises about the timescale for the  stability of the domain walls. This 
is particularly important since recently it was shown 
\cite{Blasi:2022woz} that for singlet extension of the SM the domain walls 
(if still present) will become seeds 
of the secondary phase transition $(0,v_s)\to (v_h,0)$ and will 
dominate the phase transition. We will follow closely the discussion in the section \ref{sec:qualitative} using only the tree level potential and the thermal corrections to the masses. Then the $Z_2$ breaking phase transition will occur at the temperatures
\bea
T_{Z_2}\simeq v_s\l(\frac{12 }{3+4\lambda_{hs}\frac{v_s^2}{m_s^2}}\r)^{1/2}\simeq 200-300 ~~\rm GeV,
\eea
which is a temperature of the domain wall formation. The domain wall mediated transition
 will happen at the temperature $T_w$ which is found to be  order one different
from $T_{Z_2}$.
The exact mechanism of the transition depends on the values of the couplings and can proceed either with the classical rolling or 2D bounces localized on the domain wall.
The temperature when the classical rolling can start  is reported in
Ref.\cite{Blasi:2022woz} and is equal to
\bea
T_w^{\rm rolling}\simeq T_{Z_2}\l[\frac{4 m_h^2+ m_s^2\bigg(1-\sqrt{1+\frac{8 v_s^2 \lambda_{hs}}{m_s^2}}\bigg)}{8 \Pi_h(T_{Z_2})+m_s^2\bigg(1-\sqrt{1+\frac{8 v_s^2 \lambda_{hs}}{m_s^2}}\bigg)}\r]^{1/2}. 
\eea
The nucleation temperature $(T_w^{2D})$ of $2 D$ bounces should be found numerically (Ref.\cite{Blasi:2022woz}) however it will be obviously smaller than $T_{\rm crit}$ (of $(v_s,0)\to (v_h,0)$ phase transition). At this point we can safely ignore the seeded phase transition effects if all of the domain walls annihilate in the interval of temperatures
\bea
\l[T_*,T_{Z_2}\r],~~~T_*< T_{\rm crit},
\eea
where $T_*$ is the temperature when the seeded phase transition will be completed and it is obviously less than $T_{\rm crit}$ of EW phase transition. Let us estimate how strong should be the bias $\Delta V$ between the potential energies of the two minima of $Z_2$ potential so that all of the walls can disappear. For these estimates it is sufficient to assume that  there is order one difference between $T_*$ and $T_{Z_2}$, which is generically the case.
The critical radius (above which) areas with true vacuum will start to expand is roughly
\bea
R_c\sim \frac{\sigma }{\Delta V},
\eea
where $\sigma$ is surface energy density of the wall.
So the domain walls will exist on the time scale of
\bea
\Delta t_w \sim \frac{R_c}{u}\sim \frac{\sigma }{u \Delta V},
\eea
where $u$ is velocity of the wall motion.
The change of the temperature during the wall annihilation will be roughly
\bea
\Delta T\sim T H \Delta t_w. 
\eea
 So that if $\Delta t_w H\ll 1 \Rightarrow \Delta T\ll T $ the wall annihilation happens almost instantaneously. Assuming $\sigma \sim T_{\rm crit}^3$ and $H\sim \frac{T_{\rm crit}^2}{M_p}$ we get
 \bea
 \frac{\Delta V}{T_{\rm crit}^4}\gg \frac{T_{\rm crit}}{ u M_{pl}}.
 \eea
Balancing the pressure against the friction forces $\Delta V\sim u T_{\rm crit}^4$ we can estimate the velocity and then the condition for the quick wall annihilation becomes
\bea
 \frac{\Delta V}{T_{\rm crit}^4}\gg \sqrt{\frac{T_{\rm crit}}{  M_{pl}}}\sim 10^{-8},
 \eea
which is not restrictive at all.

\bibliographystyle{JHEP}
{\footnotesize
\providecommand{\href}[2]{#2}\begingroup\raggedright\endgroup}
\end{document}